\documentclass[a4paper,11pt]{article}

\usepackage{jheppub}

\usepackage{amsmath}
\usepackage{mathtools}

\usepackage[scr=boondoxo]{mathalfa}

\usepackage{undertilde}
\usepackage{relsize}
\usepackage{arydshln}
\usepackage{cleveref}
\usepackage{slashed}
\usepackage{pdflscape}
\usepackage{dashbox}
\usepackage{afterpage}
\usepackage{pifont}
\usepackage{tabularx}
\newcolumntype{Y}{>{\centering\arraybackslash}X}

\usepackage{booktabs} 
\usepackage{bbm}
\newcommand\id{\ensuremath{\mathbbm{1}}} 

\newcommand\vev[1]{$\left\langle #1 \right\rangle$}
\newcommand\mvev[1]{\left\langle #1 \right\rangle}

\newcommand\mphant[2]{#1^{\vphantom{#2}}}
\DeclareRobustCommand{\rchi}{{\mathpalette\irchi\relax}}
\newcommand{\irchi}[2]{\raisebox{\depth}{$#1\chi$}} 
\newcommand\mand{\text{ and }}
\newcommand{\mydots}{\ifmmode\mathinner{\ldotp\kern-0.2em\ldotp\kern-0.2em\ldotp}\else.\kern-0.13em.\kern-0.13em.\fi}

\definecolor{ShamrockGreen}{rgb}{0.0, 0.62, 0.38}

\title{\boldmath Lowering the scale of Pati-Salam breaking through seesaw mixing}
\usepackage{placeins}

\author[a,b]{Matthew J. Dolan}
\author[a,c]{ Tomasz P. Dutka}
\author[a,b]{ Raymond R. Volkas}

\affiliation[a]{ARC Centre of Excellence for Particle Physics at the Terascale, School of Physics,
The University of Melbourne, Victoria 3010, Australia}
\affiliation[b]{ARC Centre of Excellence for Dark Matter Particle Physics, School of Physics,
The University of Melbourne, Victoria 3010, Australia}
\affiliation[c]{Corresponding Author}
\emailAdd{matthew.dolan@unimelb.edu.au}
\emailAdd{tdutka@student.unimelb.edu.au}
\emailAdd{raymondv@unimelb.edu.au}

\abstract{We analyse the experimental limits on the breaking scale of Pati-Salam extensions of the Standard Model.  These arise from the experimental limits on rare-meson decay processes mediated at tree-level by the vector leptoquark in the model. This leptoquark ordinarily couples to to both left- and right-handed SM fermions and therefore the meson decays do not experience a helicity suppression. We find that the current limits vary from $\mathcal{O}(80-2500)$ TeV depending on the choice of matrix structure appearing in the relevant three-generational charged-current interactions. 
We extensively analyse scenarios where additional fermionic degrees of freedom are introduced, transforming as complete Pati-Salam multiplets. These can lower the scales of Pati-Salam breaking through mass-mixing within the charged-lepton and down-quark sectors, leading to a helicity suppression of the meson decay widths which constrain Pati-Salam breaking. 
We find four multiplets with varying degrees of viability for this purpose: an $SU(2)_{L/R}$ bidoublet, a pair of $SU(4)$ decuplets and either an $SU(2)_L$ or $SU(2)_R$ triplet all of which contain heavy exotic versions of the SM charged leptons. We find that the Pati-Salam limits can be as low as $\mathcal{O}(5-150)$ TeV with the addition of these four multiplets. We also  identify an interesting possible connection between the smallness of the neutrino masses and a helicity suppression of the Pati-Salam limits for three of the four multiplets.}

\begin{document} 
\maketitle
\flushbottom
\parskip 5mm
\section{Introduction}
\label{sec:intro}

The quark-lepton unifying Pati-Salam (PS) gauge symmetry~\cite{Pati:1974yy,Pati:1983zp} is an interesting modification of the Standard Model (SM) for a number of reasons. For example, it requires the introduction of a right-handed neutrino state and therefore can incorporate neutrino mass in a number of ways. It unifies the now six seemingly disparate multiplets in each generation of the SM into two which, in the minimal variant of PS, only leads to two Yukawa couplings. It also appears as a subgroup of a number of possible grand unified theories (GUTs). An important property of the model is that it unifies quarks and leptons of the same $SU(2)$ isospin. 

Although the PS symmetry is usually considered to be broken at high scales, the theory naturally can accommodate a conserved global baryon number.  Therefore, and unlike many models of gauge coupling unification, the PS breaking can occur at scales only a few orders of magnitude above the electroweak scale. Low-scale and high-scale variants of PS differ in very few ways, traditionally through a modification of the scalar sector and the inclusion of fermion singlets. However as with all models which unify the SM multiplets, the PS symmetry in its minimal form necessarily predicts two important mass equalities not observed experimentally:
\begin{equation}
m_\nu^{\text{Dirac}} = m_u \qquad\text{ and }\qquad m_d = m_e.
\label{eq:massrelations}
\end{equation}
The scalar content required in high-scale PS models to break the PS symmetry naturally leads to a seesaw\footnote{For clarity, the requirement for a seesaw in a mixing matrix is that the singular values of the blocks making up the matrix satisfy $\sigma_i(m_X) < \sigma_1(m_Y)$ for all $i$, where $m_Y$ and $m_X$ are the dominant and non-dominant block(s) respectively and we use the usual notation that $\sigma_i(\dots)$ corresponds to the singular values of given matrix sorted into ascending order.} within the neutrino sector and not the up-quark sector, breaking the first mass relation; see e.g.~\cite{Mohapatra:1986uf}. Low-scale variants can be modified very simply in order to achieve a similar effect, although this requires extending the matter sector with, at a minimum, fermionic singlets. The second mass equality $m_d = m_e$ between the down-isospin partners also needs to be modified to produce a realistic theory

Any explanation of the broken mass degeneracy in the down-isospin sector requires a modification of the particle content of the theory. By far the most commonly considered modification is to introduce an additional scalar whose couplings to the down-quarks and charged-leptons differs by group theoretic factors; see e.g.~\cite{Pati:1983zp}. This introduces enough free parameters such that all the masses of SM can arise without issue. An alternative idea, which we pursue further in this work, was first proposed in~\cite{Foot:1997pb,Foot:1999wv} where additional fermionic states are introduced which mix with the charged-leptons inducing additional seesaw mixing. This similarly allows for a viable mass spectrum for all particles but additionally can attractively lead to phenomenologically viable PS models at much lower breaking scales than usually considered.

The limits on the PS breaking scale arise from rare meson decay processes mediated by leptoquarks. In particular the theory requires the existence of a gauge-boson leptoquark, $X_\mu$, which mediates these rare meson decays at tree level and with coupling strength similar to the strong coupling constant $g_c$. As $X_\mu$ couples quarks and fermions of the same isospin, the dominant decay modes that constrain PS breaking arise from the interactions of $X_\mu$ to the down-quarks and charged-leptons. Signals from up-isospin interactions result in neutrino (or missing energy) final states which are more difficult to constrain. 

Interestingly, the introduction of new physics to break the down-isospin mass degeneracy can further modify the meson decay rates and hence also the limits on the PS breaking scale. Ordinarily the PS breaking limits vary between $\mathcal{O}(100-1000)$ TeV.  However these mixing effects can reduce these limits down to $\mathcal{O}(10-100)$ TeV. Lower PS limits are of obvious interest as they allow for potential experimental probes of these models at scales lower than previously anticipated.

Additionally there has been a recent resurgence in interest in low-scale PS models as the leptoquarks predicted by the theory are promising candidates as explanations of anomalies in low-energy flavour-physics experiments~\cite{Aaij:2015yra,Aaij:2017vbb,Aaij:2014ora,Aaij:2019wad}. The vector leptoquark $X_\mu$ itself is an attractive candidate to explain a portion of these anomalies. However, this requires a mass around $\Lambda \simeq 30$ TeV~\cite{Heeck:2018ntp} which is na\"ively ruled out from rare meson-decay experiments. 
Modifications to the PS gauge group itself have been proposed~\cite{DiLuzio:2017vat,Calibbi:2017qbu,Bordone:2017bld} in order to allow for lighter masses of $X_\mu$ by modifying the gauge coupling to the different generations or by embedding the theory in a Randall-Sundrum background allowing for PS breaking in the TeV range~\cite{Blanke:2018sro}. Alternatively the scalar leptoquark content of the theory has been considered as a candidate to explain the anomalies~\cite{Heeck:2018ntp} in the standard PS scenario, by assuming some scalars develop significantly smaller masses than the PS breaking scale.  This  potentially leads to a hierarchy problem.

The use of charged-lepton mixing in order to both break the down-isospin mass degeneracy and reduce the allowed scale of PS breaking has already been considered in the context of the B anomalies for a specific model~\cite{Balaji:2018zna,Balaji:2019kwe} . The aim of this paper is to thoroughly analyse which PS multiplets are viable candidates in breaking the aforementioned mass degeneracy, and to evaluate the requirements on the couplings introduced in each case such that this also leads to lower experimentally allowed PS breaking scales. Therefore while we are motivated by different appealing reasons for low-scale PS, we will only focus on the requirements such that a reduction in the limits occurs.

\Cref{sec:standardPS} is an overview of the minimal PS scenario including an examination of the gauge boson and fermion masses. \Cref{sec:mesonPS} evaluates the experimental limits on PS breaking as a function of the  free parameters in the theory and determines the impact that fermionic mixing can have on the limits. \Cref{sec:exoticPS} identifies all possible PS multiplets of low dimensionality that contain  states such that mixing is induced. Finally, \cref{sec:fermionspec} evaluates the requirements on the couplings involving viable multiplets which achieve both the  desired suppression in the PS limits  and a viable values for all relevant SM masses.

\section{Pati-Salam models}
\label{sec:standardPS}
\subsection{Basic Setup}
\label{subsec:basicPS}
The Pati-Salam gauge group $G_{\text{PS}}$ extends the SM by identifying the $SU(3)$ colour group as a subgroup of an $SU(4)$ gauge group and extending the electroweak sector to be left-right symmetric:
\begin{equation}
\label{eqn:PSgaugegroup}
G_{\text{PS}} = SU(4)_{\text{c}^{\vphantom{T}}} \otimes SU(2)_{\text{L}^{\vphantom{T}}} \otimes SU(2)_{\text{R}^{\vphantom{T}}}.
\end{equation}
Often a discrete symmetry between the $SU(2)_{\text{L}}$ and $SU(2)_{\text{R}}$ sectors is imposed but is not necessary. Under this gauge symmetry the five SM fermion multiplets of each generation can be unified into two simple multiplets\footnote{With the necessary addition of a right-handed neutrino.} under $G_{\text{PS}}$
\begin{equation}
\label{eqn:fermionPS}
f_{L} \sim (\textbf{4},\textbf{2},\textbf{1})\quad\text{and} \quad f_{R} \sim (\textbf{4},\textbf{1},\textbf{2}),
\end{equation}
where $L/R$ indicates both which $SU(2)$ the fields are charged under as well as the chirality.
The breaking of $SU(4)$ to the maximal subgroup $SU(3)\times U(1)_X$ is phenomenologically required and under this breaking the fundamental of $SU(4)$ decomposes as
\begin{equation}
\label{eqn:PS-4breaking}
\textbf{4} \rightarrow \textbf{1}_{-1} \oplus \textbf{3}_{1/3}
\end{equation}
indicating that the SM quarks and leptons can be unified by identifying $U(1)_X$ with a gauged $B-L$. The SM fermions are embedded into the multiplets given in \cref{eqn:fermionPS} as\footnote{For simplicity we choose to identify the first generation of leptons with the first generation of quarks and so forth, however alternative assignments are possible~\cite{Valencia:1994cj}.}
\begin{equation}
\label{eqn:fermionPSvanillaemb}
f_{L} = \begin{pmatrix}
u_r & d_r \\
u_b & d_b \\
u_g & d_g \\
\nu_e & e \end{pmatrix}_{L}\quad\text{and}\quad f_{R} = \begin{pmatrix}
u_r & d_r \\
u_b & d_b \\
u_g & d_g \\
\nu_e & e \end{pmatrix}_{R},
\end{equation}
with similar embeddings for the other generations. Therefore the gauge transformation rules for the fields, written as matrix multiplication, are
\begin{equation}
f_{L/R} \rightarrow \mphant{U_{4}}{T} \left( f_{L/R} \right) U_{L/R}^T
\end{equation}
where $U_{4,L,R}$ are special unitary matrices for the groups $SU(4)_\text{c}$, $SU(2)_{\text{L}}$ and $SU(2)_{\text{R}}$ respectively. There are multiple  choices of scalars  which give the correct breaking patterns. A common and near-minimal choice for electroweak symmetry breaking is a complex bidoublet
\begin{equation}
\label{eqn:PSscalarbi}
\phi \sim (\textbf{1},\textbf{2},\textbf{2}) = \begin{pmatrix}
\phi_1^0 & \phi_2^+\\
\phi_1^- & \phi_2^0\end{pmatrix}, \quad \mvev{\phi} = \begin{pmatrix}
v_1 & 0\\
0 & v_2
\end{pmatrix}
\end{equation} 
where the superscripts indicate the electric charge $Q$ of each field and $\phi$ is written such that it transforms as
\begin{equation}
\phi \rightarrow \mphant{U_L}{\dagger} \phi \, U_R^\dagger.
\end{equation}
Two different combinations of scalar multiplets are usually considered for the breaking of $G_{\text{PS}}$ down to $G_{\text{SM}}$. Firstly, and most commonly, two scalars $\Delta_L \sim (\textbf{10},\textbf{3},\textbf{1})$ and $\Delta_R \sim (\textbf{10},\textbf{1},\textbf{3})$ are employed, where
\begin{equation}
\label{eqn:PSscalar1}
\Delta_{L/R}^\alpha = \frac{1}{\sqrt{2}}\begin{pmatrix}
\sqrt{2}\Delta_{11}^{\alpha+1/3} & \hphantom{\sqrt{2}}\Delta_{12}^{\alpha+1/3} & \hphantom{\sqrt{2}}\Delta_{13}^{\alpha+1/3} & \hphantom{\sqrt{2}}\Delta_{14}^{\alpha-1/3}\\
\hphantom{\sqrt{2}}\Delta_{12}^{\alpha+1/3} & \sqrt{2}\Delta_{22}^{\alpha+1/3} & \hphantom{\sqrt{2}}\Delta_{23}^{\alpha+1/3} & \hphantom{\sqrt{2}}\Delta_{24}^{\alpha-1/3}\\
\hphantom{\sqrt{2}}\Delta_{13}^{\alpha+1/3} & \hphantom{\sqrt{2}}\Delta_{23}^{\alpha+1/3} & \sqrt{2}\Delta_{33}^{\alpha+1/3} & \hphantom{\sqrt{2}}\Delta_{34}^{\alpha-1/3}\\
\hphantom{\sqrt{2}}\Delta_{14}^{\alpha-1/3} & \hphantom{\sqrt{2}}\Delta_{24}^{\alpha-1/3} & \hphantom{\sqrt{2}}\Delta_{34}^{\alpha-1/3} & \sqrt{2}\Delta_{44}^{\alpha-1\hphantom{/3}}
\end{pmatrix}_{L/R}
\end{equation}
are symmetric matrices in $SU(4)$ space (written in the defining representation), $\alpha = (-1,0,1)$ corresponds to the non-trivial $SU(2)$ charge of the scalar (in the adjoint representation) and the superscripts on each component again corresponds to its electric charge\footnote{Unless stated otherwise superscripts on fields making up a PS multiplet will correspond to its electric charge $Q$ under $G_{\text{SM}}$.}. The breaking $G_{\text{PS}} \rightarrow G_{\text{SM}}$ occurs with a non-zero vacuum expectation value (vev) in the following components
\begin{equation}
\mvev{\left(\Delta_{L/R}^{\alpha=1}\right)_{44}} = v_{L/R}^{\vphantom{T}}.
\end{equation}
Alternatively, the scalars $\rchi_L \sim (\textbf{4},\textbf{2},\textbf{1})$ and $\rchi_R \sim (\textbf{4},\textbf{1},\textbf{2})$ can be used:
\begin{equation}
\label{eqn:PSscalar2}
\rchi_{L/R}^{\vphantom{T}} = \begin{pmatrix}
\rchi^{2/3}_{r} & \rchi^{-1/3}_{r}\\
\rchi^{2/3}_{b} & \rchi^{-1/3}_{b}\\
\rchi^{2/3}_{g} & \rchi^{-1/3}_{g}\\
\rchi^{0} & \rchi^{-}
\end{pmatrix}_{L,R}, \qquad \mvev{\rchi_{L/R}^{\vphantom{T}}} = \begin{pmatrix}
0 & 0\\
0 & 0\\
0 & 0\\
\mphant{v_{L/R}}{T} & 0
\end{pmatrix}
\end{equation}
where the gauge transformation rules for these scalars are the same as $f_{L/R}$.

In both cases, $\Delta_R^{\vphantom{T}} \text{ and }\,\rchi_R^{\vphantom{T}}$ would be sufficient without their $SU(2)_L$ counterparts as long as the bidoublet $\phi$ is included. Their inclusion however allows for the possibility of imposing a discrete symmetry between the left and right gauge sectors which would lead to partial unification of the gauge couplings, $g_L = g_R$, at the relevant breaking scale. An analysis of the renormalisation group running of the gauge coupling constants has shown that the inclusion of a parity symmetry requires a PS breaking scale of $\mathcal{O}(10^{12}) \text{ GeV}$ in order for consistency with low-energy measurements of electroweak observables~\cite{Mohapatra:1986uf}, unless the parity breaking scale is decoupled from the scale of PS breaking~\cite{Chang:1984qr}. In order to allow for the scale of PS breaking to be as low as possible we assume that if such a discrete symmetry exists, then its breaking occurs independently at a higher scale to the breaking of $G_{\text{PS}}$ which allows for the scale of PS breaking to be significantly lowered. However for generality we will include the $SU(2)_L$ scalars which may be predicted by specific GUTs or lead to unique phenomenology.

In addition to the scalars in \cref{eqn:PSscalar2} often $\Phi \sim (\textbf{15},\textbf{1},\textbf{1})$ is included where
\begin{align}
\label{eqn:PSscalar3}
&\Phi = \frac{1}{2}\begin{pmatrix}
\Phi_\pi^0 + \frac{1}{\sqrt{3}} \Phi_ \eta^0 + \frac{1}{\sqrt{6}} \Phi_{15}^0 & \sqrt{2} \Phi_{12}^0 & \sqrt{2} \Phi_{13}^0 & \sqrt{2} \Phi^{2/3}_r\\
\sqrt{2}\left(\Phi_{12}^0\right)^* & - \Phi_\pi^0 +\frac{1}{\sqrt{3}}\Phi_ \eta^0 + \frac{1}{\sqrt{6}} \Phi_{15}^0 & \sqrt{2} \Phi_{23}^0 & \sqrt{2}\Phi^{2/3}_b \\
\sqrt{2}\left(\Phi_{13}^0\right)^* & \sqrt{2} \left( \Phi_{23}^0 \right)^* & -\frac{2}{\sqrt{3}} \Phi_\eta^0 + \frac{1}{\sqrt{6}} \Phi_{15}^0 & \sqrt{2}\Phi^{2/3}_g \\
\sqrt{2} \Phi^{-2/3}_{\overline{r}} & \sqrt{2} \Phi^{-2/3}_{\overline{b}} & \sqrt{2} \Phi^{-2/3}_{\overline{g}} & -\frac{3}{\sqrt{6}} \Phi_{15}^0
\end{pmatrix}\nonumber\\
&\qquad\qquad\qquad\qquad\qquad\mvev{\Phi} = \mphant{v_\Phi}{T}\,\text{diag}\left(\frac{1}{2\sqrt{6}},\frac{1}{2\sqrt{6}},\frac{1}{2\sqrt{6}},-\sqrt{\frac{3}{8}}\right)
\end{align}
such that $\Phi$ transforms as
\begin{equation}
\Phi \rightarrow \mphant{U_4}{\dagger} \Phi\, U_4^\dagger.
\end{equation}
As this scalar transforms in the adjoint representation of $SU(4)_{\text{c}}$, a non-zero vev for $\Phi$ will break this symmetry down to one of its maximal subgroups such as $SU(3)_{\text{c}} \times U(1)_{B-L}$. The scalar content described above leads to the following symmetry breaking chain
\begin{eqnarray}
\label{eqn:PSbreakingpattern}
&SU(4)_{c}^{\vphantom{T}} \times SU(2)_{L}^{\vphantom{T}} \times SU(2)_{R}^{\vphantom{T}} \nonumber\\
&\Big{\downarrow}_{\mvev{\Phi}}  \nonumber\\
&SU(3)_{c}^{\vphantom{T}} \times SU(2)_{L}^{\vphantom{T}}\times SU(2)_{R}^{\vphantom{T}} \times U(1)_{B-L}\nonumber\\
&\,\,\,\,\,\,\,\,\,\,\,\,\,\,\Big{\downarrow_{\mvev{\rchi_R^{\vphantom{T}}/\Delta_R^{\vphantom{T}}}}}   \nonumber\\
&SU(3)_c^{\vphantom{T}} \times SU(2)_{L}^{\vphantom{T}} \times U(1)_{Y}^{\vphantom{T}} \nonumber\\
&\quad\,\,\,\,\,\,\,\,\,\,\,\,\Big{\downarrow_{\mvev{\phi,\,\rchi_L^{\vphantom{T}}/\Delta_L^{\vphantom{T}}}}}  \nonumber\\
&SU(3)_c^{\vphantom{T}}\times U(1)_Q^{\vphantom{T}}
\end{eqnarray}
where $Y = T_{3R} + \frac{B-L}{2}$, $Q = T_{3L} + Y$ and the order of breaking is determined by the relative size of each vev. The inclusion of $\Phi$ therefore allows for all possible scales of symmetry breaking\footnote{The vevs of the scalars $\rchi_R$ and $\Delta_R$ directly break ${SU(4)_\text{c} \otimes SU(2)_{\text{L}} \otimes SU(2)_{\text{R}}}$ down to ${SU(3)_\text{c}\otimes SU(2)_\text{L}\otimes U(1)_\text{Y}}$.} starting from $G_{\text{PS}}$ down to the broken SM but is only necessary in scenarios where the $SU(2)_R$ gauge boson masses are desired to be smaller than the PS breaking scale.

We now turn to how the mass relations of Eq.~(\ref{eq:massrelations}) can be avoided. The equality between down-isospin partners will be addressed further below, however the equality between up-isospin partners allows us to restrict the scalar content of the theory. If $\Delta_{L/R}$ are present, a Majorana mass term for the neutrinos will be generated via $y_{L/R} \overline{f_{L/R}} \,(f_{L/R})^c \Delta_{L/R}$ and will lead to a see-saw mechanism between the neutral fermions as the hierarchy $\langle \Delta _L \rangle \ll \langle \phi \rangle \ll \langle \Delta_R \rangle$ is required due to electroweak precision tests and the Yukawa couplings are fixed by the masses of the up-type quarks. This leads to light, predominantly left-handed neutrinos, with masses given by
\begin{equation}
\label{eqn:PSneut}
m_{\nu} \simeq  y_L \langle \Delta_L \rangle + \frac{m_u^2}{y_R \langle \Delta_R \rangle}.
\end{equation}
Considering the third generation of fermions alone, $m_u = m_t \simeq 175\text{ GeV}$, requires PS breaking at large scales in order to achieve a viable low-energy neutrino mass spectrum. Setting $\langle \Delta_L \rangle = 0$ gives a rough lower-bound \vev{\Delta_R} $\gtrsim 10^{12}\text{ GeV}$ and therefore the scalars $\Delta_{L/R}$ are not viable as models of low-scale PS.\footnote{A viable neutrino mass spectrum with $\langle \Delta_R \rangle \ll 10^{12}$ GeV may be possible with a fine-tuned cancellation between the two terms appearing in \cref{eqn:PSneut} if $y_{L/R}$ have opposite sign, although we will not consider this possibility further.}

If $\rchi_{L/R}$ are present, a viable neutrino mass spectrum is only possible with the inclusion of additional particles, for example a left-handed gauge-singlet fermion $S_L \sim (\textbf{1},\textbf{1},\textbf{1})$ for each fermion generation, as otherwise $\nu_{L/R}$ are predicted to be Dirac particles with masses similar to those of the up-type quarks. Light neutrinos can arise due to the inverse and linear see-saw mechanisms~\cite{Mohapatra:1986bd,Deppisch:2004fa,Wyler:1982dd,Ma:1987zm,GonzalezGarcia:1988rw,GonzalezGarcia:1990fb,Barr:2003nn,Malinsky:2005bi,Akhmedov:1995vm} if small but non-zero lepton number violating mass terms are included. This scenario can allow for the PS breaking scale to be much lower, as we describe below, while allowing for light neutrino masses and therefore we restrict ourselves to the scalar content described in \cref{table0}.

\begin{table}[t]
\begin{center}
{\renewcommand{\arraystretch}{1.25}
\scalebox{1.0}{
\begin{tabular}{ccccc}
\toprule
 & $\Phi$ & $\phi$	& $\rchi_L$ & $\rchi_R$ \\ 
\midrule 
\midrule
$SU(4)_\text{c}$ & \textbf{15} & \textbf{1} & \textbf{4} & \textbf{4} \\
$SU(2)_\text{L}$ & \textbf{1} & \textbf{2} & \textbf{2} & \textbf{1} \\
$SU(2)_\text{R}$ & \textbf{1}  & \textbf{2} & \textbf{1} & \textbf{2}\\
\bottomrule
\end{tabular}
}
}
\end{center}
\caption[Scalar content assumed such that low-scale Pati-Salam is phenomenologically viable.]{The scalar content which we utilise and their respective dimensions under each PS gauge group.} 
\label{table0}
\end{table}

\subsection{Yukawa sector}
\label{subsec:YukPS}

The full Yukawa Lagrangian for the fermions $f_L,\,f_R\text{ and }S_L$ and the scalars $\phi,\rchi_L\text{ and }\rchi_R$ is
\begin{equation}
\label{eqn:PSYuk}
\mathcal{L}_{\textsc{yuk}^{\vphantom{T}}} = \text{Tr} \left[ y_1^{\vphantom{T}} \overline{f_L} \phi \,(f_R)^T + y_2^{\vphantom{T}} \overline{f_L} \,\phi^c (f_R)^T + y_R^{\vphantom{T}} \overline{S_L^{\vphantom{T}}} \rchi_R^{\dagger} f_R^{\vphantom{\dagger}} +y_L^{\vphantom{T}} \overline{f_L^{\vphantom{T}}} \rchi_L^{\vphantom{\dagger}} (\mphant{S_L}{\dagger})^c\right] + \frac{1}{2}\mu_S^{\vphantom{T}} \overline{S_L} (S_L)^c + \text{H.c}
\end{equation}
where generational indices are suppressed, $\phi^c = \tau_2 \,\phi^* \tau_2$ and $\tau_2 = \epsilon^{ab}$ is the two-dimensional Levi-Civita symbol. 

After spontaneous symmetry breaking, charged-fermion masses arise from \vev{\phi}
\begin{align}
\label{eqn:PSchargfermionmass}
m_u^{\vphantom{T}} &= y_1^{\vphantom{T}} v_1^{\vphantom{T}} + y_2^{\vphantom{T}} v_2^* \nonumber\\
m_d^{\vphantom{T}} &= y_1^{\vphantom{T}} v_2^{\vphantom{T}} + y_2^{\vphantom{T}} v_1^* \nonumber\\
m_e^{\vphantom{T}} &= m_d,
\end{align}
whereas, within the neutrino sector, mixing between the neutral fermions leads to
\begin{equation}
\label{eqn:PSneutralmix}
\frac{1}{2} \begin{pmatrix}
\overline{\nu_L} & \overline{\nu_R^c} & \overline{S_L}
\end{pmatrix} \begin{pmatrix}
0 & \mphant{m_u}{T} & y_L v_L\\
m_u & 0 & y_R^{\vphantom{T}} v_R^* \\
y_L v_L & y_R v_R^*	 & \mphant{\mu_S}{T}
\end{pmatrix}\begin{pmatrix}
\nu_L^c\\
\nu_R^{\vphantom{c}}\\
S_L^c
\end{pmatrix}.
\end{equation}
Adopting the hierarchy $\vert \mu_S,\,y_L\, v_L \vert < \vert m_u \vert< \vert y_R\, v_R \vert$ and setting all parameters to be real for simplicity leads to
\begin{align}
\label{eqn:PSneutralmass}
m_{\nu_1} &\simeq \mu_S\left(\frac{m_u}{y_R v_R}\right)^2 + 2 \frac{y_L v_L}{y_R v_R} m_u\nonumber\\
m_{\nu_{2,3}} &\simeq y_R v_R \pm \mu_S
\end{align}
if only one generation is considered, where the light state is predominantly made up of $\nu_L$ and the two heavy states predominantly made up of $\nu_R$ and $S_L$. A viable neutrino mass spectrum is possible for sufficiently small values of the lepton number violating mass terms $\mu_S$ and $y_L v_L$. These choices are technically natural and allow for the breaking scale $v_R$ (which breaks PS) to be lowered to $\mathcal{O}(1000)$ TeV or lower.

\subsection{Gauge Sector}
\label{subsec:gaugePS}

The kinetic term for each scalar is given by
\begin{equation}
\label{eqn:gaugePS}
\mathcal{L}_{\textsc{kin}}^{\textsc{s}} = \left(D_\mu \phi\right)^\dagger \left(D^\mu \phi\right) + \left(D_\mu \mphant{\rchi_R}{T}\right)^\dagger \left(D^\mu \mphant{\rchi_R}{T}\right) + \left(D_\mu \mphant{\rchi_L}{T}\right)^\dagger \left(D^\mu \mphant{\rchi_L}{T}\right) + \frac{1}{2}\textrm{Tr}\left(D_\mu \Phi\right) \left(D^\mu \Phi\right),
\end{equation}
the covariant derivatives are given by
\begin{alignat}{2}
\label{eqn:covderivPS}
D_\mu \phi &= \,\,\partial_\mu \phi &&+ i g_L \hat{W}_{L\mu} \phi - i g_R \phi\,\hat{W}_{R\mu}\nonumber\\
D_\mu \mphant{\rchi_L}{T} &= \partial_\mu \mphant{\rchi_L}{T} &&+ i g_4 \hat{G}_\mu \mphant{\rchi_L}{T} \,+ i g_L \mphant{\rchi_L}{T} (\hat{W}_{L\mu})^T\nonumber\\
D_\mu \mphant{\rchi_R}{T} &= \partial_\mu \mphant{\rchi_R}{T} &&+ i g_4 \hat{G}_\mu \mphant{\rchi_R}{T} \,+ i g_R \mphant{\rchi_R}{T} (\hat{W}_{R\mu})^T\nonumber\\
D_\mu \Phi &=  \,\,\partial_\mu \Phi &&+ i g_4 \left[\hat{G}_\mu,\,\Phi \right]
\end{alignat}
and $\hat{W}_{L[R]\mu}$/$\hat{G}_\mu$ are the $SU(2)_{L[R]}$/$SU(4)$ gauge fields respectively, written as matrices transforming in their defining representation.

After spontaneous symmetry breaking the spectrum of masses and mixings for the different gauge fields can be calculated. We find
\begin{equation}
\label{eqn:XmassPS}
m_X^2 = g_4^2 \left(\frac{1}{3} v_\Phi^2 + \frac{1}{2}v_L^2 + \frac{1}{2}v_R^2\right)
\end{equation}
where $X_\mu$ corresponds to a colour-triplet vector leptoquark with electric charge $2/3$,
\begin{align}
\label{eqn:WmassPS}
\mathcal{L}_{\textsc{kin}}^{\textsc{s}} &\supset \frac{1}{2}\begin{pmatrix}
W_L^+ & W_R^+
\end{pmatrix}_\mu \underbrace{\begin{pmatrix}
g_L^2\left(v_\phi^2 +v_L^2\right) & -2 g_L g_R v_1 v_2\\
-2 g_L g_R v_1 v_2 & \hphantom{-}g_R^2\left(v_\phi^2 + v_R^2\right)
\end{pmatrix}} \begin{pmatrix}
W_L^-\\
W_R^-
\end{pmatrix}^\mu\nonumber\\
&\qquad\qquad\qquad\qquad\qquad\qquad\qquad\,\, M_{W^{\pm}}^2 
\end{align}
which corresponds to the mixing matrix between the two colour-neutral, electrically-charged gauge bosons with $v_\phi^2 = v_1^2 + v_2^2$ and
\begin{align}
\label{eqn:ZmassPS}
\mathcal{L}_{\textsc{kin}}^{\textsc{s}} \supset \frac{1}{2}\begin{pmatrix}
G_{15} & W_{3L} & W_{3R}
\end{pmatrix}_\mu &\underbrace{\begin{pmatrix}
\frac{3}{2} g_4^2 \left( v_L^2 + v_R^2\right) & -\sqrt{\frac{3}{2}}g_4 g_L v_L^2 & -\sqrt{\frac{3}{2}}g_4 g_R v_R^2 \\
-\sqrt{\frac{3}{2}} g_4 g_L v_L^2 & \hphantom{-}g_L^2\left(v_\phi^2 + v_L^2\right) & -g_L g_R v_\phi^2\\
-\sqrt{\frac{3}{2}} g_4 g_R v_R^2 & -g_L g_R  v_\phi^2 & \hphantom{-}g_R^2\left(v_\phi^2 + v_R^2\right)
\end{pmatrix}}\begin{pmatrix}
G_{15} \\
W_{3L}\\
W_{3R}
\end{pmatrix}^\mu\nonumber\\
&\qquad\qquad\qquad\qquad\qquad\,\, M_0^2
\end{align}
which corresponds to the mixing matrix between the three colour- and electrically-neutral gauge bosons.

While the above equations can be solved numerically, simple analytic expressions can be derived in certain limits. Assuming a hierarchy in the scales of symmetry breaking, $v_{\text{EW}} = \sqrt{v_\phi^2 + v_L^2} < \mphant{v_R}{T}$, leads to the following spectrum of gauge boson masses:
\begin{align}
\label{eqn:gaugespectrumPS}
&m_\gamma^2 = 0, \quad m_Z^2 \simeq \frac{1}{2}\frac{3 g_R^2 g_4^2 + 3 g_L^2 g_4^2 + 2 g_R^2 g_L^2}{3 g_4^2 + 2 g_R^2}\,v_{\textsc{EW}}^2 , \quad m_{Z^{'}}^2 \simeq \frac{1}{2}\left(g_R^2 + \frac{3}{2}g_4^2 \right) v_R^2\nonumber\\
&\qquad\qquad\qquad\qquad\,\,\,\, m_{W}^2 \simeq \frac{1}{2}g_L^2v_{\textsc{EW}}^2,\quad m_{W^{'}}^2 \simeq \frac{1}{2}g_R^2 \left(v_R^2+v_{\textsc{EW}}^2\right)\nonumber\\
&\qquad\qquad\qquad\qquad\qquad m_X^2 = \frac{1}{2} \left(\frac{2}{3} v_\Phi^2 + v_L^2 + v_R^2 \right) g_4^2.
\end{align}
where the hypercharge gauge coupling is given by 
\begin{equation}
g_Y^2 = \frac{3 g_R^2 g_4^2}{3 g_4^2 + 2 g_R^2}.
\end{equation}
Note that \vev{\Phi} only contributes to the mass of the vector leptoquark $X_\mu$ and otherwise is completely decoupled from the remaining gauge bosons and fermions. 

\begin{figure}[t]
\centering
{
  \includegraphics[width=0.41\linewidth]{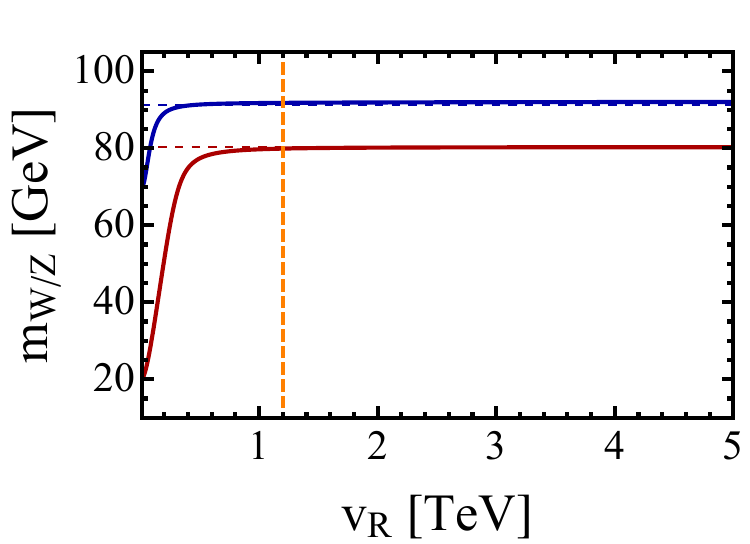}\hfill
}
{
  \hspace*{0.1mm}\includegraphics[width=0.15\linewidth]{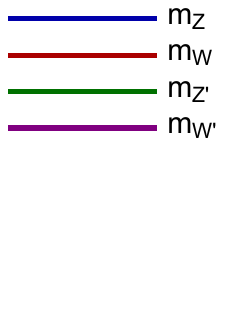}\hfill
}
{\hspace{-3mm}
  \includegraphics[width=0.4\linewidth]{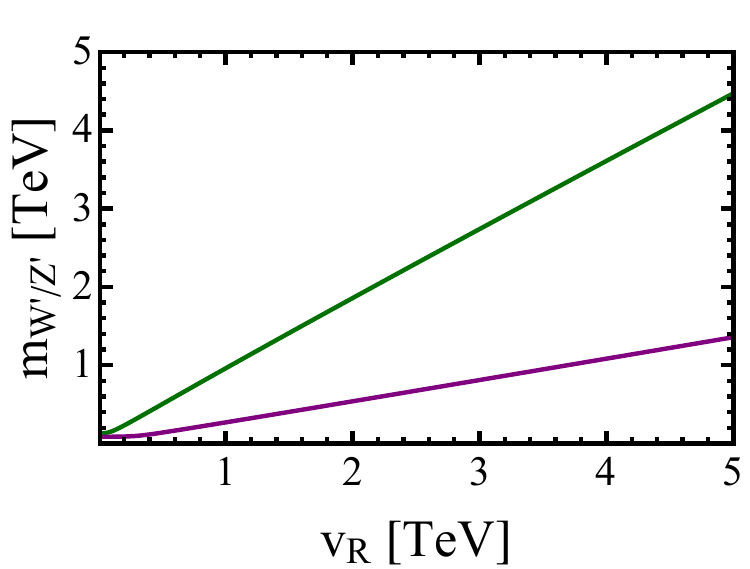}
}
\caption[Plot of the SM-like and heavy gauge bosons in PS as a function of the $SU(2)_R$ breaking scale $v_R$.]{Plot of the masses of the SM-like gauge bosons $m_Z$ and $m_W$ \textbf{(left)} and the heavy gauge bosons $m_{Z'}$ and $m_{W'}$ \textbf{(right)} as a function of the $SU(2)_R$ breaking scale $v_R$. Here the masses were found by numerically solving~\cref{eqn:WmassPS,eqn:ZmassPS} with gauge couplings given by~\cref{eqn:PStoSMgauge} assuming the normal running of the SM gauge couplings at these energy scales described in \cref{sec:Appendix C}. We roughly find that $v_R > 1.2\text{ TeV}$ is required for \cref{eqn:gaugespectrumPS} to be valid and to ensure the light gauge fields remain SM-like as indicated by the dashed vertical line, the dashed horizontal lines correspond to the measured masses of the electroweak gauge fields. We find that the $Z'$ gauge field will be significantly heavier than the $W'$ field and a ratio of $m_{Z'}/m_{W'} \simeq 3.5$. This is due to the additional mixing effects occurring between the neutral gauge bosons compared to the charged ones.}
\label{figure:LRSMgaugebosonmasses}
\end{figure}

The gauge couplings of $G_{\text{PS}}$ ($g_4$, $g_L$ and $g_R$) are related to those of the SM ($g_c$, $g_w$ and $g_Y$) at the scale of PS breaking:
\begin{equation}
\label{eqn:PStoSMgauge}
g_4^2(\mu) = g_c^2(\mu), \quad g_L^2(\mu) = g_w^2(\mu) \quad \text{and} \quad g_R^2(\mu) = \frac{3 g_Y^2(\mu) g_c^2(\mu)}{3 g_c^2(\mu) - 2g_Y^2(\mu)}.
\end{equation}
As the $SU(4)$ coupling constant is given by the usual colour gauge coupling, at low scales $g_c(\mu) > g_Y(\mu)$ and therefore $g_R(\mu) \simeq g_Y(\mu)$. \Cref{figure:LRSMgaugebosonmasses} plots the masses of the electroweak gauge bosons of the theory as a function of the $SU(2)_R$ breaking scale $v_R$. In order to decouple $v_R$ from significantly impacting the masses of the light, SM-like gauge bosons we roughly find that $\mphant{v_R}{T} > 1200\text{ GeV}$ is required in order for \cref{eqn:gaugespectrumPS} to be a valid approximation. The heavy neutral gauge boson $Z'$ is significantly heavier compared to the heavy charged gauge boson $W'$ and we roughly find a ratio of $m_{Z'}/m_{W'} \simeq 3.5$ at TeV scales. Adoption of the hierarchy $v_\Phi \gg v_R$ would imply that the leptoquark $X$ is heavier than both the $Z'$ and $W'$ fields such that $m_{W'} < m_{Z'} \ll m_X$. However, were $\Phi$ to be absent or have $v_\Phi < v_R$, the spectrum of heavy gauge bosons masses would be $m_{W'} < m_X < m_{Z'}$ with the ratio $m_{Z'} / m_X \simeq 1.3$ at TeV scales.

\section{Probing Pati-Salam models through rare meson decays}
\label{sec:mesonPS}

A promising probe of the scale of Pati-Salam breaking is through the contributions of the gauge and scalar leptoquarks predicted by different realisations of the model to low-energy hadronic processes. Of particular significance is the gauge boson leptoquark $X_\mu$ which must couple universally to all three generations of fermions. This is unlike the various possible scalar leptoquarks for which the Yukawa couplings to the lighter generations could be suppressed, perhaps through a flavour symmetry. Additionally as $SU(3)_\text{c}$ is a subgroup of the $SU(4)$ appearing in $G_\text{PS}$, the coupling strength of the leptoquark $X_\mu$ to the SM quarks is related to that of $g_c$ and cannot be treated as a free parameter. Therefore precision flavour experiments involving the lighter generations provide stringent limits\footnote{If a scalar leptoquark is present which also mediates the same decay, destructive interference between the gauge and scalar contribution to the hadronic process can somewhat lower the limits on the mass of $X$ by up to a factor of 2~\cite{Smirnov:2008zzb} depending on the mass(es) and couplings of the relevant scalar(s). In our analysis we will focus solely on the contribution from the gauge leptoquark and neglect possible regions of destructive interference with scalar contributions by assuming the masses of the scalars to be heavier than the gauge leptoquark.} on the mass of the gauge leptoquark $X_\mu$ which directly limits the scale of PS breaking (either $v_\Phi$ or $v_R$) through~\cref{eqn:gaugespectrumPS}.

\subsection{Neutral Pseudoscalar meson decays induced by $X_\mu$}
\label{subsec:mesonPSformula}

\begin{figure}[t]
\centering
\begin{center}
{
  \includegraphics[width=0.7\linewidth]{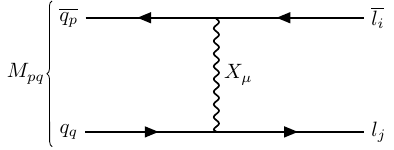}\hfill
}
\end{center}
\caption[Rare meson decay diagram mediated by $X_\mu$ at tree level.]{Tree-level Feynman diagram of the leptonic decay of a scalar meson $M_{pq}$, where $q_p$ and $q_q$ correspond to the valence quarks of the meson, to the lepton pair $l_{i,j}$ ($i,j=e,\,\mu,\,\tau$) mediated by the PS gauge leptoquark $X_\mu$. PS unifies quarks and leptons of similar isospin, therefore the processes mediated by $X_\mu$ at tree-level are easily predicted, e.g. if $q_q$ corresponds to an up-type quark then $l_j$ must be a neutrino. The strongest limits on the mass of $X_\mu$ arise in the case where $p$ and $q$ are both down-type quarks, where $M_{pq} = (K_L^0,\,B_d^0,\,B_s^0,\dots)$, and therefore the final state comprises of charged leptons with opposite sign.}
\label{figure:pseudoscalarmesondecay}
\end{figure}

The most stringent constraint on the mass of $X_\mu$ arises from limits on rare leptonic decays of pseudoscalar mesons, mediated at tree level by $X_\mu$ as shown in \cref{figure:pseudoscalarmesondecay}. As Pati-Salam unifies quarks and leptons with the same $SU(2)_{L/R}$ isospin into the same $SU(4)$ multiplet, as shown by \cref{eqn:fermionPSvanillaemb}, it couples up-type quarks to neutrinos and down-type quarks to the charged leptons. Therefore the leptonic decay channels induced by $X_\mu$ at tree-level can be predicted based on the valence quark content of the relevant meson. For example, $X_\mu$ will mediate the leptonic decay $D^0 \rightarrow \overline{\nu}\nu$ as both valence quarks are up-type but not $D^0 \rightarrow \ell^+ \ell^-$. Unsurprisingly the meson decay channels which lead to the most stringent limits on the mass of $X_\mu$ arise from opposite-sign charged-lepton final states where there is no missing energy, which only occurs in PS for neutral mesons with down-type valence quarks. As a result, the mass of $X_\mu$ is most constrained by measurements of the decays of $K_L^0,\,B_d^0$ and $B_s^0$ mesons whose leptonic decay channels are the most well measured. 

In \cref{sec:Appendix A} the decay widths for the rare, purely leptonic decays of the relevant pseudoscalar mesons are calculated. The contribution to the decay widths from the PS gauge leptoquark in \cref{eqn:twobodydecaywidth} depends on both the mass of the leptoquark $m_X$ as well as to the different elements of the CKM-like mixing matrices between $d$ and $e$, $K^{d e}_{L/R}$. These physical mixing matrices are currently unconstrained and different textures within each matrix can significantly change the relative strength of specific lepton flavour decay channels over others for a given meson. As limits and measurements of different processes differ in sensitivity, the limits on the leptoquark mass (and therefore the scale of PS breaking) can vary significantly for different choices of the mixing matrices $K^{d e}_{L/R}$.

To demonstrate this, we define general unitary matrices for $K_{L/R}^{de}$,
\begin{align}
\label{eqn:genunit}
K_{L/R}^{de} &= K_{23}(\theta_{23}^{L/R},\delta^{L/R})\,K_{13}(\theta_{13}^{L/R})\,K_{12}(\theta_{12}^{L/R})\nonumber\\[2pt]
             &= \begin{pmatrix}
             c_{{12}}\, c_{13} & c_{13}\,s_{12} & s_{13}\\
             -c_{23}\,s_{12}-e^{i\delta}c_{12}\,s_{13}\,s_{23} & c_{12}\,c_{23}-e^{i\delta}s_{12}\,s_{13}\,s_{23} & e^{i\delta} c_{13}\,s_{23}\\
             -c_{12}\,c_{23}\,s_{13}+e^{-i\delta}s_{12}\,s_{23} & \,\,\, -c_{23}\,s_{12}\,s_{13}-e^{-i\delta}c_{12}\,s_{23}\,\,\, & c_{13}\,c_{23}
             \end{pmatrix}^{L/R}
\end{align}
where $c_x \coloneqq \cos(x)\mand s_x \coloneqq \sin(x)$ as usual. In principle the matrix $K_L^{de}$ should contain all six possible phases as by convention rephasing of quark and lepton fields fixes the phases appearing in $V^{\textsc{ckm}}_L$ and $U^\textsc{lept}_L$ and therefore no rephasing\footnote{In contrast $K_R^{de}$ could be chosen to contain one complex phase if the other right-handed physical mixing matrices are left as general unitary matrices. However a more natural choice of basis would be to rephase the fields appearing in $V^{\textsc{ckm}}_R$ and $U^\textsc{lept}_R$ in analogy with the left-handed mixing matrices.} can occur for $K_L^{de}$ or $K_L^{u\nu}$. However only one complex phase is important for the discussion below and therefore we ignore all other possible complex phases which can appear in $K_{L/R}^{de}$. 

\begin{table}[t]
\begin{center}
{\renewcommand{\arraystretch}{1.15}
\scalebox{1.0}{
\begin{tabular}{ccc}
\toprule
Hadronic Process & \,\,Measured value or upper limit \,\, & $(p,i)(q,j)$	 \\ 
\midrule 
\midrule
$\mathcal{B}(K_L^0 \rightarrow e e)$ & $9.0^{+6.0}_{-4.0} \times 10^{-12}\,\,$ & (1,1)(2,1) \\
$\mathcal{B}(K_L^0 \rightarrow \mu \mu )$ & $6.84^{+0.11}_{-0.11} \times 10^{-12}\,\,\,\,\,\,\,$ & (1,2)(2,2) \\
$\mathcal{B}(K_L^0 \rightarrow \mu e )$ & $\,\,< 4.7 \times 10^{-12}$  & (1,2)(2,1)/(1,1)(2,2)\\
\midrule
$\mathcal{B}(B_d^0 \rightarrow ee)$ & $< 8.3 \times 10^{-8}$ & (1,1)(3,1)\\
$\mathcal{B}(B_d^0 \rightarrow \mu\mu)$ & $\,\,< 3.6 \times 10^{-10}$ & (1,2)(3,2)\\
$\mathcal{B}(B_d^0 \rightarrow \tau\tau)$ & $<2.1\times 10^{-3}$ & (1,3)(3,3)\\
$\mathcal{B}(B_d^0 \rightarrow \mu e)$ & $<1.0\times 10^{-9}$ & (1,2)(3,1)/(1,1)(3,2)\\
$\mathcal{B}(B_d^0 \rightarrow \tau e)$ & $<2.8\times 10^{-5}$ & (1,3)(3,1)/(1,1)(3,3)\\
$\mathcal{B}(B_d^0 \rightarrow \tau \mu)$ & $<2.2\times 10^{-5}$ & (1,3)(3,2)/(1,2)(3,3)\\
\midrule
$\mathcal{B}(B_s^0 \rightarrow ee)$ & $<2.8\times 10^{-7}$ & (2,1)(3,1)\\
$\mathcal{B}(B_s^0 \rightarrow \mu\mu)$ & $3.0^{+0.4}_{-0.4} \times 10^{-9}\,\,\,\,$ & (2,2)(3,2)\\
$\mathcal{B}(B_s^0 \rightarrow \tau\tau)$ & $<6.8\times 10^{-3}$ & (2,3)(3,3)\\
$\mathcal{B}(B_s^0 \rightarrow \mu e)$ & $<5.4\times 10^{-9}$ & (2,2)(3,1)/(2,1)(3,2)\\
$\mathcal{B}(B_s^0 \rightarrow \tau e)$ & $-$ & (2,3)(3,1)/(2,1)(3,3) \\
$\mathcal{B}(B_s^0 \rightarrow \tau \mu)$ & $-$ & (2,2)(3,3)/(3,2)(2,3)\\
\bottomrule
\end{tabular}
}
}
\end{center}
\caption[Experimental measurements or upper limits for rare meson decays which most strongly constrain PS breaking.]{Experimental measurements and upper limits on rare leptonic decays of various pseudo-scalar mesons which form the dominant constraint on the scale of Pati-Salam breaking. The third column represents which entries of the matrices $(K^{d e}_{L/R})_{pi}$ and $(K^{d e}_{L/R})_{qj}$ need to be non-zero for the given decay channel to occur via $X_\mu$ where $p,\,q$ correspond to the valence down-type quarks, $(d,\,s,\,b) = (1,\,2,\,3)$ and $i,\,j$ to the final state charged leptons $(e,\,\mu,\,\tau) = (1,\,2,\,3)$ for the diagram in \cref{figure:pseudoscalarmesondecay}. Lepton-flavour violating decay modes can occur via two different possible diagrams. For example the process $B_d^0 \rightarrow \mu e$ can arise from the $(3,1)$ and $(1,2)$ entries which corresponds to the couplings $\overline{b} \slashed{X} e$ and $\overline{d} \slashed{X} \mu$ leading to $B_d^0 \rightarrow \mu^- e^+$, but can also have a contribution from $(3,2)$ and $(1,1)$ which corresponds to $\overline{b} \slashed{X} \mu$ and $\overline{d} \slashed{X} e$ and leads to $B_d^0 \rightarrow e^- \mu^+$.} 
\label{table1}
\end{table}

\Cref{table1} lists the meson decay channels of interest\footnote{All neutral-pesudoscalar mesons will receive a similar contribution however in the cases of other mesons not listed (such as the $\pi^0$, $\eta$, $\eta '$, $\Upsilon$ etc) we find the limits on $m_X$ are subdominant compared to the mesons listed in \cref{table1} when comparing the predicted branching fraction induced by $X_\mu$ compared to the experimental sensitivity.} alongside either their measured branching fractions or the current experimental upper limit. Additionally we indicate which two matrix elements of $K_{L/R}^{de}$ are required to be non-zero in order for $X_\mu$ to mediate this process through \cref{figure:pseudoscalarmesondecay}. Note that for lepton-flavour violating decays there are two possible combinations of non-zero matrix elements which mediate a decay which is taken into account in our calculations. It is clear from \cref{table1} that the decays of $K_L^0$ are currently the most experimentally probed channels, with the lepton-flavour violating decay $K_L^0 \rightarrow \mu e$ having the most sensitive upper limit of any process.

In contrast the current experimental precision for the decays of $B_{d/s}^0$ is, for most channels, significantly weaker and will lead to weaker limits. Therefore the lower mass bound on $X_\mu$ will vary most depending on whether the structure of $K_{L/R}^{de}$ leads to decays of $K_L^0$ or not. A thorough analysis for general unitary matrices similar to \cref{eqn:genunit} for the PS gauge leptoquark has been performed~\cite{Smirnov:2018ske} to find what conditions are required on $K_{L/R}^{de}$ in order to suppress the decays of $K_L^0$, either preventing them completely or significantly reducing the predicted decay width by removing the helicity-unsuppressed contribution. Two possible scenarios were found. Firstly if
\begin{equation}
\quad\theta_{23}^L = \theta_{23}^R = \frac{\pi}{2},\quad \theta_{13}^L = \theta_{13}^R = \theta, \quad \delta^L = \delta \quad \text{and}\quad \delta^R = \pi - \delta
\end{equation}
is satisfied by \cref{eqn:genunit}, this leads to
\begin{align}
\label{eqn:struct1}
K_L^{de} = \begin{pmatrix}
c_{12}^L \,c_\theta & s_{12}^L \,c_\theta & s_\theta\\[2pt]
-e^{i\delta}c_{12}^L\,s_\theta & \,\,-e^{i\delta}s_{12}^L\,s_\theta\,\, & e^{i\delta}c_\theta\\[2pt]
e^{\mbox{-}i\delta}s_{12}^L & -e^{\mbox{-}i\delta}c_{12}^L & 0
\end{pmatrix}\quad\text{and}\quad K_R^{de} = \begin{pmatrix}
c_{12}^R \,c_\theta & s_{12}^R \,c_\theta & s_\theta\\[2pt]
e^{\mbox{-}i\delta}c_{12}^R\,s_\theta & \,\,e^{\mbox{-}i\delta}s_{12}^R\,s_\theta\,\, & -e^{\mbox{-}i\delta}c_\theta\\[2pt]
-e^{i\delta}s_{12}^R & e^{i\delta}c_{12}^R & 0
\end{pmatrix}.
\end{align}
These matrix structures will not completely prevent the decays of $K_L^0$. However, they will prevent the helicity-unsuppressed terms from contributing and therefore will be suppressed by the final-state lepton masses similar to weak decays in the SM. Note that all the entries of the upper-left $2 \times 2$ block of $K_{L/R}^{de}$ are non-zero, but an important cancellation in the amplitude arises as $\delta^{L/R}$ are exactly out of phase. Note that the $(3,3)$ entry of both matrices is zero and therefore for this scenario the decays $B_d^0 \rightarrow \tau \tau$ and $B_s^0 \rightarrow \tau \tau$ cannot occur. It is interesting to note that if $\delta$ is maximally CP violating then the helicity unsuppressed contribution completely disappears and therefore there is no contribution to the decay of $K_L^0$, whereas the helicity-suppressed contribution to this decay is maximised if $\delta$ is CP conserving.

Alternatively the desired decays can be completely prevented with the simple condition
\begin{equation}
\theta_{13}^L = \theta_{13}^R = \frac{\pi}{2}
\end{equation}
which leads to
\begin{align}
\label{eqn:struct2}
K_{L/R}^{de} = \begin{pmatrix}
0 & 0 & 1\\[2pt]
-c_{23}\,s_{12}-e^{i\delta}c_{12}\,s_{23} & c_{12}\,c_{23}-e^{i\delta}s_{12}\,s_{23} & 0\\[2pt]
-c_{12}\,c_{23}+e^{\mbox{-}i\delta}s_{12}\,s_{23} & \,\,-c_{23}\,s_{12}-e^{\mbox{-}i\delta}c_{12}\,s_{23}\,\, & 0
\end{pmatrix}^{L/R}.
\end{align}
Here the decays of $K_L^0$ trivially do not occur as both the $(1,1)$ and $(1,2)$ entries of $K_{L/R}^{de}$ are exactly zero and as can be seen in \cref{table1}, $X_\mu$ will not mediate the relevant decays. Similar to the previous scenario the decays $B_d^0 \rightarrow \tau \tau$ and $B_s^0 \rightarrow \tau \tau$ do not occur. Additionally, the processes $B_d^0 \rightarrow (ee,\,\mu\mu,\,\tau e)$ and $B_s^0 \rightarrow (\tau e,\,\tau \mu)$ will not occur as the $(2,3)$ entry is also zero. Therefore, for low-scale PS, the suppressed decay channels for $B_{d/s}^0 \rightarrow \tau \tau$ are directly correlated with suppressed decays of $K_L^0$. Increasing the experimental sensitivity of these channels could therefore provide a powerful test of PS, especially if a non-SM signal was detected in these channels as no such signal has been seen in the decays of $K_L^0$.

In both scenarios above which lead to the lowest possible limits on the PS gauge leptoquark, the matrices $K_L^{de}$ and $K_R^{de}$ are required to have a similar structure to each other. The first scenario allows for only four free parameters between the two matrices whereas in the second scenario six free parameters exist. In order for both these matrices (which are na\"{i}vely unrelated) to have the same structure suggests that a parity symmetry must be enforced at some scale. The two matrices do not have to be exactly equal to each other, although this is also possible. However in both scenarios at least one angle has to be identical for both matrices. As discussed previously it has been found that enforcing a parity symmetry alongside the PS gauge groups requires parity to be broken at $\mathcal{O}(10^{12})$~GeV or higher~\cite{Chang:1984qr,Mohapatra:1986uf}. As we are interested in minimising the scale of PS breaking it would seem quite coincidental for both $K_L^{de}$ and $K_R^{de}$ to be so similar at low scales (and rather unlikely for them to be exactly equal) with such a high scale of parity breaking. However, this requires a full analysis of the running of the relevant mixing angles which we will not explore further. It may be possible that at some high scale $K_{L/R}^{de}$ are equal to each other and after parity breaking some (all) angles are insensitive to running effects whereas others (none) run significantly, leading to the desired form for the mixing matrices. 

In addition to the two scenarios above which completely prevent $X_\mu$ from inducing decays of $K_L^0$ we identify two additional scenarios which would prevent the LFV decay $K_L^0 \rightarrow \mu e$ but not necessarily the decays $K_L^0 \rightarrow ee$ or $K_L^0 \rightarrow \mu \mu$. For example if the conditions
\begin{equation}
\theta_{12}^L = \theta_{12}^R = 0\quad\text{and} \quad \theta_{23}^L=\theta_{23}^R = \frac{\pi}{2}
\end{equation}
are satisfied, this leads to 
\begin{align}
\label{eqn:struct3}
K_{L/R}^{de} = \begin{pmatrix}
c_{13} & 0 & s_{13}\\[2pt]
-e^{i\delta}s_{13} & 0 & e^{i\delta}c_{13}\\[2pt]
0 & -e^{\mbox{-}i\delta} & 0
\end{pmatrix}^{L/R},
\end{align}
and the decays $K_L^0 \rightarrow \mu e$ and $K_L^0 \rightarrow \mu \mu$ do not occur via $X_\mu$. However the decay $K_L^0 \rightarrow ee$  does occur as suggested by \cref{table1}. Similarly if 
\begin{equation}
\theta_{12}^L = \theta_{12}^R = \frac{\pi}{2}\quad\text{and} \quad \theta_{23}^L=\theta_{23}^R = \frac{\pi}{2}
\end{equation}
is satisfied, this leads to 
\begin{align}
\label{eqn:struct4}
K_{L/R}^{de} = \begin{pmatrix}
0 & c_{13}  & s_{13}\\[2pt]
0 & -e^{i\delta}s_{13} & e^{i\delta}c_{13}\\[2pt]
e^{\mbox{-}i\delta} & 0 & 0
\end{pmatrix}^{L/R}.
\end{align}
The decays $K_L^0 \rightarrow \mu e$ and $K_L^0 \rightarrow e e$ will not occur, but $K_L^0 \rightarrow \mu \mu$ will, forming a strong constraint on the mass of $X_\mu$.

Finally as the decay channel $K_L^0 \rightarrow \mu e$ is currently the most precisely constrained of all relevant meson decay channels the \textit{largest} lower bound on the mass of $X_\mu$ will occur for $K_{L/R}^{de}$ which have a form that maximises this particular decay channel. \Cref{table1}, which indicates which entries of $K_{L/R}^{de}$ mediate this decay, suggests that if for example the mixing matrices were given by
\begin{equation}
\label{eqn:struct5}
K_{L/R}^{de} = \begin{pmatrix}
1 & 0 & 0\\
0 & 1 & 0\\
0 & 0 & 1
\end{pmatrix}
\end{equation}
then the contribution to the decay channel would be maximised. Therefore the current mass limits on the PS gauge leptoquark (and the breaking scale) can fluctuate between the mass limits obtained if the mixing matrices are given by \cref{eqn:struct1,eqn:struct2} up to the limits obtained if the matrices are given by \cref{eqn:struct5}.

\Cref{table2} shows the calculated lower bound on the gauge leptoquark masses for the relevant decay channel with different choices of $K^{d e}_{L/R}$ corresponding to the five scenarios above. Maximising the leptoquark's contribution to $K_L^0 \rightarrow \mu e$ as in \cref{eqn:struct5} leads to limits on the gauge leptoquark mass of roughly $2500$ TeV, similar to previous studies~\cite{Valencia:1994cj,Kuznetsov:1994tt,Smirnov:2007hv}. The other non-zero decays in this case indicate which other channels will occur in this scenario: $B_d^0 \rightarrow \tau e$ and $B_s^0 \rightarrow \tau \mu$. More realistically we would expect the matrices $K_{L/R}^{de}$ to be approximately diagonal rather than exactly, and therefore other processes would also be mediated by $X_\mu$, but the three listed would have the strongest signals.

If $K_{L/R}^{de}$ is of the form described in \cref{eqn:struct1}, decays of $K_L^0$ still occur but are helicity suppressed. This reduces the limits obtained from $2500$ TeV down to $\mathcal{O}(100)$ TeV (with some sensitivity to the free mixing angles) where notably the channel $K_L^0 \rightarrow e e $ leads to limits on the gauge leptoquark mass of $\mathcal{O}(10)$ TeV. This significant reduction compared to other channels of $K_L^0$ is due to the helicity-suppression of electron final states being significantly larger than for muon final states. For this scenario neglecting the electron and muon mass, as was done in~\cite{Smirnov:2018ske}, would suggest that $X_\mu$ does not mediate $K_L^0$ decays. However, we find when included they lead to comparable limits for $m_X$ to the helicity-unsuppressed $B_d^0$ and $B_s^0$ decays. This is due to the larger experimental precision obtained for $K_L^0$ decays and therefore we find that the muon mass cannot be ignored. As mentioned previously, if $\delta$ is maximally violating the helicity-unsuppressed decays of $K_L^0$ are forbidden and in this limit the results of~\cite{Smirnov:2018ske} remain valid.

\afterpage{
\begin{landscape}
\begin{table}[t]
\begin{center}
{\renewcommand{\arraystretch}{1.25}
\scalebox{1.0}{
\begin{tabular}{cccccc}
\toprule
 & \,$K^{d e}_L = K^{d e}_R=\id_{3\times3}$ \, & \,\textsc{scenario} 1 \, & \textsc{scenario} 2 \,& \,\textsc{scenario} 3 \,& \,\textsc{scenario} 4 \,	 \\ 
\midrule 
\midrule
$\mathcal{B}(K_L^0 \rightarrow e e)$ & $0$ & $13\, \kappa^{K^{ee}}_1 \text{ TeV}$ & $0$ & $1817\,\kappa^{K^{ee}}_3 \text{ TeV}$ & $0$ \\
$\mathcal{B}(K_L^0 \rightarrow \mu \mu )$ & $0$ & $177\, \kappa^{K^{\mu\mu}}_1 \text{ TeV}$ & $0$ & $0$ & $1900 \,\kappa^{K^{\mu\mu}}_4 \text{ TeV}$ \\
$\mathcal{B}(K_L^0 \rightarrow \mu e )$ & $\fbox{2467\text{ TeV}}$ & $230\, \kappa^{K^{\mu e}}_1 \text{ TeV}$ & $0$ & $0$ & $0$\\
\midrule
$\mathcal{B}(B_d^0 \rightarrow ee)$ & $0$ & $39.7\, \kappa^{B^{ee}}_1 \text{ TeV}$ & $0$ & $0$ & $0$\\
$\mathcal{B}(B_d^0 \rightarrow \mu\mu)$ & $0$ & $151\, \kappa^{B^{\mu \mu}}_1 \text{ TeV}$ & $0$ & $0$ & $0$\\
$\mathcal{B}(B_d^0 \rightarrow \tau\tau)$ & $0$ & $0$ & $0$ & $0$ & $0$\\
$\mathcal{B}(B_d^0 \rightarrow \mu e)$ & $0$ & $140\, \kappa^{B^{\mu e}}_1 \text{ TeV}$ & $0$ & $140\, \kappa^{B^{\mu e}}_3 \text{ TeV}$ & $140\, \kappa^{B^{\mu e}}_4 \text{ TeV}$\\
$\mathcal{B}(B_d^0 \rightarrow \tau e)$ & $12.1\text{ TeV}$ & $10.6\, \kappa^{B^{\tau e}}_1 \text{ TeV}$ & $10.6\, \kappa^{B^{\tau e}}_2 \text{ TeV}$ & $0$ & $10.6\, \kappa^{B^{\tau e}}_4 \text{ TeV}$\\
$\mathcal{B}(B_d^0 \rightarrow \tau \mu )$ & $0$ & $11.3\, \kappa^{B^{\tau \mu}}_1 \text{ TeV}$ & $11.3\, \kappa^{B^{\tau \mu}}_2 \text{ TeV}$ & $11.3\, \kappa^{B^{\tau \mu}}_3 \text{ TeV}$ & $0$\\
\midrule
$\mathcal{B}(B_s^0 \rightarrow ee)$ & $0$ & $29.5\, \kappa^{B^{e e}_s}_1 \text{ TeV}$ & $29.5\, \kappa^{B^{e e}_s}_2 \text{ TeV}$ & $0$ & $0$ \\
$\mathcal{B}(B_s^0 \rightarrow \mu\mu)$ & $0$ & $90.0\, \kappa^{B^{\mu \mu}_s}_1 \text{ TeV}$ & $90.0\, \kappa^{B^{\mu \mu}_s}_2 \text{ TeV}$ & $0$ \\
$\mathcal{B}(B_s^0 \rightarrow \tau\tau)$ & $0$ & $0$ & $0$ & $0$ & $0$ \\
$\mathcal{B}(B_s^0 \rightarrow \mu e )$ & $0$ & $92.3\, \kappa^{B^{\mu e}_s}_1 \text{ TeV}$ & $92.3\, \kappa^{B^{\mu e}_s}_2 \text{ TeV}$ & $92.3\, \kappa^{B^{\mu e}_s}_3 \text{ TeV}$ & $92.3\, \kappa^{B^{\mu e}_s}_4 \text{ TeV}$\\
$\mathcal{B}(B_s^0 \rightarrow \tau e )$ & $0$ & $-$ & 	$0$ & $0$ & $-$ \\
$\mathcal{B}(B_s^0 \rightarrow \tau \mu )$ & $-$ & $-$ & $0$ & $-$ & $0$ \\
\bottomrule
\end{tabular}
}
}
\end{center}
\caption[Limits on the gauge leptoquark mass $m_X$ compared to current measurements without a chiral suppression.]{Limits on the gauge leptoquark mass $m_X$ compared to current measurements (or upper limits) for the mesons $K_L^0$, $B_d^0$ and $B_s^0$. Each column represents different choices for the matrices $K^{d e}_{L/R}$ where scenarios 1-4 are given by \cref{eqn:struct1,eqn:struct2,eqn:struct3,eqn:struct4} respectively. Each scenario corresponds to possible structures of the mixing matrices which would in some way suppress the decays of $K_L^0$. In scenario 1 the decays to $K_L^0$ are non-zero albeit helicity-suppressed, leading to mass limits similar to what is obtained from decays of $B_d^0$ and $B_s^0$. Scenario 2 completely suppresses the decays of $K_L^0$ and the most significant mass limits now come from the decays of $B_s^0$. Scenario 3 and 4 correspond to examples of mixing matrices which would suppress the decay channel $K_L^0 \rightarrow \mu e$ but not $K_L^0 \rightarrow e e$ or $K_L^0 \rightarrow \mu \mu$ which would form the dominant limit. The first column represents a scenario which maximises the rate of $K_L^0 \rightarrow \mu e$ decays leading to the largest mass limit on $X_\mu$. In all the scenarios above the parameter $\kappa^{X}_{\alpha}$ corresponds to the combination of mixing angles relevant to that decay chain. We find the lower bound on the PS breaking scale to be no smaller than $\mathcal{O}(100)$ TeV. For the final two channels $B_s^0 \rightarrow \tau e / \tau \mu$ there are currently no measured upper bounds, however non-zero contributions to these channels are indicated by `$-$' and their future measurement will form a constraint for a given choice of $K^{de}_{L/R}$.} 
\label{table2}
\end{table}
\end{landscape}
}

If the mixing matrices are given by \cref{eqn:struct2}, decays of $K_L^0$ do not occur via $X_\mu$ either helicity-suppressed or -unsuppressed. Only five decay channels are non-zero in this scenario and the dominant mass limit will arise from decays of $B_s^0$ and are of similar order to the previous scenario: $m_X \sim \mathcal{O}(100)$ TeV. The final two scenarios described by \cref{eqn:struct3,eqn:struct4} completely suppress the channel $K_L^0 \rightarrow \mu e$ but does not suppress the lepton-flavour conserving channels of $K_L^0$. Here the resulting mass limits are $\mathcal{O}(1900)$ TeV unless the mixing angles in $\kappa$ were significantly tuned (e.g. $\theta_{12}^L \simeq -\theta_{12}^R$) in order to suppress these channels. In all cases the last two channels $B_s^0 \rightarrow \tau e$ and $B_s^0 \rightarrow \tau \mu$ are currently unconstrained by measurement and therefore do not lead to limits on $m_X$. In cases were a scenario predicts a contribution to these channels, future measurement will be a relevant constraint which we indicate in \cref{table2} by `$-$'.

For simplicity, the interference between the SM and PS contributions to the decay channels with lepton-flavour conserving final states has not been calculated. As the PS contribution to the decay rate is inversely proportional to the fourth power of $m_X$, the decay rate will decrease by orders of magnitude for small increases in $m_X$. The derived lower bound on $m_X$ from such channels should be understood as a conservative lower bound and will slightly change by order one factors once the SM contribution is incorporated. Obviously when the dominant limit arises from a LFV decay channel no such interference occurs and the derived limit can be considered even more robust. 

In all cases, non-zero decay channels listed in \cref{table2} come with factors of $\kappa$ corresponding to the combination of mixing angles arising from $K_{L/R}^{de}$. These expressions, particularly in the cases of scenario 1 and 2, are quite complicated due to the large number of free parameters allowed. Therefore minimising the decay widths as a function of the free mixing angles is difficult. Instead, we perform a numerical scan of the free parameters in order to estimate the maximum and minimum lower bound on the mass of $X_\mu$ for a given scenario.

\Cref{table11,table12} explicitly calculate the leptoquark mass limits for different choices of the mixing angles appearing in scenario 1 and 2 ordered from largest to smallest limits. The limit on the leptoquark mass varies for different choices but not significantly. The decay channel which forms the dominant constraint also varies for different choices of angles, which we indicate. A numerical scan over the parameter space shows that for scenario 1 the mass limits on $X_\mu$ vary from $81-177$ TeV for different values of the mixing angles. The fourth entry of \cref{table11} is a benchmark taken from~\cite{Smirnov:2018ske} who found in their case a lower bound mass limit of $86$ TeV compared to the $117$ TeV we find. In their analysis they neglected the final-state muon mass and therefore assumed no induced decays of $K_L^0$ from $X_\mu$, whereas we find, when included, it forms the dominant constraint. When neglected we find a limit of $84$ TeV from the process $B_{d/s}^0 \rightarrow \mu \mu$ in full agreement with~\cite{Smirnov:2018ske}. For \cref{table11} we find that, though helicity-suppressed, decays of $K_L^0$ are important to consider for PS breaking limits. For the case of scenario 2 we conducted a similar scan of parameters and found the mass limits on $X_\mu$ varies roughly from $84-102$ TeV as indicated by \cref{table12}.

\begin{table}[t]
\begin{center}
{\renewcommand{\arraystretch}{1.35}
\begin{tabularx}{\linewidth}{c Y Y}
\textsc{scenario 1} & Limit on $m_X$ & Dominant channel\\
\toprule
$\theta_{12}^L=\frac{3\pi}{2},\,\theta_{12}^R=2\pi,\,\theta=\frac{7\pi}{4},\,\delta=\pi$ & $177$ TeV & $K_L^0 \rightarrow \mu \mu$\\
$\theta_{12}^L=\frac{\pi}{4},\,\theta_{12}^R=\frac{\pi}{4},\,\theta=\frac{\pi}{4},\,\delta=0$ & $164$ TeV & $K_L^0 \rightarrow \mu e$\\
$\theta_{12}^L=\frac{\pi}{2},\,\theta_{12}^R=\frac{\pi}{8},\,\theta=0,\,\delta=\frac{\pi}{2}$ & $145$ TeV & $B_d^0 \rightarrow \mu \mu$\\
$\theta_{12}^L=0,\,\theta_{12}^R=0.81,\,\theta=1.183,\,\delta=0$ & $117$ TeV & $K_L^0 \rightarrow \mu e$\\
$\theta_{12}^L=\frac{\pi}{4},\,\theta_{12}^R=\frac{\pi}{4},\,\theta=\frac{\pi}{4},\,\delta=\frac{\pi}{2}$ & $107$ TeV & $B_d^0 \rightarrow \mu \mu$\\
$\theta_{12}^L=2.06,\,\theta_{12}^R=2.4,\,\theta=5.11,\,\delta=4.58$ & $81$ TeV & $B_s^0 \rightarrow \mu e$\\
\end{tabularx}
}
\end{center}
\caption[Limits on $m_X$ for different benchmark values in scenario 1 without a chiral suppression.]{Limits obtained for the gauge leptoquark mass $X_\mu$ for different choices of the mixing angles appearing in \cref{eqn:struct1} ordered from largest to smallest. The decay process from which the dominant limit arises is also listed. In some cases, even though it is helicity-supressed, the dominant limit will still arise from $K_L^0$ decays. The angles in the fourth row were first used in~\cite{Smirnov:2018ske} from which they obtained a limit of $86$ TeV from $B_{d/s}^0 \rightarrow \mu \mu$ when the muon and electron mass were neglected. We find similar limits however we highlight the importance of including the muon mass as the channel $K_L^0 \rightarrow \mu e$ still forms the dominant limit for this scenario. A numerical scan finds the limits in this scenario can vary from $81-177$ TeV.}
\label{table11}
\begin{center}
{\renewcommand{\arraystretch}{1.35}
\begin{tabularx}{\linewidth}{c Y Y}
\textsc{scenario 2} & Limit on $m_X$ & Dominant channel\\
\toprule
$\theta_{12}^L=2.4,\,\theta_{12}^R=2.3,\,\theta_{23}^L=\frac{\pi}{2},\,\theta_{23}^R=0,\,\delta^{L}=2\pi,\,\delta^R=2.77$ & $102$ TeV & $B_s^0 \rightarrow \mu e$\\
$\theta_{12}^L=\frac{\pi}{9},\,\theta_{12}^R=\frac{\pi}{2},\,\theta_{23}^L=1,\,\theta_{23}^R=0,\,\delta^{L/R}=0$ & $100$ TeV & $B_s^0 \rightarrow \mu e$\\
$\theta_{12}^L=\frac{\pi}{3},\,\theta_{12}^R=\frac{\pi}{6},\,\theta_{23}^L=\frac{\pi}{2},\,\theta_{23}^R=1,\,\delta^{L/R}=0$ & $92$ TeV & $B_s^0 \rightarrow \mu \mu$\\
$\theta_{12}^L=\frac{\pi}{3},\,\theta_{12}^R=\frac{\pi}{6},\,\theta_{23}^L=\frac{\pi}{2},\,\theta_{23}^R=1,\,\delta^{L/R}=\frac{\pi}{2}$ & $86$ TeV & $B_s^0 \rightarrow \mu \mu$\\
$\theta_{12}^L=\frac{\pi}{4},\,\theta_{12}^R=\frac{\pi}{8},\,\theta_{23}^L=0,\,\theta_{23}^R=0,\,\delta^{L/R}=0$ & $85$ TeV & $B_s^0 \rightarrow \mu e$\\
$\theta_{12}^L=0.72,\,\theta_{12}^R=3.05,\,\theta_{23}^L=4.02,\,\theta_{23}^R=2\pi,\,\delta^{L}=0,\,\delta^R=\frac{3\pi}{2}$ & $84$ TeV & $B_s^0 \rightarrow \mu e$\\
\end{tabularx}
}
\end{center}
\caption[Limits on $m_X$ for different benchmark values in scenario 2 without a chiral suppression.]{Limits obtained for the gauge leptoquark mass $X_\mu$ for some different choices of mixing angles appearing in \cref{eqn:struct2} ordered from largest to smallest. The decay process from which the dominant limit arises is also listed. Here the decays of $K_L^0$ are completely forbidden and the dominant limit will always arise either from $B_s^0 \rightarrow \mu \mu$ or $B_s^0 \rightarrow \mu e$. Through a numerical scan we find the mass of the PS leptoquark varies between roughly $84-102$ TeV depending on different choices of mixing angles, a much closer range compared to the results of Scenario 2.}
\label{table12}
\end{table}

\subsection{Fermion mass degeneracy}
\label{subsec:fermionmassdeg}

Although the dominant constraint on the PS breaking scale arises from pseudoscalar meson decays, a secondary requirement for a viable Pati-Salam model is to address the lack of mass degeneracy between fermion pairs with the same $SU(2)_{L/R}$ isospin. As indicated by~\cref{eqn:PSchargfermionmass} the simplest PS models lead to the fermion mass relations
\begin{equation}
m_d = m_e \qquad \text{  and  }\qquad m_u = m_\nu^\text{Dirac}
\end{equation} 
for all three generations of SM fermions.

Comparing this prediction to the measured masses of the down-isospin fermions shown in \cref{table3} for the different generations at different energy scales demonstrates that this tree-level relation must be broken. As discussed previously, the mass relation between the up-isospin components can be easily broken due to seesaw mixing in the neutral fermion sector as demonstrated in~\cref{subsec:YukPS}. In the case of the down-isospin components, with no additional particle content, the mass relation is unbroken at tree level and holds at the scale of PS breaking. In scenarios where PS is broken at high scales, this relation could potentially be viable as threshold effects as well as renormalisation group running can potentially be sufficient to explain the observed mass differences of the different generations. PS breaking scales as low as $\mathcal{O}(1000)$ TeV can explain the bottom and tau lepton mass differences~\cite{Volkas:1995yn} as the two Yukawa couplings unify at around this scale. It therefore could be possible for high-scale PS models to explain the mass differences between all three generations in the same way. \Cref{table3} shows that for PS breaking scales below $1000$ TeV the different generations of down-quark and charged-lepton Yukawa couplings cannot be equal at the PS breaking scale and therefore there should be some tree-level explanation for their difference.

\begin{table}[t]
\begin{center}
{\renewcommand{\arraystretch}{1.15}
\scalebox{1.0}{
\begin{tabular}{ccccccc}
\toprule
 & \,$\mu = m_Z$\, &$\mu = 1\text{ TeV}$ &$\mu = 10\text{ TeV}$ & $\mu = 100\text{ TeV}$& $\mu = 1000\text{ TeV}$	 \\ 
\midrule 
\midrule
$m_e/m_d$ & $0.177$  & $0.205$ &$0.230$ & $0.251$ &$0.271$\\
$m_{\mu}/m_s$ & $1.891$  & $2.195 $ & $2.454 $ & $2.688 $ & $ 2.902$\\
$m_{\tau}/m_b$ & $0.612$  & $0.724 $ & $ 0.823$ & $0.913$ & $0.997$\\
\bottomrule
\end{tabular}}}
\end{center}
\caption[Mass ratios $m_e/m_d$ for each generation at different energy scales.]{Measured mass ratios $m_e/m_d$ for each generation at fixed energy scales $\mu$ at one-loop and assuming SM running of the Yukawas. Additional details of the running calculations performed can be found in \cref{sec:Appendix C}.} 
\label{table3}
\end{table}

The mass relations can be broken at tree level by the existence of additional particle content, either scalar or fermion, necessarily transforming as complete PS multiplets above the scale of PS breaking. For example, the inclusion of a scalar $(\textbf{15},\textbf{2},\textbf{2})$ Higgs particle, sometimes referred to as the Minimal Quark-Lepton Symmetric Model (MQLS), which has a non-zero vev will induce a Georgi-Jarlskog like texture~\cite{Georgi:1979df} lifting the degeneracy between the down quark and charged lepton masses~\cite{Pati:1974yy,Pati:1983zp}. The additional Yukawa couplings results in enough freedom such that the mass ratios measured and shown in~\cref{table3} can arise.

We explore an alternative possibility first noted in~\cite{Foot:1997pb,Foot:1999wv} where additional anomaly-free fermion multiplets transforming under the PS gauge group are introduced. If the additional multiplets contain components with the same quantum numbers as the down quarks or charged leptons, mixing effects could induce a see-saw which can decouple the down-quark and charged-lepton Yukawas, breaking the tree-level mass relations obtained without their inclusion. In this scenario, both the up- and down-isospin components have their PS mass relations broken due to see-saw effects, so the breaking of the mass relations between all SM fermions is explained by a similar mechanism.

An additional consequence of introducing extra fermionic states is that they can cause the gauge boson leptoquark $X_\mu$ to couple in a chiral-like way to the light SM-like fermions. As an example, consider the introduction of fermion multiplets $F_{L/R}$ that contain components $E_{L/R}^-$ that have the same quantum numbers as the SM charged leptons:
\begin{equation}
f_{L/R} = \begin{pmatrix}
u_r & d_r \\
u_b & d_b \\
u_g & d_g \\
\nu_e & e \end{pmatrix}_{L/R} \,\,\quad \oplus \quad\,\, F_{L/R} = \begin{pmatrix} 
     & \dots &  & \\
    \vdots & \ddots & & \\
      &  & E_{L/R}	^-  & \\
      &   &    & \,\,\ddots\,\,\,
    \end{pmatrix}.
\end{equation}
Here both $E_L^-$ and $E_R^-$ are required phenomenologically such that no massless charged fermion states appear and the exotic multiplets $F_L$ and $F_R$ need not transform in the same way. However, the combination must be anomaly free.

If Yukawa interactions connecting the multiplets $F_{L/R}$ and $f_{L/R}$ exist, mass mixing will be induced as per
\begin{equation}
\mathcal{L}_{eE} = \begin{pmatrix}
\overline{e_L} & \overline{E_L}
\end{pmatrix} \begin{pmatrix}
m_{ee} & m_{eE}\\
m_{Ee} & m_{EE}
\end{pmatrix}\begin{pmatrix}
e_R\\
E_R
\end{pmatrix} + \text{H.c.}
\end{equation}
Diagonalising into the mass basis for the charged fermions leads to 
\begin{equation}
\label{eqn:genericemixing}
(e'_{L/R})_{\mathscr{l}}^{\vphantom{T}} = c_{\theta_{L/R}} e_{L/R} + s_{\theta_{L/R}} E_{L/R} \qquad\qquad
(E'_{L/R})_{\mathscr{h}}^{\vphantom{T}} = -s_{\theta_{L/R}} e_{L/R} + c_{\theta_{L/R}} E_{L/R}
\end{equation}
where the subscripts $\mathscr{l}$ and $\mathscr{h}$ indicate the light and heavy eigenstates respectively. The pseudoscalar meson decays discussed above are induced by the gauge leptoquark interactions between the colour triplet $d$ and charged lepton $e$ from the multiplet $f_{L/R}$. 

Expanding \cref{eqn:elldXgauge} with \cref{eqn:genericemixing} leads to
\begin{equation}
\label{eqn:ellXmassmix}
\mathcal{L}_{Xd e} = \frac{g_4}{\sqrt{2}}\left( \overline{d'}\, K^{d e}_L\, \slashed{X} P_L (c_{\theta_L} e' - s_{\theta_L} E') ) + \overline{d'} \,K^{d e}_R \,\slashed{X}P_R (c_{\theta_R} e' - s_{\theta_R} E') \right) + \text{ H.c}
\end{equation}
for the gauge interactions in the mass basis where generational indices have been suppressed for simplicity. Because of phenomenological constraints, any fermions with SM quantum numbers must be significantly heavier than the pseudoscalar mesons whose decays supply the dominant constraint on PS breaking; therefore, processes such as $K_L^0 \rightarrow E E$ or $K_L^0 \rightarrow e E$ are kinematically forbidden. The decay $K_L^0 \rightarrow e e$ will exist as before, but now suppressed by the relevant mass mixing angles. Therefore the only relevant interactions in \cref{eqn:ellXmassmix} for meson decay are
\begin{equation}
\label{eqn:ellXmassmix1}
\mathcal{L}_{Xd e} \supset \frac{g_4}{\sqrt{2}}\left( \overline{d'}\, (c_{\theta_L} K^{d e}_L)\, \slashed{X} P_L e' + \overline{d'} \,(c_{\theta_R} K^{d e}_R) \,\slashed{X} P_R e'\right) + \text{ H.c}
\end{equation}
which will lead to a decay rate given by \cref{eqn:HScont,eqn:HUcont} with the replacement $K^{de}_{L/R} \rightarrow \mathcal{K}^{de}_{L/R} = c_{\theta_{L/R}} K^{de}_{L/R}$. As the mixing angles $\theta_L$ and $\theta_R$ can significantly differ, this can effectively lead to a chiral coupling between $X_\mu$ and the fermions $d$ and $e$ causing a suppression in the helicity-unsupressed contribution of the total meson decay rates in \cref{eqn:twobodydecaywidth}. This therefore allows for an overall weaker lower bound on the leptoquark mass compared to those in \cref{table2} and therefore the scale of PS breaking. As noted previously, however, in order to significantly helicity-suppress the decays mediated by $X_\mu$, one of the angles $\theta_{L/R}$ is required to be very small e.g.\ $\theta_R \lesssim 10^{-4}$ in the case of $K_L^0 \rightarrow \mu e$ decays.

\afterpage{
\begin{landscape}
\begin{table}[t]
\begin{center}
{\renewcommand{\arraystretch}{1.25}
\scalebox{1.0}{
\begin{tabular}{cccccc}
\toprule
 & \,$K^{d e}_L = \id_{3\times3}$ \, & \,\textsc{scenario} 1 \, & \textsc{scenario} 2 \,& \,\textsc{scenario} 3 \,& \,\textsc{scenario} 4 \,	 \\ 
\midrule 
\midrule
$\mathcal{B}(K_L^0 \rightarrow e e)$ & $0$ & $13\, \kappa^{K^{ee}}_1 \text{ TeV}$ & $0$ & $13\,\kappa^{K^{ee}}_3 \text{ TeV}$ & $0$ \\
$\mathcal{B}(K_L^0 \rightarrow \mu \mu )$ & $0$ & $177\, \kappa^{K^{\mu\mu}}_1 \text{ TeV}$ & $0$ & $0$ & $177 \,\kappa^{K^{\mu\mu}}_4 \text{ TeV}$ \\
$\mathcal{B}(K_L^0 \rightarrow \mu e )$ & $194\text{ TeV}$ & $230\, \kappa^{K^{\mu e}}_1 \text{ TeV}$ & $0$ & $0$ & $0$\\
\midrule
$\mathcal{B}(B_d^0 \rightarrow ee)$ & $0$ & $0.2\, \kappa^{B^{ee}}_1 \text{ TeV}$ & $0$ & $0$ & $0$\\
$\mathcal{B}(B_d^0 \rightarrow \mu\mu)$ & $0$ & $10.8\, \kappa^{B^{\mu \mu}}_1 \text{ TeV}$ & $0$ & $0$ & $0$\\
$\mathcal{B}(B_d^0 \rightarrow \tau\tau)$ & $0$ & $0$ & $0$ & $0$ & $0$\\
$\mathcal{B}(B_d^0 \rightarrow \mu e)$ & $0$ & $10.0\, \kappa^{B^{\mu e}}_1 \text{ TeV}$ & $0$ & $10.0\, \kappa^{B^{\mu e}}_3 \text{ TeV}$ & $10.0\, \kappa^{B^{\mu e}}_4 \text{ TeV}$\\
$\mathcal{B}(B_d^0 \rightarrow \tau e)$ & $2.7\text{ TeV}$ & $3.2\, \kappa^{B^{\tau e}}_1 \text{ TeV}$ & $3.2\, \kappa^{B^{\tau e}}_2 \text{ TeV}$ & $0$ & $3.2\, \kappa^{B^{\tau e}}_4 \text{ TeV}$\\
$\mathcal{B}(B_d^0 \rightarrow \tau \mu )$ & $0$ & $3.2\, \kappa^{B^{\tau \mu}}_1 \text{ TeV}$ & $3.2\, \kappa^{B^{\tau \mu}}_2 \text{ TeV}$ & $3.2\, \kappa^{B^{\tau \mu}}_3 \text{ TeV}$ & $0$\\
\midrule
$\mathcal{B}(B_s^0 \rightarrow ee)$ & $0$ & $0.2\, \kappa^{B^{e e}_s}_1 \text{ TeV}$ & $0.2\, \kappa^{B^{e e}_s}_2 \text{ TeV}$ & $0$ & $0$ \\
$\mathcal{B}(B_s^0 \rightarrow \mu\mu)$ & $0$ & $6.5\, \kappa^{B^{\mu \mu}_s}_1 \text{ TeV}$ & $6.5\, \kappa^{B^{\mu \mu}_s}_2 \text{ TeV}$ & $0$ \\
$\mathcal{B}(B_s^0 \rightarrow \tau\tau)$ & $0$ & $0$ & $0$ & $0$ & $0$ \\
$\mathcal{B}(B_s^0 \rightarrow \mu e )$ & $0$ & $6.7\, \kappa^{B^{\mu e}_s}_1 \text{ TeV}$ & $6.7\, \kappa^{B^{\mu e}_s}_2 \text{ TeV}$ & $6.7\, \kappa^{B^{\mu e}_s}_3 \text{ TeV}$ & $6.7\, \kappa^{B^{\mu e}_s}_4 \text{ TeV}$\\
$\mathcal{B}(B_s^0 \rightarrow \tau e )$ & $0$ & $-$ & 	$0$ & $0$ & $-$ \\
$\mathcal{B}(B_s^0 \rightarrow \tau \mu )$ & $-$ & $-$ & $0$ & $-$ & $0$ \\
\bottomrule
\end{tabular}
}
}
\end{center}
\caption[Limits on $m_X$ compared to current measurements or upper limits with a full chiral suppression.]{Limits on the gauge leptoquark mass $m_X$ compared to current measurements (or upper limits) for the mesons $K_L^0$, $B_d^0$ and $B_s^0$ in the chiral limit (e.g. $c_{\theta_R}=0$) where all decays are now helicity suppressed. Eeach scenario is given by \cref{eqn:struct1,eqn:struct2,eqn:struct3,eqn:struct4} respectively. As \textit{all} decays are now helicity-suppressed, the dominant decay channel in scenario 1 will arise from $K_L^0 \rightarrow \mu e$ unless forbidden by a specific choice of $\kappa$. As Scenario 2 completely suppresses the decays of $K_L^0$ and now the decays of $B_{d/s}^0$ are helicity-suppressed, the limits on the leptoquark mass quite substantially decrease, similarly the limits from scenarios 3 and are significantly reduced. In particular scenario 3 allows for incredibly low mass scales due to the large helicity-suppression present for electron final-states. The first column represents a scenario which maximises the rate of $K_L^0 \rightarrow \mu e$ decays leading to the largest mass limit on $X_\mu$ which in this case is roughly $200$ TeV. In all the scenarios above the parameter $\kappa^{X}_{\alpha}$ corresponds to the combination of mixing angles relevant to that decay chain. We find the lower bound on the PS breaking scale to be no smaller than $\mathcal{O}(10)$ TeV. For the final two channels $B_s^0 \rightarrow \tau e / \tau \mu$ there are currently no measured upper bounds, however non-zero contributions to these channels are indicated by `$-$' and their future measurement will form a constraint for a given choice of $K^{de}_{L/R}$.} 
\label{table4}
\end{table}
\end{landscape}
}

\Cref{table4} demonstrates the impact this can have on the mass limits in the extreme case of an exactly chiral theory ($m_{E} \rightarrow \infty$) where for example $c_{\theta_L} =1$ and $c_{\theta_R} = 0$ for all three generations\footnote{These limits are also valid for scenarios where quark-lepton unification occurs for only one chirality of fermions, e.g. a gauge group given by $SU(4)_L \times SU(3)_{R} \times SU(2)_L \times SU(2)_R \rightarrow SU(3)_c \times SU(2)_L \times SU(2)_R$ where the vector leptoquark $X$ couples to only one chirality of fermions.}, assuming the same matrix textures for $K^{de}_{L/R}$ as in \cref{table2}. The leptoquark mass limits are significantly lowered compared to \cref{table2} as the helicity unsuppressed contribution from \cref{eqn:HUcont} no longer contributes and therefore the decay rate is suppressed by the charged lepton masses. Scenario 1, which already had helicity-suppressed $K_L^0$ decays, will have its limits largely unchanged except for when the dominant channel arises from $B_{d/s}^0$. If there are no contributions to the decays of $K_L^0$, as in scenario 2, then the helicity suppression on the other decay channels allows for PS breaking scales as low as $\mathcal{O}(10)$ TeV. Interestingly, in these scenarios with a chiral-like coupled $X_\mu$, if there is a significant contribution to $K_L^0 \rightarrow ee$, then the limits are significantly smaller compared to $K_L^0 \rightarrow \mu\mu[\mu e]$ due to the large helicity-suppression present for the electron. Therefore with the presence of exotic fermion multiplets, a signficant contribution to $K_L^0$ decays can be possible with small leptoquark masses provided it only couples $K_L^0$ to electrons. This is unlike the case without mass mixing where any induced decay for $K_L^0$ by $X_\mu$ causes mass limits larger than $1000$ TeV irrespective of the final decay product. In such a chiral scenario only one of $K_L^{de}$ and $K_R^{de}$ is required to have a matrix structure given by each indicated scenario as the other is significantly suppressed through seesaw effects. Therefore for scenarios involving charged-lepton or down-quark seesaws, $K_{L/R}^{de}$ do not need to be related and therefore no parity symmetry needs to be imposed at a high scale.

In \cref{table13,table14} the limits on $m_X$ are re-evaluated for the same benchmark scenarios in \cref{table11,table12} now assuming an exact helicity suppression. In both scenarios a significant reduction in the mass limits can occur. This reduction occurs for any choice of mixing angles in scenario 2, whereas in scenario 1, a significant reduction occurs only when the dominant decay channel comes from either $B_d^0$ or $B_s^0$. In general a reduction in the limits on $m_X$ by a factor of $0.05-0.07$ naturally occurs compared to the scenario without additional mass mixing. We find that the limits on $m_X$ can be as low as $5$ TeV potentially allowing for discovery signals of different PS particles comfortably within reach of current and possible future hadron colliders.

Exotic fermion multiplets are therefore an attractive feature for low scale Pati Salam models. It is curious that they allow for significantly lighter PS breaking scales whilst also potentially breaking the mass degeneracy amongst the down-isospin fermions. This effect is possible not only for mixing between the charged leptons as discussed above, but also if heavy versions of the down quarks, $D$, were to be included which mix with the light $d$ quarks. This would lead to a similar reduction in the mass limits on $X_\mu$ as above for the same reasons.

\begin{table}[t]
\begin{center}
{\renewcommand{\arraystretch}{1.35}
\begin{tabularx}{\linewidth}{c Y Y Y}
\textsc{scenario 1} & Limit on $m_X$ & Reduction & Dominant channel\\
\toprule
$\theta_{12}^L=\frac{3\pi}{2},\,\theta=\frac{7\pi}{4},\,\delta=\pi$ & $177$ TeV & 1 & $K_L^0 \rightarrow \mu \mu$\\
$\theta_{12}^L=\frac{\pi}{4},\,\theta=\frac{\pi}{4},\,\delta=0$ & $139$ TeV & 0.85 &$K_L^0 \rightarrow \mu e$\\
$\theta_{12}^L=2.06,\,\theta=5.11,\,\delta=4.58$ & $51$ TeV &0.63 & $K_L^0 \rightarrow \mu \mu$\\
$\theta_{12}^L=0,\,\theta=1.183,\,\delta=0$ & $11$ TeV & 0.09 &$K_L^0 \rightarrow e e$\\
$\theta_{12}^L=\frac{\pi}{2},\,\theta=0,\,\delta=\frac{\pi}{2}$ & $8.5$ TeV & 0.06 &$B_d^0 \rightarrow \mu e$\\
$\theta_{12}^L=\frac{\pi}{4},\,\theta=\frac{\pi}{4},\,\delta=\frac{\pi}{2}$ & $7.8$ TeV & 0.07 &$B_d^0 \rightarrow \mu \mu$\\
\end{tabularx}
}
\end{center}
\caption[Limits on $m_X$ for different benchmark values in scenario 1 with a chiral
suppression.]{Limits obtained for the gauge leptoquark mass $X_\mu$ for different choices of the mixing angles appearing in \cref{eqn:struct1} with an exact helicity suppression where we have chosen $K_R^{de} = 0_{3\times 3} $. The decay process from which the dominant limit arises is also listed as well as the percentage reduction from the non-helicity suppressed scenario given in \cref{table11}. In some cases only a small reduction in the mass limits occurs, this is a result of the Kaon decay channels being helicitiy suppressed as a result of the structure of the mixing matrices instead of due to a chiral coupling of $X_\mu$. Only the benchmark scenarios which initially had dominant decay channels arising from $B_d$ or $B_s$ experience a significant reduction in their limits as they were initially not helicity suppressed.}
\label{table13}

\begin{center}
{\renewcommand{\arraystretch}{1.35}
\begin{tabularx}{\linewidth}{c Y Y Y}
\textsc{scenario 2} & Limit on $m_X$ & Reduction & Dominant channel\\
\toprule
$\theta_{12}^L=2.4,\,\theta_{23}^L=\frac{\pi}{2},\,\delta^{L}=2\pi$ & $6.2$ TeV & 0.06 & $B_s^0 \rightarrow \mu e$\\
$\theta_{12}^L=\frac{\pi}{9},\,\theta_{23}^L=1,\,\delta^{L}=0$ & $6.0$ TeV& 0.06 & $B_s^0 \rightarrow \mu e$\\
$\theta_{12}^L=\frac{\pi}{4},\,\theta_{23}^L=0,\,\delta^{L}=0$ & $6.0$ TeV& 0.07 & $B_s^0 \rightarrow \mu e$\\
$\theta_{12}^L=0.72,\,\theta_{23}^L=4.02,\,\delta^{L}=0$ & $5.9$ TeV & 0.07& $B_s^0 \rightarrow \mu e$\\
$\theta_{12}^L=\frac{\pi}{3},\,\theta_{23}^L=\frac{\pi}{2},\,\delta^{L}=0$ & $5.6$ TeV& 0.06 & $B_s^0 \rightarrow \mu \mu$\\
$\theta_{12}^L=\frac{\pi}{3},\,\theta_{23}^L=\frac{\pi}{2},\,\delta^{L}=\frac{\pi}{2}$ & $5.6$ TeV & 0.065& $B_s^0 \rightarrow \mu \mu$\\
\end{tabularx}
}
\end{center}
\caption[Limits on $m_X$ for different benchmark values in scenario 1 with a chiral
suppression.]{Limits obtained for the gauge leptoquark mass $X_\mu$ for some different choices of mixing angles appearing in \cref{eqn:struct2} with an exact helicity suppression where we have chosen $K_R^{de} = 0_{3 \times 3} $. The decay process from which the dominant limit arises is also listed as well as the percentage reduction from the non helicity-suppressed scenarios in \cref{table12}. Each benchmark experiences a significant reduction in their mass limits of over an order of magnitude allowing for extremely light masses of $X_\mu$ and significantly lower scales of PS breaking.}
\label{table14}
\end{table}

\section{Exotic PS fermion multiplets}
\label{sec:exoticPS}

As discussed above, exotic PS fermion multiplets which contain states with the correct quantum numbers to mix with the charged leptons or down quarks can simultaneously explain the experimentally observed mass non-degeneracy between these two types of fermions as well as lower the phenomenologically-allowed scale of PS breaking. A number of different viable PS multiplets are possible; for consistency we require that the multiplets added do not introduce anomalies and that they allow for a phenomenologically valid mass spectrum. For simplicity we focus on small multiplets which transform with no more than two indices total under all three gauge groups when written in their defining representations. All possible combinations satisfying this requirement are listed in \cref{table5}. We take the scalar content of the theory to remain unchanged but will note where additional exotic scalars may be required for a viable model in some cases. Although the PS gauge group is an attractive subgroup of some GUTs e.g.\ $SO(10)$ or $E_6$, we will not require successful gauge-coupling unification or partial unification. We will not consider the location of Landau poles\footnote{These would further motivate considering multiplets of small dimensionality as large multiplets will have a much more significant impact on RGE evolution relevant for viable GUT theories.} or require that the extra exotic fermions fit into complete GUT multiplets.  We simply focus on the specific particle content required at low scales sufficient to lift the mass degeneracy between $\ell$ and $d$ and simultaneously lower the scale of PS breaking. 

\begin{table}[t]
\begin{center}
{\renewcommand{\arraystretch}{1.15}
\scalebox{1.0}{
\begin{tabular}{ccccc}
\toprule
 & PS multiplet & $A(R)$ & \,\,$SU(3)\otimes U(1)_Q$ decomposition\,\, & Candidate 	 \\ 
\midrule 
\midrule
$f^a$ & $(\textbf{1},\textbf{2},\textbf{1})$ & $0$ & $\textbf{1}_{-1/2} \oplus \textbf{1}_{+1/2}$ & \ding{54}\\
$f^\alpha$ & $(\textbf{1},\textbf{1},\textbf{2})$ & $0$ & $\textbf{1}_{-1/2} \oplus \textbf{1}_{+1/2}$ & \ding{54}\\
$f^A$ & $(\textbf{4},\textbf{1},\textbf{1})$ &$\pm 1$ & $\textbf{1}_{-1/2} \oplus \textbf{3}_{+1/6}$ & \ding{54}\\
\midrule
$f^a_{\hphantom{a}b}$ & $(\textbf{1},\textbf{3},\textbf{1})$ & $0$ & $\textbf{1}_{-1} \oplus \textbf{1}_0 \oplus \textbf{1}_{+1}$ & \ding{52}\\
$f^\alpha_{\hphantom{\alpha}\beta}$ & $(\textbf{1},\textbf{1},\textbf{3})$ &$0$  & $\textbf{1}_{-1} \oplus \textbf{1}_0 \oplus \textbf{1}_{+1}$ & \ding{52}\\
$f^{AB}$ & $(\textbf{6},\textbf{1},\textbf{1})$ & $0$ & $\textbf{3}_{-1/3} \oplus \overline{\textbf{3}}_{+1/3}$ & \ding{52}\\
$f^{AB}$ & $(\textbf{10},\textbf{1},\textbf{1})$ & $\pm 8$ & $\textbf{1}_{-1} \oplus \textbf{3}_{-1/3} \oplus \textbf{6}_{+1/3}$ & \ding{52}\\
$f^A_{\hphantom{A}B}$ & $(\textbf{15},\textbf{1},\textbf{1})$ & $0$ & $\overline{\textbf{3}}_{-2/3} \oplus \textbf{1}_0 \oplus \textbf{8}_0 \oplus \textbf{3}_{+2/3}$ & \ding{54}\\
\midrule
$f^{a\alpha}$ & $(\textbf{1},\textbf{2},\textbf{2})$ & $0$ & $\textbf{1}_{-1} \oplus \textbf{1}_0 \oplus \textbf{1}_0 \oplus \textbf{1}_{+1}$ & \ding{52}\\
$f^{A a}$ & $(\textbf{4},\textbf{2},\textbf{1})$ & $\pm 2$ & $\textbf{1}_{-1} \oplus \textbf{3}_{-1/3} \oplus \textbf{1}_0 \oplus \textbf{3}_{+2/3}$ & \ding{52}\\
$f^{A \alpha}$ & $(\textbf{4},\textbf{1},\textbf{2})$ &$\pm 2$ & $\textbf{1}_{-1} \oplus \textbf{3}_{-1/3} \oplus \textbf{1}_0 \oplus \textbf{3}_{+2/3}$ & \ding{52}\\
\bottomrule
\end{tabular}
}
}
\end{center}
\caption[Different dimensional representations of possible PS fermions which lead to charged-lepton or down-quark mixing.]{Different dimensional representations of fermions $f$ (with indices in their defining representation) under PS where $a,\,\alpha$ and $A$ correspond to $SU(2)_L,\,SU(2)_R$ and $SU(4)_c$ indices respectively. Multiplets are considered good candidates if they contain states with the same quantum numbers as either the down quarks or charged leptons, which will mix with its SM counterpart, when broken to $SU(3)_c \otimes U(1)_Q$ . Additionally the $SU(4)$ anomaly coefficient $A(R)$ for each fermion is indicated where the sign depends on whether the fermion is left- or right-handed. } 
\label{table5}
\end{table}

\subsection{Fermion extensions}
\label{sec:exoticPS1}

Considering the relevant exotic multiplets indicated in \cref{table5}, below we study basic phenomenological implications of introducing each given multiplet. We indicate heavy, exotic versions of down quarks, charged leptons and neutrinos by $D$, $E$ and $N$ respectively. We will also briefly comment on the implications for baryon number violation in each case. As we are only introducing exotic fermions, the Yukawa sector of the Lagrangian is the only possible area where additional violation could arise.

\subsubsection{$SU(2)_{L/R}$ Triplets}
\label{sec:tripPS}

If a triplet fermion is added, for example $\Psi_{3}^R \sim (\textbf{1},\textbf{1},\textbf{3})$, the extra terms appearing in the Yukawa Lagrangian are
\begin{equation}
\label{eqn:YUK3}
\mathcal{L}_{\textsc{yuk}^{\vphantom{T}}} \supset \text{Tr}\left[ \sqrt{2}\, \mphant{\mathlarger{y}_{\Psi_3}}{T} \overline{(\Psi_3^R)^c} \rchi_R^\dagger f_R + \frac{1}{2}\mphant{\mathlarger{\mu}_{\Psi_3}}{T}	 \overline{(\Psi_3^R)^c} \,(\Psi_3^R)^T \right] + \text{H.c.}
\end{equation}
As $\Psi_3^R$ is uncharged under the colour group $SU(4)_c	$, it contains no quark states. However, it contains components which can mix with both the charged and neutral SM leptons
\begin{equation}
\label{eqn:tripemb}
f_{L/R} = \begin{pmatrix}
\cdot & \cdot \\
\cdot & \cdot \\
\cdot & \cdot \\
\nu & e \end{pmatrix}_{L/	R}, \qquad \Psi_3^R = \frac{1}{\sqrt{2}}\begin{pmatrix}
N_R & \sqrt{2}\,(E^-_L)^c \\
\sqrt{2}\, E_R^- & -N_R\\
\end{pmatrix}
\end{equation}
where the dotted components of $f_{L/R}$ do not mix with $\Psi_3^R$ but the undotted do and $\Psi_3^R \rightarrow \mphant{U_R}{\dagger}\, \Psi_3^R \,U_R^\dagger$. Here we have assigned $\Psi_3^R$ to be a right-handed fermion without loss of generality. Expanding out \cref{eqn:PSYuk,eqn:YUK3} with the parameterisation given in \cref{eqn:tripemb} leads to a mixing matrix given by
\begin{equation}
\label{eqn:tripchargelepmixR}
\mathcal{L}_{eE} = \begin{pmatrix}
\overline{e_L} & \overline{E_L}
\end{pmatrix} \begin{pmatrix}
m_e & 0 \\
\sqrt{2}\, y_{\Psi_3}^R v_R^* & \,\,\,\mu_{\Psi_3^R}
\end{pmatrix}
\begin{pmatrix}
e_R\\
E_R
\end{pmatrix}
\end{equation}
for the charged lepton states and
\begin{equation}
\label{eqn:tripneutlepmixR}
\mathcal{L}_{\nu N}=\frac{1}{2} \begin{pmatrix}
\overline{\nu_L} & \overline{\nu_R^c} & \overline{N_R^c} 
\end{pmatrix}\begin{pmatrix}
0 & m_u & 0 \\
m_u & 0 & y_{\Psi_3}^R v_R^*\\
0 & y_{\Psi_3}^R v_R^* & \mu_{\Psi_3^R} \\
\end{pmatrix}\begin{pmatrix}
\nu_L^c\\
\nu_R\\
N_R
\end{pmatrix}
\end{equation}
for the neutral states. If $f_L$, $f_R$ and $\Psi$ are the only fermions introduced, the PS symmetry enforces that the singular values of $m_d$ and $m_u$ are given by the down and up quark masses respectively. \Cref{eqn:tripneutlepmixR} assumes that the fermion singlets $S_L$ discussed in \cref{sec:standardPS}, appearing in the usual low-scale PS scenario, are not included. As can be seen from the neutrino mass mixing matrix, $S_L$ may no longer be necessary for light neutrino masses as an inverse seesaw arises naturally with the triplet. We therefore neglect $S_L$, and if it were to be included \cref{eqn:tripneutlepmixR} would be trivially extended by the last row and column of \cref{eqn:PSneutralmix}.

The terms introduced in \cref{eqn:YUK3} do not violate the global $U(1)_J$ symmetry which \cref{eqn:PSYuk} obeys and is the only global accidental symmetry of this Yukawa Lagrangian. These terms unsurprisingly enforce the assignment
\begin{equation}
J(\Psi_3^R) = 0 
\end{equation}
as $\Psi_3^R$ transforms as a real representation. Therefore baryon number remains an unbroken global symmetry of the Yukawa sector similar to the vanilla scenario described in \cref{sec:Appendix Z}. Interestingly if the singlet $S_L$ is not included, the scalar $\rchi_L$ no longer Yukawa couples to any of the fermions and the relevant term of the scalar potential given in \cref{eqn:scalarpotbaryonV} no longer violates the $U(1)_J$ global symmetry as we are free to choose $J(\rchi_L) = -1$.  The proton would therefore remain absolutely stable (assuming no additional particle content) after breaking of the PS symmetry, whereas lepton number would be broken.

If instead an $SU(2)_L$ triplet fermion $\Psi_3^L$ was introduced, similar conclusions are reached related to the leptonic mass mixing matrices and baryon number violation\footnote{However if both $\Psi_3^L$ and $\Psi_3^R$ are included simultaneously, the Yukawa Lagrangian enforces $J(\rchi_L) = J(\rchi_R) = 1$ and the relevant quartic term of the scalar potential would once again violate $U(1)_J$ and therefore $B$ regardless of the existence of $S_L$.}. The Yukawa Lagrangian in this case is similar to the one appearing in \cref{eqn:YUK3} with the replacements $\Psi_3^R \rightarrow (\Psi_3^L)^c$, $\rchi_R \rightarrow (\rchi_L)^*$ and $f_R \rightarrow (f_L)^c$ leading to the mixing matrix
\begin{equation}
\label{eqn:tripchargelepmixL}
\mathcal{L}_{eE} = \begin{pmatrix}
\overline{e_L} & \overline{E_L}
\end{pmatrix} \begin{pmatrix}
m_e & \sqrt{2}\, y_{\Psi_3}^L v_L \\
0 & \,\,\,\mu_{\Psi_3^L}
\end{pmatrix}
\begin{pmatrix}
e_R\\
E_R
\end{pmatrix}
\end{equation}
for the charged lepton states and
\begin{equation}
\label{eqn:tripneutlepmixL}
\mathcal{L}_{\nu N}=\frac{1}{2} \begin{pmatrix}
\overline{\nu_L} & \overline{\nu_R^c} & \overline{N_L} 
\end{pmatrix}\begin{pmatrix}
0 & m_u & y_{\Psi_3}^L v_L \\
m_u & 0 & 0 \\
y_{\Psi_3}^L v_L & 0 & \mu_{\Psi_3^L} \\
\end{pmatrix}\begin{pmatrix}
\nu_L^c\\
\nu_R\\
N_L^c
\end{pmatrix}
\end{equation}
for the neutral states.

If both $\Psi_3^L$ and $\Psi_3^R$ are included simultaneously the mass mixing matrices for the charged and neutral sectors are simply a combination of \cref{eqn:tripchargelepmixR,eqn:tripchargelepmixL,eqn:tripneutlepmixR,eqn:tripneutlepmixL}. Explicitly writing out the mass mixing matrices gives
\begin{equation}
\label{eqn:tripchargelepmixL&R}
\mathcal{L}_{eE\mathcal{E}} = \begin{pmatrix}
\overline{e_L} & \overline{E_L} & \overline{\mathcal{E}_L}
\end{pmatrix} \begin{pmatrix}
m_e & 0 & \sqrt{2}\, y_{\Psi_3}^L v_L \\[2pt]
\sqrt{2}\, y_{\Psi_3}^R v_R & \,\,\,\mu_{\Psi_3^R} & 0 \\[2pt]
0 & 0 & \,\,\,\mu_{\Psi_3^L}
\end{pmatrix}
\begin{pmatrix}
e_R\\
E_R\\
\mathcal{E}_R
\end{pmatrix}
\end{equation}
for the charged leptons, where $E_{L/R}$ and $\mathcal{E}_{L/R}$ correspond to the charged lepton states appearing in $\Psi_3^R$ and $\Psi_3^L$ respectively, and
\begin{equation}
\label{eqn:tripneutlepmixL&R}
\mathcal{L}_{\nu N}=\frac{1}{2} \begin{pmatrix}
\overline{\nu_L} & \overline{\nu_R^c} & \overline{N_L} & \overline{N_R^c}
\end{pmatrix}\begin{pmatrix}
0 & m_u & y_{\Psi_3}^L v_L & 0\\
m_u & 0 & 0 &  y_{\Psi_3}^R v_R^*\\
y_{\Psi_3}^L v_L & 0 & \mu_{\Psi_3^L} & 0 \\
0 &  y_{\Psi_3}^R v_R^* & 0 & \mu_{\Psi_3^R}
\end{pmatrix}\begin{pmatrix}
\nu_L^c\\
\nu_R\\
N_L^c\\
N_R
\end{pmatrix}
\end{equation}
for the neutral states.

\subsubsection{$SU(2)_L/SU(2)_R$ Bi-doublet}
\label{sec:biPS}

\begin{sloppypar} 
If a fermion transforming as a bi-doublet under the two $SU(2)$ gauge groups of PS ${\Psi_{22}\sim(\textbf{1},\textbf{2},\textbf{2})}$ is added, the Yukawa Lagrangian is extended by
\begin{align}
\label{eqn:YUKbi}
\mathcal{L}_{\textsc{yuk}^{\vphantom{T}}} \supset \text{Tr} \Big[ \mathlarger{y}_{\Psi_{22}}^R 	\overline{f_L} \mphant{\rchi_R}{\dagger} (\Psi_{22}^T)^c (i \tau_2) + \mathlarger{y}_{\Psi_{22}}^L	  \overline{\Psi_{22}} (i \tau_2) \rchi_L^\dagger f_R  + \mathlarger{\mu}_{\Psi_{22}}^{\vphantom{T}} \overline{\Psi_{22}} &(i \tau_2) (\Psi_{22})^c (i \tau_2) \Big] \nonumber\\
 & + \text{H.c}.
\end{align}
As before, $\Psi_{22}$ is uncharged under the $SU(4)_c$ colour group and therefore only contains uncoloured states which can mix with the SM leptons:
\begin{equation}
\label{eqn:biemb}
f_{L/R} = \begin{pmatrix}
\cdot & \cdot \\
\cdot & \cdot \\
\cdot & \cdot \\
\nu & e \end{pmatrix}_{L/	R}, \qquad\Psi_{22} = \begin{pmatrix}
-(E_R^-)^c & N_L^0\\
\hphantom{-}(N_R^0)^c & E_L^-
\end{pmatrix}
\end{equation}
where $\Psi_{22}$ is written with two raised indices and therefore transforms as ${\Psi_{22} \rightarrow U_L^{\vphantom{T}} \mphant{\Psi_{22}}{T} U_R^T}$. Expanding out \cref{eqn:PSYuk,eqn:YUKbi} with \cref{eqn:biemb} leads to
\begin{equation}
\label{eqn:bidoubchargelepmix}
\mathcal{L}_{eE} = \begin{pmatrix}
\overline{e_L} & \overline{E_L}
\end{pmatrix} \begin{pmatrix}
m_e  &  \mathlarger{y}_{\Psi_{22}}^R v_R\\
\mathlarger{y}_{\Psi_{22}}^L v_L^*  & \mu_{\Psi_{22}} 
\end{pmatrix}
\begin{pmatrix}
e_R\\
E_R
\end{pmatrix}
\end{equation}
for the charged lepton mass mixing matrix and
\begin{equation}
\label{eqn:bidoubneutlepmix}
\mathcal{L}_{\nu N}=\frac{1}{2}\begin{pmatrix}
\overline{\nu_L} & \overline{\nu_R^c} & \overline{N_L} & \overline{N_R^c} 
\end{pmatrix} \begin{pmatrix}
0 & m_u & 0 & \mathlarger{y}_{\Psi_{22}}^R v_R\\
m_u & 0 & 0 & \mathlarger{y}_{\Psi_{22}}^L v_L^*\\
0 & 0 & 0 & \mu_{\Psi_{22}}  \\
\mathlarger{y}_{\Psi_{22}}^R v_R & \mathlarger{y}_{\Psi_{22}}^L v_L^* & \mu_{\Psi_{22}}  & 0
\end{pmatrix}
\begin{pmatrix}
\nu_L^c\\
\nu_R\\
N_L^c\\
N_R
\end{pmatrix}
\end{equation}
for the neutral mass mixing.
\end{sloppypar}

Similar to the case of the triplet, \cref{eqn:YUKbi} does not violate the global $U(1)_J$ symmetry (which, recall, is the only accidental symmetry in this Yukawa Lagrangian) and implies
\begin{equation}
J(\Psi_{22}) = 0.
\end{equation}
Therefore $U(1)_J$ is not violated in the Yukawa Lagrangian and baryon number violation proceeds through the scalar sector as before. Unlike the scenarios involving only one triplet, in the absence of the fermion $S_L$, \cref{eqn:YUKbi} still enforces the choice $J(\rchi_L) = J(\rchi_R) = 1 $ and therefore baryon number will remain violated if both scalars are included and will proceed via similar diagrams to those presented in \cref{sec:Appendix Z}.

\subsubsection{$SU(4)_c/SU(2)_{L/R}$ Bi-fundamentals}
\label{sec:vectorPS}

\begin{sloppypar}
In the case where additional copies of the usual PS fermions exist, there are a number of different possible scenarios depending on the chirality of the fermions introduced. An even number of fermion bi-fundamentals needs to be introduced, as can be seen in \cref{table5}, for anomaly cancellation and therefore we focus on the minimal scenario where a pair of bi-fundamentals are added. If two exact copies of $f_L$ and $f_R$ are added, that is a left-handed fermion transforming as $(F_L)_\alpha \sim (\textbf{4},\textbf{2},\textbf{1})$ under PS and a right-handed fermion transforming as $(F_R)_{\dot{\alpha}} \sim (\textbf{4},\textbf{1},\textbf{2})$ where we explicitly show the dotted and undotted Lorentz indices for clarity, then the Yukawa Lagrangian is extended by the terms
\begin{align}
\label{ch4:eqn:YUK42}
\mathcal{L}_{\textsc{yuk}^{\vphantom{T}}} = \text{Tr} \Big[\Big\{ \mathlarger{y}_{{f_L}{f_R}}^{\vphantom{\dagger}} \overline{f_L} \,\phi \,(\mphant{f_R}{T})^T +\,\, &\mathlarger{\widetilde{y}}_{{f_L}{f_R}}^{\vphantom{\dagger}}\overline{f_L} \,\phi^c (\mphant{f_R}{T})^T + \mathlarger{y}_{f_R}^{\vphantom{\dagger}} \overline{S_L} \,\rchi_R^{\dagger} \,f_R^{\vphantom{T}} +\mathlarger{y}_{f_L}^{\vphantom{\dagger}} \overline{f_L} \,\rchi_L^{\vphantom{T}} (\mphant{S_L}{\dagger})^c \nonumber\\
&\,\,+(f_L \rightarrow F_L) + (f_R \rightarrow F_R)\Big\}\Big]+ \frac{1}{2}\mu_S^{\vphantom{T}} \overline{S_L}\, S_L^c  + \text{H.c}
\end{align}
with the additional fermion multiplets
\begin{equation}
\label{ch4:eqn:bi42emb}
f_{L/R} = \begin{pmatrix}
u_r & d_r \\
u_b & d_b \\
u_g & d_g \\
\nu & e \end{pmatrix}_{L/R}, \qquad F_{L/R} = \begin{pmatrix}
U_r & D_r \\
U_b & D_b \\
U_g & D_g \\
N^0 & E^- \end{pmatrix}_{L/	R}
\end{equation}
where $U_i$ and $D_i$ ($i = r,b,g$) have electric charge $+2/3$ and $-1/3$ respectively. 
Note that no bare mass terms are present in \cref{ch4:eqn:YUK42} and therefore the expansion will be similar to that of \cref{eqn:PSYuk} in the vanilla PS scenario. In particular all mass terms between the SM and exotic charged fermions will be related to the electroweak vevs $v_1$ and $v_2$ implying that the heavy exotic charged fermions $E$, $D$ and $U$ would be expected to appear near the electroweak scale which is ruled out phenomenologically.

Alternatively two fermions with the same PS representations but \textit{opposite} chirality to the bi-fundamentals already included in PS could be added, that is a right-handed fermion transforming as $(F_L)_{\dot{\alpha}} \sim (\textbf{4},\textbf{2},\textbf{1})$ under PS and a left-handed fermion transforming as $(F_R)_{\alpha} \sim (\textbf{4},\textbf{1},\textbf{2})$. In this case the Yukawa Lagrangian will be similar, albeit with additional bare mass and Yukawa terms:
\begin{align}
\label{ch4:eqn:YUK422}
\mathcal{L}_{\textsc{yuk}^{\vphantom{T}}} &= \text{Tr} \Big[\Big\{ \mathlarger{y}_{{f_L}{f_R}}^{\vphantom{\dagger}} \overline{f_L} \,\phi \,(\mphant{f_R}{T})^T +\,\, \mathlarger{\widetilde{y}}_{{f_L}{f_R}}^{\vphantom{\dagger}}\overline{f_L} \,\phi^c (\mphant{f_R}{T})^T + \mathlarger{y}_{f_R}^{\vphantom{\dagger}} \overline{S_L} \,\rchi_R^{\dagger} \,f_R^{\vphantom{T}} +\mathlarger{y}_{f_L}^{\vphantom{\dagger}} \overline{f_L} \,\rchi_L^{\vphantom{T}} (\mphant{S_L}{\dagger})^c \nonumber\\
&+(f_{L/R} \rightarrow F_{R/L} \,\,\&\,\, \rchi_R \rightarrow \rchi_L)\Big\}+\mu_L\, \overline{f_L}\, F_L + \mu_R \,\overline{F_R}\, f_R + y_{\Phi}^L\, \overline{f_L}\Phi F_L +y_{\Phi}^R\, \overline{F_R}\Phi f_R  \Big]\nonumber\\
&\qquad\qquad\qquad\qquad\qquad\qquad\qquad\qquad\qquad\qquad\qquad+ \frac{1}{2}\mu_S^{\vphantom{T}} \overline{S_L}\, S_L^c  + \text{H.c.}
\end{align}
Now the exotic charged fermions can have their masses decoupled from the electroweak scale for large bare mass terms. However as the exotic fermion multiplets transform in the exact same way as the multiplets containing the SM fields, large values for the bare mass terms $\mu_L$ and $\mu_R$ will not decouple the masses of the down-quarks and charged-leptons. However, if the $SU(4)$ adjoint scalar, $\Phi$, is included in the scalar spectrum, its allowed Yukawa couplings between $f_{L/R}$ and $F_{L/R}$ will lead to a breaking of the mass degeneracy since $\langle \Phi \rangle \propto (1,1,1,-3)$. Therefore, in order for such a scenario to be even remotely feasible, additional scalars beyond that of $\phi,\,\rchi_L$ and $\rchi_R$ are required so that the down-isospin mass degeneracy can be broken: a seesaw mechanism \textit{as well as} the addition of scalars which induce a Georgi-Jarlskog-like texture are simultaneously required.

The charged-lepton mixing matrix will then be given by
\begin{equation}
\label{ch4:eqn:bifundmasseE}
\mathcal{L}_{eE} = \begin{pmatrix}
\overline{e_L} & \overline{E_L}
\end{pmatrix} \begin{pmatrix}
m_f & \mu_L - \sqrt{\frac{3}{2}} \mathlarger{y}_{\Phi}^L v_\Phi\\
\mu_R - \sqrt{\frac{3}{2}} \mathlarger{y}_{\Phi}^R v_\Phi &  M_F 
\end{pmatrix}
\begin{pmatrix}
e_R\\
E_R
\end{pmatrix}
\end{equation}
where $m_f = \widetilde{y}_{f_L f_R} v_1 +y_{f_L f_R} v_2^*$ and $M_F = \widetilde{y}_{F_R F_L} v_1 + y_{F_R F_L} v_2^*$. The down quark mixing matrix is given similarly and only differs by group theoretic factors introduced by the allowed Yukawa couplings of the fermions to $\Phi$:
\begin{equation}
\label{ch4:eqn:bifundmassdD}
\mathcal{L}_{dD} = \begin{pmatrix}
\overline{\vphantom{D_L}d_L} & \overline{D_L}
\end{pmatrix} \begin{pmatrix}
m_f & \mu_L + \sqrt{\frac{1}{6}} \mathlarger{y}_{\Phi}^L v_\Phi\\
\mu_R + \sqrt{\frac{1}{6}} \mathlarger{y}_{\Phi}^R v_\Phi &  M_F 
\end{pmatrix}
\begin{pmatrix}
d_R\\
D_R
\end{pmatrix}.
\end{equation}
Similar matrices can be written down for the mass mixing between the neutral fermions and between $u_{L/R}$ and $U_{L/R}$:
\begin{equation}
\label{ch4:eqn:bifundmassnuNS}
\hspace*{-3.5mm}\mathcal{L}_{\nu N} = \begin{pmatrix}
\overline{\nu_L} \\
 \overline{\nu_R^c} \\
  \overline{N_L} \\
   \overline{N_R^c} \\
    \overline{S_L}
\end{pmatrix}^T \begin{pmatrix}
0 & m_x & 0 & \mu_L - \sqrt{\frac{3}{2}} y_{\Phi}^L v_\Phi & y_{f_L} v_L\\
m_x & 0 & \mu_R - \sqrt{\frac{3}{2}} y_{\Phi}^R v_\Phi & 0 & y_{f_R} v_R^*\\
0 & \mu_R - \sqrt{\frac{3}{2}} y_{\Phi}^R v_\Phi & 0 & M_X & y_{F_R} v_L\\
\mu_L - \sqrt{\frac{3}{2}} \mathlarger{y}_{\Phi}^L v_\Phi & 0 & M_X & 0 & y_{F_L} v_R^*\\
y_{f_L} v_L & y_{f_R} v_R^* & y_{F_R} v_R^* & y_{F_L} v_R^* & \mu_S
\end{pmatrix}
\begin{pmatrix}
\nu_L^c\\
\nu_R\\
N_L^c\\
N_R\\
S_L^c
\end{pmatrix}
\end{equation}
and
\begin{equation}
\label{ch4:eqn:bifundmassuU}
\mathcal{L}_{uU} = \begin{pmatrix}
\overline{u_L} & \overline{U_L}
\end{pmatrix} 
\begin{pmatrix}
m_x & \mu_L + \sqrt{\frac{1}{6}} \mathlarger{y}_{\Phi}^L v_\Phi\\
\mu_R + \sqrt{\frac{1}{6}} \mathlarger{y}_{\Phi}^R v_\Phi &  M_X 
\end{pmatrix} 
\begin{pmatrix}
u_R\\
U_R
\end{pmatrix}
\end{equation}
respectively. The only difference between the up-quark and down-quark mass mixing matrices being $m_x = y_{f_L f_R} v_1 + \widetilde{y}_{f_L f_R} v_2^*$ and $M_X =y_{F_R F_L} v_1  + \widetilde{y}_{F_R F_L} v_2^*$ compared to $m_f$ and $M_F$ defined above.


\begin{sloppypar}
Let us briefly comment on the implications for such models where mass mixing matrices occur for all SM fermion types as in \cref{ch4:eqn:bifundmasseE,ch4:eqn:bifundmassdD,ch4:eqn:bifundmassnuNS,ch4:eqn:bifundmassuU}. Due to the additional mixing within the up-isospin sector, which have very similar mixing matrices to the down-isospin sector, we find it very difficult to achieve a viable mass spectrum for all SM fermions in this scenario. It appears highly unlikely that the inclusion of a pair of bi-fundamental fermions is able to achieve a chiral-suppression on the experimental limits on $m_X$ as well as a phenomenologically-viable mass spectrum for the SM fermions. Due to the number of coupled mass mixing matrices as well as the large number of mass parameters, we do not try to prove this statement. Note, however, that if one of the off-diagonal entries of $M_{dD}$ is made large (required such that the $D'$ states do not gain electroweak masses), this also implies the same thing for $M_{uU}$. Assuming that this occurs for the bottoem-left entry, we have that
\begin{equation}
    m_{u/d} \simeq (\mu_L + \sqrt{\frac{1}{6}} y_\Phi^L v_\Phi) - m_{x/f}\left(\mu_R + \sqrt{\frac{1}{6}} y_\Phi^R v_\Phi\right)^{-1} M_{X/F}
\end{equation}
which would imply that the up-quarks and down-quarks have similar masses if the first term is dominant compared to the second. If instead the second term was dominant compared to the first, at a minimum, it would be difficult to generate the correct top mass considering that $m_{x/X}$ are tied to the electroweak scale.

A chiral suppression along with a viable mass spectrum of SM fermions is possible with the addition of \textit{two pairs} of $SU(4)\otimes SU(2)_{L/R}$ PS bi-fundamental fermions on top of the usual fermions $f_L$ and $f_R$ as has already been shown in~\cite{Calibbi:2017qbu}, and more recently in \cite{Iguro:2021kdw}. As two pairs of bi-fundamentals are introduced, more freedom in the mixing matrices allows for a viable mass spectrum as well as allowing for chiral couplings of $X_\mu$ to the light SM-like fermions. As a non-minimal variant of this model has already been studied within the literature we do not consider this possibility any further and simply conclude that this model, in its minimal form, does not appear viable. 

It is easy and unsurprising to see that for both \cref{ch4:eqn:YUK42,ch4:eqn:YUK422} the accidental global symmetry $U(1)_J$ is unbroken and therefore for these scenarios all baryon number violating interactions will proceed similarly to \cref{sec:Appendix Z}.
\end{sloppypar}

\subsubsection{$SU(4)_c$ Sextet}
\label{sec:sexPS}
\begin{sloppypar}
Adding a fermion which transforms as an $SU(4)$ sextet $\Psi_6 \sim (\textbf{6},\textbf{1},\textbf{1})$, extends the Yukawa Lagrangian by
\begin{align}
\label{eqn:YUK6}
\mathcal{L}_{\textsc{yuk}^{\vphantom{T}}} \supset \text{Tr}\Big[ \mathlarger{y}_{\Psi}^L \overline{\mphant{f_L}{T}} (i \tau_2 ) \rchi_L^{\dagger} \Psi_6  + \mathlarger{y}_{\Psi}^R  \overline{( \utilde{\Psi}_6 )^c}\,\mphant{\rchi_R}{\dagger} (i \tau_2) (f_R)^T + \frac{1}{2} \mathlarger{\mu}_{\Psi_6} \overline{(\utilde{\Psi}_6)^c} \,\Psi_6 \Big] + \text{H.c}
\end{align}
where $(\utilde{\Psi}_6)_{mn} = (\Psi_6)^{ij} \epsilon_{ijmn}$. As this multiplet is uncharged under $SU(2)_{L/R}$, the electric charge of each component is given by the generator of $SU(4)$ identified with $B-L$. The branching rule for this multiplet is 
\begin{equation}
(\textbf{6},\textbf{1},\textbf{1}) \rightarrow \overline{\mathbf{3}}_{+1/3} \oplus \mathbf{3}_{-1/3}
\end{equation}
when broken to $SU(3)_c \times U(1)_{Q}$ of the SM. This irreducible representation therefore contains the correct states to mix with the down quarks of the standard model:
\begin{equation}
\label{eqn:sexemb}
	f_{L/R} = \begin{pmatrix}
\cdot & d_r \\
\cdot & d_b \\
\cdot & d_g \\
\cdot & \cdot \end{pmatrix}_{L/R}, \qquad \Psi_{6} = \left(
	\renewcommand\arraystretch{1.35}
	\begin{array}{cccc}
	0 & \hphantom{-}(D_{L,g})^c & -(D_{L,b})^c & \,D_{R,r}\\
	-(D_{L,g})^c & 0 & \hphantom{-}(D_{L,r})^c & \,D_{R,b}\\
    \hphantom{-}(D_{L,b})^c & -(D_{L,r})^c & 0 & \,D_{R,g} \\ 
	-D_{R,r} & -D_{R,b} & -D_{R,g} & 0	
	\end{array}
	\right)
\end{equation}
where the components $D_{(L/R),i}$ ($i = {r,g,b}$) have electric charge $-1/3$ and ${\Psi_6 \rightarrow \mphant{U_4}{T}\,\Psi_6 \, U_4^T}$.
\end{sloppypar}

The resulting mixing matrix for the down quark and its partners can be found by expanding \cref{eqn:PSYuk,eqn:YUK6} with \cref{eqn:sexemb}, giving
\begin{equation}
\label{eqn:sexmassmix}
\mathcal{L}_{dD} = \begin{pmatrix}
\overline{d_L} & \overline{D_L}
\end{pmatrix} \begin{pmatrix}
m_d  &  \mathlarger{y}_{\Psi_6}^L v_L^*\\
\mathlarger{y}_{\Psi_6}^R v_R &  \mu_{\Psi_{6}}
\end{pmatrix}
\begin{pmatrix}
d_R\\
D_R
\end{pmatrix},
\end{equation}
and, in the absence of additional fermionic states which mix with $\ell$, the singular values of $m_d$ are given by the charged-lepton masses.

The additional Yukawa interactions involving $\Psi_6$ lead to explicit violation of the global symmetry $U(1)_J$: the bare mass term requires $J(\Psi_6) = 0$, while the two Yukawa terms of \cref{eqn:YUK6} imply $J(\Psi_6) = 2$ and $J(\Psi_6) = -2$ respectively, assuming the singlet $S_L$ is included for neutrino mass and therefore $J(\rchi_{L/R})=J(f_{L/R})=1$ is enforced. Therefore any two of the three terms in \cref{eqn:YUK6} in combination with the Yukawa couplings of $f_{L/R}$ to $S_L$ in \cref{eqn:PSYuk} leads to baryon number violating interactions. Interestingly for a more minimal PS model where the only scalars\footnote{Therefore the Yukawa terms $y_{\Psi}^L f_L \rchi_L^{\dagger} \Psi_6$ and $y_L \overline{f_L} \rchi_L (S_L)^c$ will not be present.} present are $\rchi_R$ and $\phi$, a global baryon number is restored in the limit $\mu_{\psi_6} \rightarrow 0$ which therefore could be argued to be small from technical naturalness.

\begin{figure}[t]
\centering
\begin{center}
{
  \includegraphics[width=0.6\linewidth]{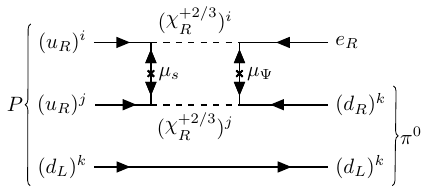}\hfill
}
\end{center}
\caption[Example box diagram of proton decay with additional $SU(4)$ sextets.]{An example box diagram of proton decay implied by a PS model extended by a fermionic $SU(4)$ sextet fermion. The $p \rightarrow \pi^0 e^+$ diagram above is similar to the coloured Higgsino mediated proton decay, $p \rightarrow K^+ \overline{\nu}$, predicted by the minimal SUSY $SU(5)$ GUT model which typically requires GUT scale masses for the Higgsinos, see e.g.~\cite{Hisano:1992jj}.}
\label{figure:sectetPdecay}
\end{figure}

Evaluating the full implications of baryon number violation with a realistic mass spectrum for the SM particles is quite complicated for the sextet and beyond the scope of this work. However, an example diagram for $p \rightarrow \pi^0 e^+$ is shown in \cref{figure:sectetPdecay} in the case where $\rchi_L$ is removed. The diagram involves the couplings
\begin{align}
\mathcal{L} \supset y_R \,\overline{S_L} (\rchi_R^{+2/3})^*_i (u_R)^i + y_\Psi^R \left((\overline{D_L})_k e_R + \epsilon_{ijk}(\overline{D_R^c})^i (d_R)^j\right)(\rchi_R^{+2/3})^k &+ \mu_{\Psi_6} (\overline{D_L})_i (D_R)^i \nonumber\\ &\qquad\,\,\,+ \text{H.c}
\end{align}
where $SU(3)$ indices are explicitly shown and the above couplings are found by expanding out \cref{eqn:PSYuk,eqn:YUK6}. This decay diagram leads to an effective four-fermion interaction very similar to the coloured Higgsino mediated $p \rightarrow K^+ \overline{\nu}$ in minimal SUSY $SU(5)$, see e.g.~\cite{Hisano:1992jj}. As this usually requires very large Higgsino masses (at or larger than the GUT scale) and the SM quantum numbers of $D_L$ and $D_R$ are the same as the coloured Higgsino, we expect that in scenarios where baryon number is not imposed (by setting $\mu_\Psi$ to be small), this diagram alone would likely lead to significant limits on the masses of the particles running in the loop which would translate into large limits on the PS breaking scale. An additional suppression factor of $\mu_S$ appears in \cref{figure:sectetPdecay} which is required to be small for neutrino mass generation which may interestingly connect the large proton lifetime to the smallness of neutrino mass in this suggested minimal model.

\subsubsection{$SU(4)_c$ Decuplets}
\label{sec:decPS}

Finally we consider the possibility that two fermions of opposite handedness\footnote{Both are required for anomaly cancellation.} transforming as $\Psi_{L/R}^{10} \sim (\textbf{10},\textbf{1},\textbf{1})$ are added. The relevant, non-zero terms added to the Yukawa Lagrangian are
\begin{align}
\label{eqn:YUK10}
\mathcal{L}_{\textsc{yuk}^{\vphantom{T}}} \supset \text{Tr} \Big[ \sqrt{2} \, \mathlarger{y}_{\Psi_{10}}^R \overline{\Psi^{10}_L}\, \mphant{\rchi_R}{\dagger} 	(i \tau_2) (\mphant{f_R}{\dagger})^T &- \sqrt{2}\, \mathlarger{y}_{\Psi_{10}}^L (\overline{f_L})^T (i \tau_2) \rchi_L^\dagger \Psi^{10}_R \nonumber\\
 &\quad\quad+ \mathlarger{\mu}_{\Psi_{10}} \overline{\Psi^{10}_L} \,\Psi^{10}_R + \sqrt{6}\, \mathlarger{y}_{\Phi} \overline{\Psi^{10}_L} \,\Phi \,\Psi^{10}_R     \Big] + \text{H.c}.
\end{align}
Similar to the case of the sextet above, these multiplets are uncharged under $SU(2)_{L/R}$ and therefore the branching rule for their breaking to $SU(3)_c \otimes U(1)_Q$ is simply given by
\begin{equation}
(\textbf{10},\textbf{1},\textbf{1}) \rightarrow \textbf{6}_{1/3} \oplus \textbf{3}_{-1/3} \oplus \textbf{1}_{-1}.
\end{equation}
These multiplets therefore contain the necessary states to mix with \textit{both} the down quarks and the charged leptons:
\begin{equation}
\label{eqn:decemb}
f_{L/R} = \begin{pmatrix}
\cdot & d_r \\
\cdot & d_b \\
\cdot & d_g \\
\cdot & e \end{pmatrix}_{L/R}, \quad\Psi^{10}_{L/R}  = \frac{1}{\sqrt{2}}\left(
\renewcommand\arraystretch{1.5}
\begin{array}{cccc}
\cdot & \cdot & \cdot & \, D_r\\
\cdot & \cdot & \cdot &  \,D_b\\
\cdot & \cdot & \cdot &  \,D_g\\
D_r & D_b & D_g & \sqrt{2} \,E^-
\end{array}
\right)_{L/R}
\end{equation}
where the components $D_{(L/R),i}$ ($i=r,b,g$) have electric charge $-1/3$. Contained in $\Psi^{10}_{L/R}$ is an exotic colour sextet Dirac fermion, which is embedded in the upper-left $3 \times 3$ block of the second multiplet in \cref{eqn:decemb}. If such a sextet was discovered alongside heavy copies of the charged leptons and down quarks it would suggest the existence of these exotic fermion multiplets within a PS symmetry.

In this scenario, the mixing matrices for the charged leptons and down quarks are related to each other as they exist in the same multiplet. The charged lepton mixing matrix is given by
\begin{equation}
\label{eqn:decupmassmixE}
\mathcal{L}_{eE} = \begin{pmatrix}
\overline{e_L} & \overline{E_L}
\end{pmatrix} \begin{pmatrix}
m_F & \sqrt{2} \, \mathlarger{y}_{\Psi_{10}}^L v_L^* \\[2pt]
\sqrt{2} \, \mathlarger{y}_{\Psi_{10}}^R v_R &   \,\,\,\,\mathlarger{\mu}_{\Psi_{10}} - \sqrt{\frac{3}{2}}\, \mathlarger{y}_{\Phi} v_{\Phi} 
\end{pmatrix}
\begin{pmatrix}
e_R\\
E_R
\end{pmatrix}
\end{equation}
and the down quark mixing matrix is of a similar form and only differs by some group theoretic factors
\begin{equation}
\label{eqn:decupmassmixD}
\mathcal{L}_{dD} = \begin{pmatrix}
\overline{\vphantom{D_L}d_L} & \overline{D_L}
\end{pmatrix} \begin{pmatrix}
m_F & \mathlarger{y}_{\Psi_{10}}^L v_L^* \\[2pt]
\mathlarger{y}_{\Psi_{10}}^R v_R &   \,\,\,\,\mathlarger{\mu}_{\Psi_{10}} - \sqrt{\frac{1}{6}}\, \mathlarger{y}_{\Phi} v_{\Phi}
\end{pmatrix}
\begin{pmatrix}
d_R\\
D_R
\end{pmatrix}
\end{equation}
where $m_F$ is enforced by the PS symmetry to appear in both mass matrices and its singular values are no longer necessarily given by the charged-lepton or down-quark masses individually. Note that, similar to the case of the bi-fundamental, a Georgi-Jarlskog like texture will occur with the inclusion of the scalar $\Phi$. Unlike the bi-fundamental scenario however, extra group theoretic factors between the two mass matrices exist even without the inclusion of $\Phi$ and therefore it is not required for a breaking of the down-isospin mass degeneracy (though it may be needed to achieve a phenomenologically viable spectrum of masses). Additionally as $E$ and $D$ are contained in the same multiplet, there will be gauge interactions between the two mediated by $X_\mu$ similar to $e$ and $d$. 

The colour sextet fermion $\psi$ develops a mass given by $m_\psi = \mathlarger{\mu}_{\Psi_{10}} + \sqrt{\frac{1}{6}} y_\Phi v_\Phi$ and therefore the mass spectrum of the exotic states depends on the relative sizes of the dimensionful parameters $\mathlarger{y}_{\Psi_{10}}^R v_R,\,\mathlarger{\mu}_{\Psi_{10}}\text{ and }y_\Phi v_\Phi$\footnote{For phenomenological reasons the parameters $m_e$ and $\mathlarger{y}_{\Psi_{10}}^L v_L^*$ are required to be smaller in size than these parameters.}. For example assuming $\mathlarger{\mu}_{\Psi_{10}} > y_\Phi v_\Phi \gg \mathlarger{y}_{\Psi_{10}}^R v_R$ leads to a mass spectrum where the colour sextet would be the heaviest of the three exotics and the heavy, charged leptons would be the lightest. 

The global symmetry $U(1)_J$ remains unbroken in the Yukawa sector and enforces the charge assignment
\begin{equation}
J(\Psi^{10}_L) = J(\Psi^{10}_R) = 2
\end{equation}	
regardless of whether $S_L$ is included or not and therefore baryon number remains an accidental symmetry of the Yukawa sector. The additional multiplets $\Psi_{L/R}^{10}$ do not contain any neutral states under $U(1)_Q$ and there are no lepton number violating terms present in \cref{eqn:YUK10}. This is easy to see as in the absence of $S_L$ the Yukawa Lagrangian obeys a secondary global symmetry 
\begin{equation}
J'(\phi,\,\Phi) = 0, \,\,J'(f_L,\,f_R) = 1, \,\, J'(\rchi_L,\,\rchi_R) = -3 \,\text{ and }\,J'(\Psi_{L}^{10},\Psi_R^{10}) = -2.
\end{equation}
Lepton number can be identified as $L = \frac{1}{4}(J' - 3 T)$ such that
\begin{equation}
L(f_{L/R}) = \begin{pmatrix}
0 & 0\\
0 & 0\\
0 & 0\\
1 & 1\\
\end{pmatrix} \,\text{ and }\, L(\Psi_{L/R}^{10}) = \begin{pmatrix}
-1 & -1 & -1 & \hphantom{,}0\\
-1 & -1 & -1 & \hphantom{,}0\\
-1 & -1 & -1 & \hphantom{,}0\\
\hphantom{-}0 & \hphantom{-}0 & \hphantom{-}0 & \hphantom{,}1
\end{pmatrix}
\end{equation}
which is unbroken by $\langle \rchi_L, \rchi_R \rangle$ and therefore $\nu_L$ and $\nu_R$ develop a Dirac mass of order $m_u$. Therefore, for a realistic model, $S_L$ or some other lepton number violating physics needs to be included in order to break $L=\frac{1}{4}(J' - 3T)$ and allow for small neutrino masses. If $S_L$ is included the neutrino mass matrix is given by \cref{eqn:PSneutralmass} as before. Baryon number violation will therefore proceed similarly to \cref{sec:Appendix Z} and therefore will lead to the same order of magnitude limits on the relevant quartic coupling of the scalar potential.

\section{Fermion mixing}
\label{sec:fermionspec}

From the results of \cref{sec:exoticPS1} only a few PS multiplets lead to mass mixing in the charged-lepton and/or down-quark sectors in the desired way. \Cref{table15} summarises viable multiplets from the ones considered and indicates in which sector mixing will occur.  Charged-lepton mixing exclusively will occur with the addition of $SU(2)_{L/R}$ triplets or bidoublets and in the down-sector exclusively with the addition of $SU(4)$ sextets\footnote{With the caveat that the additional baryon number violating Yukawa interactions do not lead to proton and neutron decay rates larger than the experimental bounds.}. Mixing will occur in both sectors either through a combination of the aforementioned multiplets or (minimally) with\footnote{Note that in this case the mass mixing between $d-D$ and $e-E$ will be `coupled' and the mass mixing matrices only differ by group theoretic factors.} the addition of a pair of $SU(4)$ decuplets or with a pair of $SU(4)\otimes SU(2)_{L/R}$ bi-fundamental fermions, though only the former can generate a viable SM fermion mas spectrum.

\begin{table}[t]
\begin{center}
{\renewcommand{\arraystretch}{1.15}
\scalebox{1.0}{
\begin{tabular}{ccc}
\toprule
PS multiplet & \,$e-E$ mixing\, &$d-D$ mixing \\ 
\midrule 
\midrule
$(\textbf{1},\textbf{1},\textbf{3})$ & \ding{52}  & \ding{54} \\
$(\textbf{1},\textbf{3},\textbf{1})$ & \ding{52}  & \ding{54} \\
$(\textbf{1},\textbf{2},\textbf{2})$ & \ding{52}  & \ding{54} \\
$(\textbf{6},\textbf{1},\textbf{1})$ & \ding{54}  & \ding{52}  \\
$(\textbf{10},\textbf{1},\textbf{1})$ & \ding{52}  & \ding{52} \\
\bottomrule
\end{tabular}}}
\end{center}
\caption[List of PS multiplets which lead to desired mixing effects.]{Summary of the results of \cref{sec:exoticPS1} which indicates which viable fermion extensions lead to a seesaw with the charged leptons by adding heavy states $E$ or with the down quarks by adding heavy states $D$, required to explain the lack of mass-degeneracy predicted by PS. Including an $SU(2)_{L/R}$ triplet or bi-doublet will induce $e-E$ mixing and including an $SU(4)$ sextet will lead to $d-D$ mixing. Including two $SU(4)$ decouplet fermions of opposite chirality or will lead to `coupled' $e-E$ and $d-D$ mixing.} 
\label{table15}
\end{table}

We analyse, where possible, the conditions necessary on the parameters of the Lagrangian which lead to the desired effects of decreasing the experimentally allowed PS breaking scale through helicity suppression and generating a viable fermion mass spectrum through mixing for all SM-like particles. We will separately consider the validity of charged-lepton mixing and down-quark mixing and finally comment on the more general scenarios where mixing occurs for both fermion types.

\subsection{$e-E$ mixing}
\label{subsubsec:fermionspec-eE}

Mixing with heavy exotics within the charged-lepton sector will arise if at least one $SU(2)_{L/R}$ triplet or $SU(2)_{L/R}$ bi-doublet is included. As the minimal low-scale PS model already requires an additional singlet $S_L$ for viable neutrino mass, it would be attractive if these more complicated multiplets could allow for a viable mass spectrum for the neutral and charged leptons without the need for the singlet $S_L$. In this situation we parametrise the general mass mixing matrix for the charged leptons as
\begin{align}
\label{eqn:geneEmix}
\mathcal{L}_{eE} = \begin{pmatrix}
\overline{e_L} & \overline{E_L}
\end{pmatrix} &\underbrace{\begin{pmatrix}
m_{ee} & m_{eE}\\
m_{Ee} & m_{EE}
\end{pmatrix}}\begin{pmatrix}
e_R\\
E_R
\end{pmatrix} + \text{H.c,}\nonumber\\
& \qquad\, M_{eE}
\end{align}
where the different elements of $M_{eE}$ in terms of Lagrangian parameters are given in \cref{sec:tripPS,sec:biPS} for each possible case we have considered. Additionally we parametrise the diagonalisation matrices for $M_{eE}$
\begin{equation}
U_L^\dagger \, M_{eE} \, \mphant{U_R}{\dagger} = \text{diag}(\dots) = M_{eE}^{\text{diag}},
\end{equation}
by 
\begin{equation}
U_{L/R} = \begin{pmatrix}
V & W \\
X & Y
\end{pmatrix}_{L/R}
\end{equation}
where the unitarity condition on $U$ implies $VV^\dagger + WW^\dagger = XX^\dagger + YY^\dagger = \id$. Therefore the relationship between the interaction and mass eigenstates for the charged leptons is given by
\begin{equation}
e_{L/R} = V_{L/R}\,e_{L/R}' + W_{L/R}\,E_{L/R}'
\end{equation}
with $e_{L/R}'\ (E_{L/R}')$ corresponding to the light (heavy) mass eigenstate. The fields $e_{L/R}$ correspond to the uncoloured components of $f_{L/R}$ charged under $SU(4)$ which couple to $d_{L/R}$ via $X_\mu$.

Therefore the relevant physical mixing matrices between the light, SM-like states and down quarks relevant for meson decays is given similarly to the vanilla PS scenario by
\begin{equation}
K^{d e}_L = (U^d_L)^\dagger \,V_L, \,\,\,\,\, K^{d e}_R = (U^d_R)^\dagger \,V_R
\end{equation}
where now the matrices $V_{L/R}$ are no longer unitary. The condition required for a chiral suppression to occur in the relevant pseudoscalar meson decays is for one of $K_{L/R}^{de}$ to satisfy
\begin{equation}
\Vert K_{L/R}^{de} \Vert \ll 1
\end{equation}
such that the helicity-unsuppressed contribution to each decay is sufficiently reduced as discussed in \cref{subsec:fermionmassdeg}. 

The limits on heavy charged-leptons are model-dependent but in all cases far exceed the masses of the SM charged leptons. For example for an $SU(2)_L$ triplet the limits are roughly $800$ GeV~\cite{Sirunyan:2017qkz} and reduce down to $300$ GeV in the case of vector-like lepton doublets~\cite{Kumar:2015tna}. We will conservatively assume that the masses of the heavy charged-lepton states all exceed $1$ TeV, this obviously requires at least one block appearing in $M_{eE}$ to have all its singular values larger than $1$ TeV.

Comparing the general form of $M_{eE}$ in \cref{eqn:geneEmix} to those derived for the cases of either $SU(2)_{L/R}$ triplet or the bi-doublet given in \cref{eqn:tripchargelepmixR,eqn:tripchargelepmixL,eqn:bidoubchargelepmix} respectively, we see that two possible mass terms can be significantly larger than the electroweak scale: either the Yukawa couplings $y_{\Psi_{3}}^R v_R$ or $y_{\Psi_{22}}^R v_R$ proportional to the scale of $SU(2)_R$ breaking (which is absent in the case of the $SU(2)_L$ triplet), or the bare mass terms $\mu_{\Psi_3}$ and $\mu_{\Psi_{22}}$ which are completely unconstrained. Therefore the correct phenomenological masses for the heavy charged leptons are possible in the scenarios where $\Vert y_{\Psi_{3}}^R\,v_R,\,y_{\Psi_{22}}^R\,v_R \Vert \geq 1$ TeV or $\Vert \mu_{\Psi_3},\,\mu_{\Psi_{22}} \Vert\geq 1$ TeV. In all cases the mass term generated by $v_R$ appears on the off-diagonal of $M_{eE}$, either in the top-right entry in the case of the bi-doublet extension and in the bottom-left entry for the case of an $SU(2)_R$ triplet. Although this term is absent in the case of the $SU(2)_L$ triplet, it may be generated in the bottom-left entry of $M_{eE}$ by assuming a modified scalar sector, which we discuss further below, and therefore the discussion below remains relevant for this exotic fermion choice. The bare mass terms $\mu_{\Psi_3}$ and $\mu_{\Psi_{22}}$ appear in the bottom-right entry in all three cases. We suppress the labels $\Psi_3$ and $\Psi_{22}$ below  and write $y_{\Psi}^R v_R$ and $\mu_\Psi$ for simplicity.

Consider first the scenario where $\mu_\Psi$ is taken to be larger than all other mass terms, e.g. $\mu_\Psi = m_{EE} > m_{eE},\,m_{Ee},\,m_{ee}$. Following the results from \cref{subsec:oneDsingval} the resultant masses for the seesaw states are approximated by\footnote{$m_{\mathscr{h}}$ is the mass of the heavy fermion, not the Higgs boson.}
\begin{equation}
m_\mathscr{l} \simeq \left\vert m_{ee} - \frac{m_{eE}\,m_{Ee}}{m_{EE}} \right\vert \qquad \text{and} \qquad m_{\mathscr{h}} \simeq m_{EE}.
\end{equation}
Using \cref{eqn:2x2singval} the mass eigenstates relate to the interaction eigenstates by
\begin{align}
e_{L} &\simeq \underbrace{\left(1 - \frac{1}{2}\left(\frac{m_{eE}}{m_{EE}}\right)^2 \right)} e_L' + \underbrace{\left(\frac{m_{eE}}{m_{EE}}\right)} E_L'\nonumber\\
&\qquad\qquad\quad V_L\qquad\qquad\qquad\quad\,\, W_L
\end{align}
for the left-handed states and
\begin{align}
e_{R} &\simeq \underbrace{\left(1 - \frac{1}{2}\left(\frac{m_{Ee}}{m_{EE}}\right)^2 \right)} e_R' + \underbrace{\left(\frac{m_{Ee}}{m_{EE}}\right)} E_R'\nonumber\\
&\qquad\qquad\quad V_R\qquad\qquad\qquad\quad\,\, W_R
\end{align}
for the right-handed states. One generation of fermions with real parameters has been assumed for simplicity in order to establish viability. Similar conclusions are reached for multi-generational scenarios where each entry of $M_{eE}$ is promoted to a block matrix of appropriate dimension. The one-dimensional equivalents of $V_{L/R}$ and $W_{L/R}$ are indicated and, as can be seen in this scenario where the bare mass is the dominant term in the charged-lepton seesaw, both the left- and right-handed light charged lepton mass states are predominantly made up of the fields $e_{L/R}$ embedded in $f_{L/R}$. Therefore, while this would lead to a very mild suppression in the PS mass limits as $V_{L/R}$ are slightly suppressed compared to the scenario without charged-lepton mass mixing (where $V_{L/R} =1$), the mass limits obtained for $X_\mu$ will be of similar size to those appearing in \cref{table11} and therefore the limits will remain at around $\mathcal{O}(100-1000)$ TeV depending on the structure of $K_{L/R}^{de}$.

Alternatively, $y_{\Psi}^R  v_R $ can be the dominant mass term within $M_{eE}$ which corresponds to one of the off-diagonal terms in our parametrisation. Following a similar procedure we find
\begin{equation}
\label{eqn:toprightdommass}
m_\mathscr{l} \simeq \left\vert m_{Ee} - \frac{m_{ee}\,m_{EE}}{m_{eE}} \right\vert
\end{equation}
and
\begin{align}
\label{eqn:toprightdommix}
e_{L} &\simeq -\underbrace{\frac{m_{EE}}{m_{eE}}} e_L' + \underbrace{\left(1-\frac{1}{2}\left(\frac{m_{EE}}{m_{eE}}\right)^2\right)} E_L'\nonumber\\
&\qquad\quad V_L\qquad\qquad\qquad\,\,\, W_L\nonumber\\
e_{R} &\simeq \underbrace{\left(1 - \frac{1}{2}\left(\frac{m_{ee}}{m_{eE}}\right)^2 \right)} e_R' + \underbrace{\left(\frac{m_{ee}}{m_{eE}}\right)} E_R'\nonumber\\
&\qquad\qquad\quad V_R \qquad\qquad\qquad\quad W_R
\end{align}
for the scenario where $m_{eE}$ is dominant and
\begin{equation}
\label{eqn:bottomleftdommass}
m_\mathscr{l} \simeq \left\vert m_{eE} - \frac{m_{ee}\,m_{EE}}{m_{Ee}} \right\vert
\end{equation}
and
\begin{align}
\label{eqn:bottomleftdommix}
e_{L} &\simeq \underbrace{\left(1 - \frac{1}{2}\left(\frac{m_{ee}}{m_{Ee}}\right)^2 \right)} e_L' + \underbrace{\left(\frac{m_{ee}}{m_{Ee}}\right)} E_L'\nonumber\\
&\qquad\qquad\quad V_L \qquad\qquad\qquad\quad W_L \nonumber\\
e_{R} &\simeq -\underbrace{\frac{m_{EE}}{m_{Ee}}} e_R' + \underbrace{\left(1-\frac{1}{2}\left(\frac{m_{EE}}{m_{Ee}}\right)^2\right)} E_R'\nonumber\\
&\qquad\quad V_R \qquad\qquad\qquad\,\,\, W_R
\end{align}
if $m_{Ee}$ is dominant. In both cases, one of $e_{L/R}$ is predominantly made up of the light mass state $e_{L/R}'$ while the opposite chirality is predominantly made up from the heavy mass state $E_{R/L}'$. $V_L$ or $V_R$ are now significantly different in size from each other, where one will be suppressed compared to the other. If $m_{eE}$ is the largest term then $V_L \ll V_R$ and the leptoquark $X_\mu$ will strongly couple $e_R'$ and $d_R'$ but not $e_L'$ and $d_L'$ and vice versa for $m_{Ee}$ dominance. Therefore the desired helicity suppression in meson decays induced by $X_\mu$ will occur for the hierarchy $y_\Psi^R v_R > \mu_\Psi$ leading to decreased PS breaking limits, potentially as low as those appearing in \cref{table4}. Both an $SU(2)_R$ triplet or the $SU(2)_{L/R}$ bidoublet are therefore viable candidates as one off-diagonal term is proportional to $v_R$ and therefore can be large in size. In the case of an $SU(2)_L$ triplet, assuming no modification to the scalar fields, the only mass term not tied to the electroweak scale is $\mu_\Psi$ and therefore helicity suppression in the relevant meson decays will not occur as $\mu_\Psi$ must be dominant phenomenologically.

For a multi-generational scenario all terms in $M_{eE}$ are promoted to block matrices and using the results in \cref{subsec:multiDsingval} we find for $\Vert m_{Ee} \Vert > \Vert m_{ee},\,m_{eE},\,m_{EE}\Vert$
\begin{equation}
m_L \simeq m_{eE} - m_{ee} \left(m_{Ee}^{-1}\right) m_{EE} 
\end{equation}
and
\begin{align}
e'_L &\simeq \underbrace{\left(\id - \frac{1}{2} \left( m_{ee} m_{Ee}^{-1} \right) \left( m_{ee}m_{Ee}^{-1}\right)^\dagger \right)O_L} \,e_L + \underbrace{m_{ee}m_{Ee}^{-1}\mathcal{O}_L}\, E_L\nonumber\\
&\qquad\qquad\qquad\qquad\,\, V_L\qquad\qquad\qquad\qquad\qquad\quad\,\, W_L\nonumber\\
e'_R &\simeq -\underbrace{m_{Ee}^{-1}m_{EE}\,O_R}\, e_R + \underbrace{\left(\id - \frac{1}{2} \left(m_{Ee}^{-1} m_{EE}\right) \left(m_{Ee}^{-1}m_{EE}\right)^\dagger \right)\mathcal{O}_R} E_R\nonumber\\
&\qquad\qquad\,\,\, V_R \qquad\qquad\qquad\qquad\qquad\qquad\,\, W_R
\end{align}
where $O_{L/R}$ and $\mathcal{O}_{L/R}$ are the unitary matrices which diagonalise the light and heavy mass blocks which appear after block diagonalisation, e.g. $O_L^\dagger m_L \mphant{O_R}{\dagger} = \text{diag}(\dots)$. For obvious reasons these expressions are only valid when $m_{Ee}$ is nonsingular. Similar expressions for $m_{eE}$ dominance can be derived quite simply and the results only differ by the same permutations of parameters as between \cref{eqn:toprightdommass,eqn:toprightdommix}.

One last potential scenario which will lead to the heavy lepton masses exceeding $1$ TeV occurs for the tuned case $y_\Psi v_R \simeq \mu_\Psi$. Consider for example when $m_{Ee} \simeq m_{EE}$ and are both dominant in $M_{eE}$. We find for one generation
\begin{equation}
m_\mathscr{l} \simeq \left\vert \frac{m_{ee}-m_{eE}}{\sqrt{2}} \right\vert
\end{equation}
and
\begin{align}
e_{L} &\simeq -\underbrace{\left(1-\frac{1}{4}\left(\frac{m_{eE}+m_{ee}}{m_{Ee}}\right)^2\right)} e_L' + \underbrace{\frac{1}{2}\left(\frac{m_{eE}+m_{ee}}{m_{Ee}}\right)} E_L'\nonumber\\
&\qquad\qquad\qquad\quad\,\, V_L\qquad\qquad\qquad\qquad\qquad\,\, W_L\nonumber\\
e_{R} &\simeq \underbrace{\frac{1}{\sqrt{2}}} e_R' + \underbrace{\frac{1}{\sqrt{2}}} E_R'\nonumber\\
&\quad\,\,\,\, V_R\qquad\quad W_R
\end{align}
with similar results if $m_{eE} \simeq m_{EE}$ except for the reassignments $L \leftrightarrow R$ on the fields above. 
Here the mass eigenstate $e'_R$ is made up of a roughly equal amount of the fields $e_R$ and $E_R$ and therefore while the mixing parameter $V_R$ is decreased, it remains an order one number. The overall strength of $K_R^{de}$ will be lowered, however the helicity-unsuppressed contribution for each meson decay will remain dominant as it is several orders of magnitude larger than the helicity-unsuppressed contribution, as discussed in \cref{subsec:fermionmassdeg}. Therefore we find that in order for helicity suppression to allow for decreased experimental limits on the PS breaking scale, the hierarchy $y_\Psi^R v_R \gg \mu_\Psi$ is required.

We will therefore adopt the hierarchy $\Vert y_\Psi^R v_R \Vert > \Vert y_\Psi v_L, \, m_e, \, \mu_\Psi \Vert$ below in order to helicity suppress the mass limits of $X_\mu$, further requiring the correct down-quark and charged-lepton masses with appropriate SM-like weak couplings as a secondary condition will further restrict the parameters in each mass mixing matrix.

Specifically for the case of an $SU(2)_R$ triplet, adopting the above hierarchy in $M_{eE}$ leads to the light-mass block approximately given by
\begin{equation}
\label{eqn:triplightmassblockR}
m_\ell \simeq -\frac{1}{v_R} m_d\, (Y_{\Psi_3})^{-1} \mu_{\Psi_3}
\end{equation}
where we have introduced three generations of the triplet $\Psi_3^R$ and  $Y_{\Psi_3}$ corresponds to  a $3 \times 3$ matrix further assumed to be nonsingular. Unless stated otherwise $y_\Psi$ and $Y_\Psi$ will distinguish between when one or three generations of exotic fermions are introduced respectively. Within \cref{eqn:triplightmassblockR} we have used the relation $m_e = m_d$ which is enforced by the PS symmetry and therefore the singular values of $m_d$ are given by the down-quark masses run up to the PS breaking scale. Similarly the singular values of the light block $m_\ell$ must be given by the masses of the charged-leptons at the same scale. An unavoidable consequence of \cref{eqn:triplightmassblockR} is that it necessarily predicts that all generations of charged leptons have masses strictly lighter than their corresponding generation of down-quark. We prove this result in \cref{subsec:gapprop}. While this is accurate for the first and third generations, this mass hierarchy is flipped in the second generation as shown in \cref{table3} where the muon is heavier than the strange quark at least for energy scales below $1000$ TeV which we are considering.

Therefore we find that \cref{eqn:triplightmassblockR} is unable to reproduce the correct charged-lepton masses if $m_e = m_d$ is enforced by the PS symmetry to give the down-quark masses. The alternative seesaw scenario where $\Vert \mu_{\Psi_3} \Vert > \Vert Y_{\Psi_3} v_R \Vert$, although unable to lead to the desired chiral suppression in the PS breaking scale, is also unable to reproduce the correct muon and strange masses as it requires $m_\ell \simeq m_d$. Therefore a potential hybrid scenario where the hierarchy in the singular values between $Y_{\Psi_3}^R v_R$ and $\mu_{\Psi_3}$ flips for the second generation compared to the first and third would also not be viable. Similar arguments apply to the $SU(2)_L$ triplet, which was already unable to give sufficiently heavy masses to the charged-lepton partners. Therefore both triplet scenarios are unable to reproduce the correct charged-lepton masses in their minimal form. This is a direct consequence of the zero entry appearing in the mass matrix $M_{eE}$ but applies to any similar scenario with quark-lepton mass unification and exotic mass mixing. Similar arguments apply to the scenario where both an $SU(2)_L$ and $SU(2)_R$ triplet are added as in \cref{eqn:tripchargelepmixL&R,eqn:tripneutlepmixL&R} and therefore we find that $SU(2)_{L/R}$ triplets require a more exotic scalar sector such that all mass terms in the Yukawa Lagrangian can be generated.

\subsubsection{Top-right dominance: Bi-doublet fermions}
\label{subsubsec:bidoubcase}

Turning to the case of the $SU(2)_{L/R}$ bi-doublet, now each entry of the charged-lepton mass matrix $M_{eE}$ is non-zero. The choice $\Vert Y_{\Psi_{22}} v_R \Vert > \Vert \mu_{\Psi_{22}} \Vert$, such that the desired suppression in the PS limits occurs, implies that the top-right entry of $M_{eE}$ is the dominant block, as can be seen in \cref{eqn:bidoubchargelepmix}. Therefore the light mass block is given by
\begin{equation}
\label{eqn:chargelepbidoub}
m_\ell \simeq Y_{\Psi_{22}}^L v_L - \frac{1}{v_R} \mu_{\Psi_{22}} (Y_{\Psi_{22}}^R)^{-1} m_d
\end{equation}
where again the PS symmetry enforces $m_e = m_d$. Now with the addition of the mass term $Y_{\Psi_{22}}^L v_L$ the correct charged-lepton masses can be generated. For example in the limit $\mu_{\Psi_{22}} \rightarrow 0_{3\times 3}$ the charged lepton masses are simply given by
\begin{equation}
m_\ell \simeq Y_{\Psi_{22}}^L v_L
\end{equation}
and therefore the effect of the seesaw in $M_{eE}$ is to completely disassociate the mass origin of the SM-like charged leptons from the down quarks as now they arise from different Yukawa couplings and vevs and implies that $v_L \gtrsim m_\tau$.

Using \cref{subsec:multiDsingval}, the mass states of the light leptons relate to the interaction states by
\begin{align}
e_L' &\simeq - (O_L^e)^\dagger \mathcal{X}  e_L + (O_L^e)^\dagger\left(\id -\frac{1}{2}\mathcal{X} \mathcal{X}^\dagger\right)E_L\nonumber\\
e_R' &\simeq (O_R^e)^\dagger\left(\id -\frac{1}{2}\mathcal{Z}^\dagger \mathcal{Z}\right)e_R - (O_R^e)^\dagger\mathcal{Z}^\dagger E_R
\end{align}
where $O^e_{L/R}$ diagonalise the light mass block and
\begin{align}
\label{eqn:bidoubpertmat}
\mathcal{X} &\simeq \frac{1}{v_R}\mu_{\Psi_{22}} \left(Y_{\Psi_{22}}^R\right)^{-1} + \frac{v_L}{v_R^2} Y_{\Psi_{22}}^L m_d^\dagger [(Y_{\Psi_{22}}^R)^\dagger]^{-1}(Y_{\Psi_{22}}^R)^{-1} ,\nonumber\\
\mathcal{Z} &\simeq \frac{1}{v_R}(Y_{\Psi_{22}}^R)^{-1} m_d + \frac{v_L}{v_R^2}(Y_{\Psi_{22}}^R)^{-1}[(Y_{\Psi_{22}}^R)^\dagger]^{-1}\mu_{\Psi_{22}}^\dagger Y_{\Psi_{22}}^L
\end{align}
and we have expanded up to second order in $\mathcal{X}$ and $\mathcal{Z}$. In this seesaw regime, the light left-handed lepton states measured in experiments are predominately made up of $E_L$ appearing in the bi-doublet whereas the right-handed light states are predominately made up of $e_R$ appearing in $f_R$ which interacts with $X_\mu$. 

For clarity, the relevant physical mixing matrices are given by
\begin{equation}
K_L^{de} \simeq -(U_L^d)^\dagger \mathcal{X}^\dagger O_L^e ,\,\, K_R^{de} \simeq (U_R^d)^\dagger  \left(\id -\frac{1}{2}\mathcal{Z}^\dagger \mathcal{Z}\right)O_R^e\text{ and  }U_\text{PMNS} \simeq N_\nu^\dagger \left(\id -\frac{1}{2}\mathcal{X} \mathcal{X}^\dagger\right) O_L^e,
\end{equation}
where $N_\nu$ corresponds to the relevant non-unitary submatrix of the neutral fermion diagonalising matrix for the active neutrinos. Therefore in this scenario the deviation from unitarity of the PMNS matrix arising from mixing in the charged lepton sector \textit{and} the suppression of the matrix $K_L^{de}$ is determined by the smallness of the matrix $\mathcal{X}$. In \cref{eqn:bidoubpertmat} we have included the second order terms in the expansion as for sufficiently small values in the matrix $\mu_{\Psi_{22}}$ this deviation will be dominated by the second-order term. As $Y_{\Psi_{22}}^L v_L$ and $m_d$ are fixed to give the correct SM charged-lepton and down-quark masses respectively, the second order term is therefore fixed for a given choice of $Y_{\Psi_{22}}^R v_R$ and therefore can place a lower bound on the scale $v_R$ required to satisfy both PMNS unitarity constraints and lead to the desired chiral suppression in the rare meson decays which constrain PS.

The neutrino mass matrix with the addition of the fermion bi-doublets is given by the block matrix equivalent of \cref{eqn:bidoubneutlepmix} which specifically for the hierarchy $\Vert Y_{\Psi_{22}}^R v_R \Vert > \Vert m_u, Y_\Psi^L v_L \Vert > \Vert \mu_{\Psi_{22}} \Vert$ leads to the following four mass blocks (after block diagonalisation):
\begin{alignat}{2}
\label{eqn:neutlepbidoub}
m_{1,2} &\simeq v_R \, Y_{\Psi_{22}}^R  &&\nu_{1/2} \simeq \frac{1/i}{\sqrt{2}}( \nu_L \pm  N_R^c) \nonumber\\
m_3 &\simeq \frac{v_L}{v_R} \left( Y_{\Psi_{22}}^L (Y_{\Psi_{22}}^R)^{-1} m_u + m_u^T \left[Y_{\Psi_{22}}^L (Y_{\Psi_{22}}^R)^{-1}\right]^T\right) &&\nu_3 \simeq \nu_R^c \nonumber\\
m_4 &\simeq \frac{1}{v_L v_R} \mu_{\Psi_{22}} \left[(Y_{\Psi_{22}}^L)^T m_u^{-1} Y_{\Psi_{22}}^R + (Y_{\Psi_{22}}^R)^T (m_u^T)^{-1} Y_{\Psi_{22}}^L \right]^{-1} \mu_{\Psi_{22}}^T\quad &&\nu_4 \simeq N_L.
\end{alignat}
Here there is a heavy pseudo-Dirac pair with degenerate masses to the heavy charged leptons, a light neutrino block with mass arising from a linear seesaw mechanism whose mass eigenstate is predominantly made up of $\nu_R$. and an even-lighter neutrino mass block which is predominantly made up of $N_L$ appearing in the bi-doublet. Therefore
\begin{equation}
m_{\nu_4} \ll m_{\nu_3} \ll m_{\nu_{1}},\,m_{\nu_2}
\end{equation}
where $m_{\nu_4}$ corresponds to the three light neutrinos observed through oscillation experiments. A more complete expression for the relationship between the neutrino mass and interaction eigenstates can be found at the end of \cref{subsec:bidoubneutmass}. In \cref{subsec:bidoubneutmass} we argue that the above scenario where $\mu_{\Psi_{22}}$ is significantly smaller than all other mass parameters, which will lower the experimental limits on PS breaking, is the only scenario which allows for an experimentally valid spectrum of active neutrino masses unless a PS breaking scale larger than $10^{11}$ GeV is adopted. It is quite striking that the only hierarchy of parameters which allows for both low scale PS breaking and appropriately light active neutrino masses, when bi-doublet fermions are introduced, is the same hierarchy that will lead to a suppression in the experimental limits on the PS breaking scale. Additionally, in the limit $\mu_{\Psi_{22}} \rightarrow 0_{3 \times 3}$ where the lightest neutrinos are massless at tree-level, there is an additional global symmetry conserved in the Yukawa Lagrangian and therefore taking this mass matrix to be small is technically natural.

\begin{figure}[t]
\centering
{
  \includegraphics[width=0.49\linewidth]{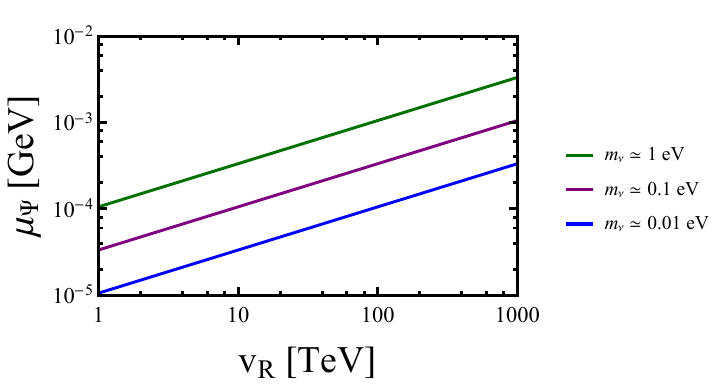}
}
{
  \includegraphics[width=0.38\linewidth]{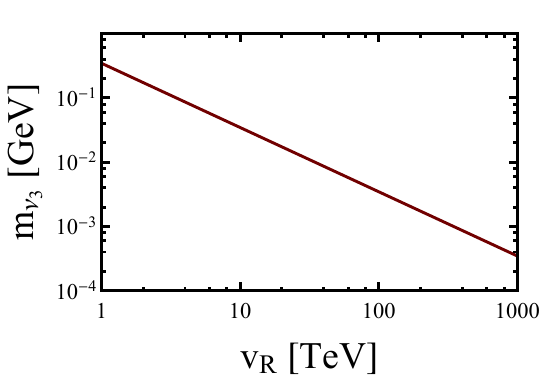}
}
\caption[Plot of $\mu_{\Psi_{22}}$ as a function of $v_R$ required for a viable neutrino mass spectrum.]{Plot of the required size of $\mu_{\Psi_{22}}$ in GeV as a function of $v_R$ in TeV in the one generational scenario where $y_{\Psi_{22}}^R =1$ has been assumed \textbf{(left)} and the masses of the intermediately heavy neutrino as a function of the same scale \textbf{(right)}. Different lines correspond to different choices for the value of the neutrino masses and we have fixed $m_\ell \simeq 1\text{ GeV}$, $m_b \simeq 4\text{ GeV}$ and $m_t \simeq 173\text{ GeV}$ in order to reproduce the correct tau, bottom- and top-quark masses. Requiring sufficiently light neutrinos is only possible when $\mu_{\Psi_{22}}$ is sufficiently small which also happens to be the requirement in order to helicity-suppress the experimental limits on PS breaking.}
\label{figure:muvsvrbidoub}
\end{figure}

The correct charged-lepton masses for each generation and the active neutrino mass limits and differences can both be achieved with the addition of the bi-doublet without needing to introduce any additional states, such as the singlet $S_L$. \Cref{eqn:chargelepbidoub,eqn:neutlepbidoub} can be solved simultaneously for the three unknown blocks $Y_{\Psi_{22}}^R,\,Y_{\Psi_{22}}^L$ and $\mu_{\Psi_{22}}$ with the assumed hierarchy between each block. In the one generational scenario,  the masses of the two lightest neutrino states and the light charged lepton state are given by
\begin{align}
m_{\nu_3} &\simeq \frac{v_L}{v_R}\frac{2 y_{\Psi_{22}}^L m_t}{y_{\Psi_{22}}^R}\\
m_{\nu_4} &\simeq \frac{1}{v_L v_R} \frac{\mu_\Psi^2 m_t}{2 y_{\Psi_{22}}^L y_{\Psi_{22}}^R}\\
m_{\ell} &\simeq v_L y_{\Psi_{22}}^L - \frac{\mu_\Psi m_b}{v_R y_{\Psi_{22}}^R}
\end{align}
where we have fixed the masses of the quarks to the top- and bottom-quark respectively. Setting $m_{\ell} \simeq 1\text{ GeV}$ in order to reproduce the correct tau mass and $m_b \simeq 4\text{ GeV}$ and $m_t \simeq 173\text{ GeV}$ leads to \cref{figure:muvsvrbidoub} where we find the required size of $\mu_{\Psi_{22}}$ as a function of $y_{\Psi_{22}}^R v_R$ for different choices of the light neutrino mass. We find for sub-eV neutrino masses, setting $v_R$ to be between $1-100$ TeV requires $\mu_{\Psi_{22}}$ to be below the MeV scale. Due to the small size of $\mu_{\Psi_{22}}$ required from neutrino mass limits, the term $\mu_{\Psi_{22}} m_d/y_{\Psi_{22}}^R v_R$ entering the charged lepton masses is negligible compared to $y_{\Psi_{22}}^L v_L$ and therefore the correct charged-lepton masses are easily possible. As discussed, such small values of $\mu_{\Psi_{22}}$ required for neutrino mass quite conveniently also lead to a chiral suppression in the PS breaking scale. The mass of the intermediately heavy neutrino, $\nu_3$, will vary only with the size of $y_{\Psi_{22}}^R v_R$ as all other parameters are related to SM masses. The predicted mass of $m_{\nu_3}$ lies roughly between the MeV and GeV scales depending on the $v_R$ breaking scale, with larger scales of $SU(2)_R$ breaking corresponding to lighter masses as demonstrated in the second plot of \cref{figure:muvsvrbidoub}. We note that for this proposed scenario, the quark masses arise from the vevs $v_1$ and $v_2$ from $\phi$ whereas the charged-lepton masses predominately arise from the vev $v_L$ appearing in $f_L$.

The lightest neutrino mass eigenstate arises predominantly from the state $N_L$ appearing in the bi-doublet fermion $\Psi_{22}$ and therefore, as phenomenologically required, will have SM-like weak interactions. The left-handed charged-lepton mass eigenstates arise predominantly from the state $E_L$ which appears in the same bi-doublet as $N_L$, as shown in \cref{eqn:biemb}, additionally they form an $SU(2)_L$ doublet with each other after $SU(2)_R$ breaking. The (pseudo-unitary) PMNS matrix is identified from the coupling of $W_L^{\pm}$ to $N_L$ and $E_L$ as
\begin{equation}
U_{L} = \left(\id - \eta' \right) U_N^\dagger \left(\id - \eta' \right) O_L^e
\end{equation} 
where 
\begin{equation}
\eta' \simeq \frac{1}{2}\mathcal{X}\mathcal{X}^\dagger
\end{equation}
measures the deviation from unitarity arising from mixing. We find the deviation to be the same in both sectors, at least at the lowest order, as explained in \cref{subsec:multiDsingval,subsec:bidoubneutmass}. As usual $U_N$ and $O_{L}^e $ are the unitary matrices which diagonalise the light mass blocks in each sector of the block-diagonal mass matrices.

Writing
\begin{align}
U_{L} &= \left[\left(\id - \eta' \right) U_N^\dagger \left(\id - \eta' \right) O_L^e\right]\left[(O_L^e)^\dagger U_N U_N^\dagger O_L^e\right]\nonumber\\
&= \left[ \left(\id - \eta' \right) U_N^\dagger \left(\id - \eta' \right)U_N\right] U_N^\dagger O_L^e\nonumber\\
&\simeq \left( \id - \eta' - U_N^\dagger \eta' \,U_N \right) U_N^\dagger O_L^e\nonumber\\
&= \left(\id - \eta\right)U_\text{PMNS}
\end{align}
allows us to estimate the deviation from unitarity by $(\id - \eta)$, as is usually done, where
\begin{equation}
\eta \simeq \eta' + U_N^\dagger \eta'\, U_N
\end{equation}
and we identify the combination $U_N^\dagger O_L^e$ with the unitary PMNS matrix as usual.

\begin{figure}[t]
\centering
{
  \includegraphics[width=0.45\linewidth]{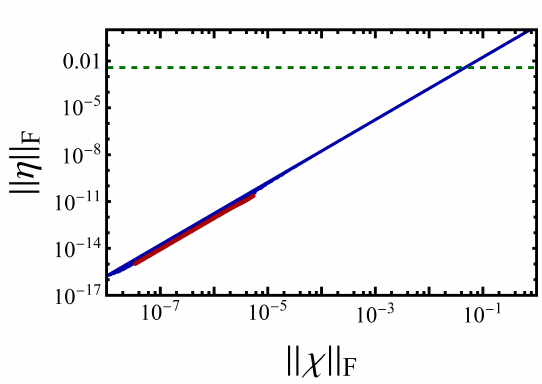}
}
{
  \includegraphics[width=0.45\linewidth]{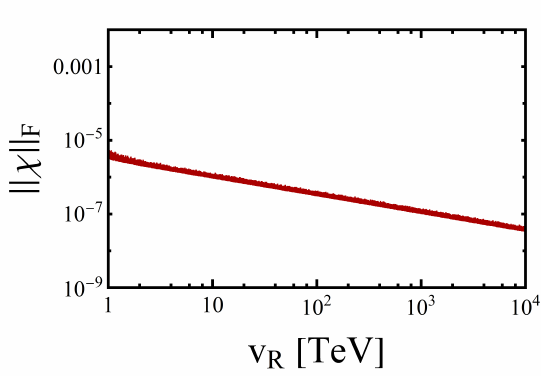}\hfill
}
\caption[Plot of the deviation of unitarity as a function of $\mathcal{X}$ and $v_R$ in the bi-doublet scenario.]{Plot of the deviation of unitarity \textbf{(left)} as a function of $\mathcal{X}$ where points in blue correspond to a random scan on the entries of $\mu_{\Psi_{22}}$ and $Y_{\Psi_{22}}^R$ whereas points in red correspond to the region which gives the correct charged-lepton and neutrino masses through a Casas-Ibarra parameterisation. We find that there are no limits on the scale $v_R$ from unitarity violation and furthermore the region with a viable neutrino mass spectrum predicts a deviation of unitarity many orders of magnitude less than the experimental precision. The \textbf{(right)} plot shows the generated $\mathcal{X}$ through the Casas-Ibarra parameterisation as a function of $v_R$ which leads to a viable charged-lepton and neutrino mass spectrum. As the deviation of unitary is given by $\eta \simeq \mathcal{X}^2$, the deviation at all scales is significantly smaller than the current experimental limits. $\Vert \cdot \Vert_F$ corresponds to the Frobenius norm of the given matrix which we use to present the data and the limits on the deviation indicated by the dashed green line is taken from~\cite{Fernandez-Martinez:2016lgt}.}
\label{fig:bidoubunit}
\end{figure}

The left plot of \cref{fig:bidoubunit} shows the deviation from unitarity as a function of $\mathcal{X} \simeq \mu_{\Psi_{22}} (Y_{\Psi_{22}}^R)^{-1}$ where we have fixed $U_N = \id$ for simplicity. We further assumed that $Y_{\Psi_{22}}^R$ was diagonal with entries close to unity such that the vev of $v_R$ can be as small as possible\footnote{For a non-degenerate spectrum of singular values in $Y_{\Psi_{22}}$, requiring all the masses of the heavy charged leptons to be larger than $1$ TeV requires larger $v_R$ breaking scales, e.g. if $\sigma_1(Y_{\Psi_{22}})=1/100$ then this requires $v_R \geq 100$ TeV.}. We then performed two scans over different values of $v_R$ ranging from $1$ TeV to $10^4$ TeV. In the first scan (in blue) we randomly scanned over the entries of $\mu_{\Psi_{22}}$ and therefore do not generate the correct neutrino mass spectrum but establishes the relationship between $\eta$ and $\mu_{\Psi_{22}}$. The second scan (in red) had $\mu_{\Psi_{22}}$ fixed for a given $v_R Y_{\Psi_{22}}^R$ through a Casas-Ibarra parametrisation in order to generate the correct charged lepton and neutrino masses. The resultant PMNS matrix was compared to the experimental limits on the deviation from unitarity, $\eta^\text{exp}$ which we take from~\cite{Fernandez-Martinez:2016lgt}, and we find that the region which leads to a viable mass spectrum for the SM leptons predicts a deviation of unitary many orders of magnitude below the current limit. This is due to the deviation being proportional to the small matrix $\mu_{\Psi_{22}}$. For alternative models where, for example, the size of $\mu_{\Psi_{22}}$ and therefore $\Vert\mathcal{X}\Vert_F$ is not fixed to be small by requiring small neutrino masses (through the introduction of additional neutral lepton mass mixing), deviation from unitarity limits constrain
\begin{equation}
\Vert \mathcal{X} \Vert_F \lesssim 10^{-1}.
\end{equation}

\begin{figure}[t]
\centering
{
  \includegraphics[width=0.45\linewidth]{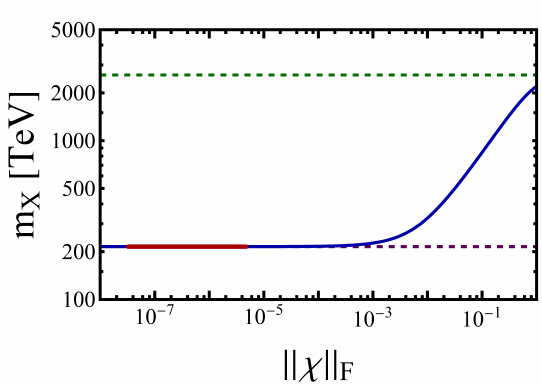}\hfill
}
{
  \includegraphics[width=0.45\linewidth]{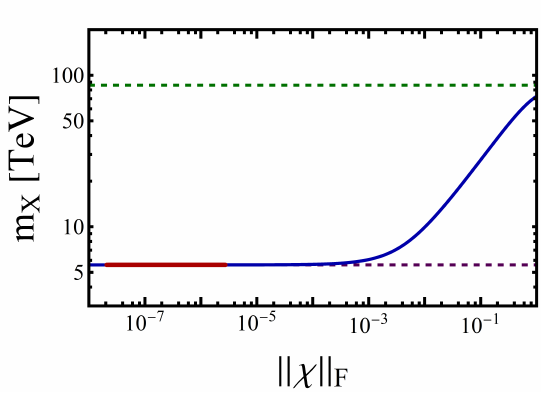}
}
\caption[Plot of the limits on $m_X$ as a function of $\mathcal{X}$ with bi-doublet fermions.]{Plot of the experimental limits on $m_X$ as a function of $\mathcal{X}$ where we have assumed $K_{L/R}^{de} = \id$ \textbf{(left)} and mixing angles given by the fifth row of \cref{table12,table14} \textbf{(right)}. Points in blue correspond to a random scan over $\mu_{\Psi_{22}}$ in order to establish its variation whereas points in red correspond to where the entries of $\mu_{\Psi_{22}}$ are fixed through a Casas-Ibarra parameterisation to give a valid neutrino mass spectrum. We find that all regions which lead to a viable neutrino mass spectrum also imply that the limits on the mass of $X_\mu$ are completely helicity suppressed. This remains true for any choices of the physical mixing matrices $K_{L/R}^{de}$. The horizontal dashed green lines corresponds to the limits obtained without helicity suppression, as in \cref{table11,table12,table2}, and the horizontal dashed purple lines corresponds to the limits calculated with exact helicity suppression as in \cref{table4,table13,table14}. For large values of $\Vert \mathcal{X} \Vert_F$ the limits on the mass of $X_\mu$ approach their usual values whereas for values where $\Vert \mathcal{X} \Vert_F \lesssim 10^{-3}$ the limits on $m_X$ are almost identical to the limits obtained with exact helicity suppression.}
\label{fig:bidoubXmasses}
\end{figure}

\Cref{fig:bidoubXmasses} plots the experimental limits on the mass of $X_\mu$ as a function of $\mathcal{X}$ where again points in blue have the entries of $\mu_{\Psi_{22}}$ randomly scanned over and points in red correspond to where $\mu_{\Psi_{22}}$ has been fixed by through a Casas-Ibarra parameterisation. We find that the only region of parameter space which allows for a low-scale seesaw in the charged lepton and neutrino masses also happens to be the region where the limits on the mass of $X_\mu$ have been completely helicity suppressed to their lowest values. The plot on the left and right represent two different choices of $K_{L/R}^{de}$ where we find the same behaviour regardless of the form of $K_{L/R}^{de}$. For some more complicated scenario where the entries of $\mathcal{X}$ are not related to the smallness of neutrino mass, deviation from unitarity limits require $\Vert \mathcal{X} \Vert_F \lesssim 10^{-1}$ and therefore would at a minimum lead to the limits decreasing by about a factor of a half compared to the limits obtained without any charged-lepton mixing. For values where $\Vert \mathcal{X} \Vert_F \lesssim 10^{-2}$ the limits on the mass of $X_\mu$ are decreased by about an order of magnitude and for $\Vert \mathcal{X} \Vert_F \lesssim 10^{-3}$ the limits are completely helicity suppressed to their lowest values.

Therefore we find that with the addition of fermionic $SU(2)_{L/R}$ bi-doublets to the usual PS fermions the singlet $S_L$ is no longer necessary for the generation of neutrino masses. The down-quark, charged-lepton and neutrino masses can all be successfully generated for low scales of PS breaking only in the region where $\Vert \mu_{\Psi_{22}} \Vert$ is smaller than all other mass parameters, and this region also leads to an order of magnitude reduction in the limits on $m_X$ through helicity suppression. The smallness of the entries of $\mu_{\Psi_{22}}$ can be justified through technical naturalness, since a new global symmetry is recovered (which can be identified with a type of lepton number) when they are taken to zero. In this limit the active neutrino become massless, however other massive Majorana neutrinos are present. 

A number of predictions for the exotic fermions can be made in the scenario where only $f_{L/R}$ and $\Psi_{22}$ are present. Firstly heavy charged-lepton partners and pseudo-Dirac pairs of heavy neutrinos are predicted to have the same masses as each other at the scale of $SU(2)_R$ breaking and additional, intermediately heavy neutrinos are predicted to exist with masses between MeV to GeV which are sufficiently heavy to not violate any cosmological bounds. For more complicated scenarios with additional fermions (but the same charged-lepton mass mixing matrix), where the size of $\mathcal{X}$ is not tied to the smallness of neutrino mass, we find that unitarity deviation constraints begin to constrain the overall size of $\mathcal{X}$ and imply that the limits on $m_X$ must be reduced by at least a factor of a half.

\subsubsection{Bottom-left dominance: $SU(2)_{L/R}$ Triplet fermions}
\label{subsubsec:modtripcase}

The analysis of the alternative scenario where the mass matrix $Y_\Psi^R v_R$ appears in the bottom-left entry of $M_{eE}$ (such as when an $SU(2)_{R}$ triplet is added) follows very similarly to the above, with only a few key differences. As mentioned previously, the scalar content we have chosen is unable to generate a viable charged lepton mass spectrum in the case of the triplets due to the lack of an $e_L E_R$ or $E_L e_R$ mass term in the Yukawa sector, as per \cref{eqn:tripchargelepmixR,eqn:tripchargelepmixL}. However, these missing mass terms can be generated through a modified scalar sector. For example, if $\rchi_L \sim (\textbf{4},\textbf{2},\textbf{1})$ is replaced with $\rchi' \sim (\textbf{4},\textbf{2},\textbf{3})$ then, as was first noted in~\cite{Balaji:2019kwe}, the missing mass term in the top-right entry of $M_{eE}$ in \cref{eqn:tripchargelepmixR}, proportional to $v_L$, is generated for the case of an $SU(2)_R$ triplet. Similarly, if $\rchi_R \sim (\textbf{4},\textbf{1},\textbf{2})$ is replaced with $\rchi'' \sim (\textbf{4},\textbf{3},\textbf{2})$ then the missing mass term in the bottom-left entry of $M_{eE}$, proportional to $v_R$, is generated for the case of an $SU(2)_L$ triplet. We therefore assume that all mass terms are generated in what follows, with a $Y_{\Psi_3}^R v_R$ mass matrix appearing in the bottom-left entry of $M_{eE}$ in \cref{eqn:tripchargelepmixL}, but remain agnostic about the scalar sector which generates it.

We shall also not undertake a detailed study of the requirements for neutrino masses in these modified scenarios, but make the following brief observations: A crucial difference between the triplet and bi-doublet scenarios is that the neutral lepton mass mixing matrix in the case of triplets is given by an inverse/linear seesaw similar to \cref{eqn:PSneutralmix}, with the exception that the terms are also related to the charged-lepton mass mixing matrix. In \cref{subsec:tripneutmass} we argue that, unlike the bi-doublet case, the mass mixing matrices in this scenario are such that low-scale seesaws for the neutrinos and charged leptons are unable to correctly reproduce the SM lepton masses. A viable setup therefore requires additional physics to further decouple the charged- and neutral-lepton mass mixing matrices, for example with the addition of singlet fermions $S_L$~\cite{Balaji:2019kwe} which also appear in the usual low-scale PS particle spectrum. Introducing $SU(2)_L$ or $SU(2)_R$ triplets therefore requires both a modification of the scalar content of the theory (to generate a viable charged-lepton mass spectrum), as well as \textit{additional} fermionic states to generate sufficiently light neutrinos for low scales of PS breaking.

We therefore assume some additional physics is included in the neutrino sector to allow for a viable mass spectrum and simply comment on the implications of a charged-lepton mass matrix of the form
\begin{equation}
\mathcal{L}_{eE} = \begin{pmatrix}
\overline{e_L} & \overline{E_L}
\end{pmatrix} \begin{pmatrix}
m_d & Y_{\Psi_3}^L v_L  \\
\sqrt{2}\, Y_{\Psi_3}^R v_R^* & \,\,\,\mu_{\Psi_3}
\end{pmatrix}
\begin{pmatrix}
e_R\\
E_R
\end{pmatrix} ,
\end{equation}
where the dominant seesaw term is in the bottom-left entry has on the limits on PS breaking through $m_X$.

Adopting the hierarchy $\Vert Y_{\Psi_3}^R \Vert > \Vert m_d,\,\mu_{\Psi_3},\,Y_{\Psi_3}^L v_L  \Vert$,  necessary for a chiral suppression to occur, leads to 
\begin{equation}
\label{eqn:bottomleftchargelepmass}
m_\ell \simeq v_L Y_{\Psi_3}^L - \frac{1}{v_R} m_d (Y_{\Psi_3}^R)^{-1} \mu_\Psi
\end{equation}
for the light states after diagonalisation. The relationship between the interaction and mass eigenstates is now given by
\begin{align}
\label{eqn:botleftintmassmixE}
e'_L &\simeq (O_L^e)^\dagger \left(\id -\frac{1}{2}\mathcal{X} \mathcal{X}^\dagger\right) e_L - (O_L^e)^\dagger\mathcal{X}\,E_L\nonumber\\
e'_R &\simeq -(O_R^e)^\dagger\mathcal{Z}^\dagger\,e_R + (O_R^e)^\dagger\left(\id -\frac{1}{2}\mathcal{Z}^\dagger \mathcal{Z}\right) E_R
\end{align}
where $O^e_{L/R}$ diagonalises the light mass block, 
\begin{align}
\mathcal{X} &\simeq \frac{1}{v_R} m_d \left(Y_{\Psi_3}^R\right)^{-1} + \frac{v_L}{v_R^2} Y_{\Psi_3}^L \mu_{\Psi_3}^\dagger [(Y_{\Psi_3}^R)^\dagger]^{-1}(Y_{\Psi_3}^R)^{-1}\nonumber\\
\mathcal{Z} &\simeq \frac{1}{v_R}(Y_{\Psi_3}^R)^{-1} \mu_{\Psi_3} + \frac{v_L}{v_R^2}(Y_{\Psi_3}^R)^{-1}[(Y_{\Psi_3}^R)^\dagger]^{-1}m_d^\dagger Y_{\Psi_3}^L
\end{align}
and we have expanded up to second order. The physical mixing matrices are now given by
\begin{equation}
\label{eqn:lepbotleftdom}
K_L^{de} \simeq -(U_L^d)^\dagger \left(\id -\frac{1}{2}\mathcal{X} \mathcal{X}^\dagger\right)O_L^e,\,\, K_R^{de} \simeq -(U_R^d)^\dagger \mathcal{Z} O_R^e \text{ and  }U_\text{PMNS} \simeq N_\nu^\dagger \left(\id -\frac{1}{2}\mathcal{X} \mathcal{X}^\dagger\right) O_L^e.
\end{equation}
The notable differences to the results in the previous section where the top-right entry of $M_{eE}$ is dominant are: (i) the leptoquark $X_\mu$ now effectively only couples to the left-handed states (compared with right-handed previously), and (ii)  the parameter controlling the deviation from unitarity for the PMNS and the parameter determining the degree of helicity suppression in meson decays are different. Unitarity deviation is determined by $\mathcal{X}$, which is given at lowest order by $\mathcal{X} \simeq  m_d (Y_{\Psi_3}^R v_R)^{-1}$, while the degree of helicity suppression in the $X_\mu$ couplings is given by $\mathcal{Z} \simeq (Y_{\Psi_3}^R v_R)^{-1} \mu_{\Psi_3}$. In the previous scenario of top-right entry dominance, both are controlled by $\mathcal{X}$ which in that case was given by $\mathcal{X} \simeq (Y_{\Psi_3}^R v_R)^{-1} \mu_{\Psi_3}$. Unlike before where $\mu_{\Psi_3}$ could be lowered, the only way to decrease the deviation from unitarity is by increasing the scale $v_R$ as $m_d$ is fixed by the SM down-quark masses. This leads to larger masses for the leptoquark $X_\mu$, and thus experimental limits on the deviation will more strongly constrain the allowed scales of $SU(2)_R$ breaking. As before, the degree of helicity suppression is controlled by $\mathcal{Z} \sim (Y_{\Psi_3}^R v_R)^{-1} \mu_{\Psi_3}$ but now there are no constraints on $\mathcal{Z}$ from unitarity deviation.

\begin{figure}[t]
\centering
{
  \includegraphics[width=0.45\linewidth]{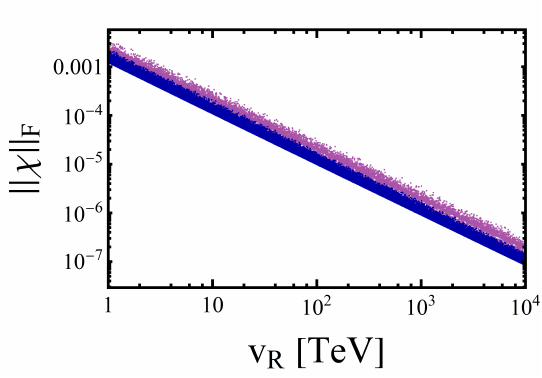}\hfill
}
{
  \includegraphics[width=0.45\linewidth]{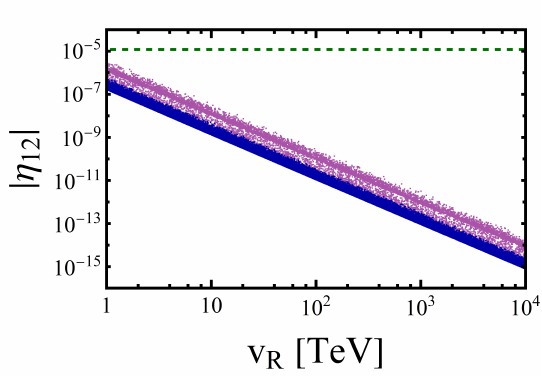}
}
\caption[Variation of $\mathcal{X}$ as a function of $v_R$ with additional triplets.]{Plots of the variation of $\mathcal{X}$ \textbf{(left)} and the variation of the most experimentally constrained entry of $\eta$ \textbf{(right)} as a function of $v_R$. Points in blue correspond to where $\Vert \mu_{\Psi_3} \Vert < \Vert m_d \Vert$ and therefore $\mathcal{X} \simeq \mathcal{X}^{(1)}$ and points in purple correspond to the opposite regime where $\mathcal{X} \simeq \mathcal{X}^{(2)}$ where we have scanned over the entries of $\mu_{\Psi_3}$ only requiring $\Vert \mu_{\Psi_3} \Vert < \Vert Y_{\Psi_3}^R v_R \Vert$ such that the seesaw assumption is satisfied. Here there are no current experimental constraints on $\mathcal{X}$ from unitarity deviation however future constraints on $\eta$ will begin to constrain $v_R$.}
\label{fig:tripdeviation}
\end{figure}

The two plots of \cref{fig:tripdeviation} show the size of $\mathcal{X}$ as a function of $v_R$ where as before we consider the singular values in $Y_{\Psi_3}^R$ to vary between $0.1$ and $1$. Points in blue correspond to $\mathcal{X}$ at first order where 
\begin{equation}
\mathcal{X}^{(1)} \simeq  m_d (Y_{\Psi_3}^R v_R)^{-1}
\end{equation}
and points in light purple correspond to where $\mathcal{X}$ has been calculated up to second order where
\begin{equation}
\mathcal{X}^{(2)} \simeq \mathcal{X}^{(1)} + \frac{v_L}{v_R^2} Y_{\Psi_3}^L \mu_{\Psi_3}^\dagger [(Y_{\Psi_3}^R)^\dagger]^{-1}(Y_{\Psi_3}^R)^{-1}.
\end{equation}
As $\mu_{\Psi_3}$ is a free parameter -- it is not fixed from the requirement of a viable neutrino mass spectrum -- for regions where $\Vert \mu_{\Psi_3} \Vert \gg \Vert m_d \Vert$ the second order term in $\mathcal{Z}$ can dominate leading to larger levels of unitarity deviation. Of course for larger values of $\mu_{\Psi_3}$ the degree of helicity suppression in the limits on $m_X$ also decreases leading to larger limits on the PS breaking scale. The right plot of \cref{fig:tripdeviation} shows the level of unitarity deviation in the most experimentally constrained entry of $\eta$ and it shows that for all values of $v_R$ the deviation is below the current experimental limit. Unlike the previous scenario, as $\mu_{\Psi_3}$ enters $\mathcal{X}$ at second order, large values of $\mu_{\Psi_3}$ are not constrained by requiring $\vert \eta_{ij} \vert < \vert \eta^{\text{exp}}_{ij} \vert$. For $v_R \sim 1$ TeV and large values in the entries of $\mu_{\Psi_3}$, the deviation from unitarity in the most constrained entry of $\eta$ is roughly one order of magnitude below the current experimental limits. For small values of $\mu_{\Psi_3}$ (which correspond to the region which helicity suppresses the limits on $m_X$) the predicted deviation in unitarity is roughly an order of magnitude smaller. Unlike the previous scenario, however, \cref{fig:tripdeviation} demonstrates that as the constraints on $\eta^{\text{exp}}$ get stronger, they will lead to constraints on the allowed magnitude of $v_R$. This will require limits several orders of magnitude stronger than the current ones.

\begin{figure}[t]
\centering
{
  \includegraphics[width=0.45\linewidth]{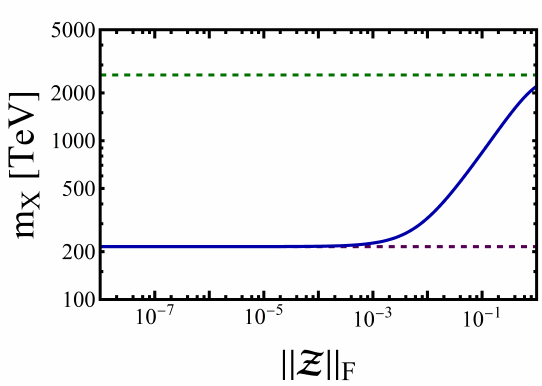}
}
\caption[Plot of $m_X$ as a function of $\mathcal{Z}$ with additional triplet fermions.]{Plot of the limits on $m_X$ as a function of $\Vert \mathcal{Z} \Vert_F$ assuming that $K_{L}^{de} = \id$. As the extent of chiral suppression in the couplings of $X_\mu$ to $e'_R$ and $d'_R$ is controlled by $\mathcal{Z}$, requiring full helicity suppression in the limits on $m_X$ requires $\Vert \mathcal{Z} \Vert_F \lesssim 10^{-3}$ as in the previous section. Unlike the previous section $\mathcal{Z}$ is not constrained by limits on unitarity deviation.}
\label{fig:tripmasslim}
\end{figure}

Similarly to the scenario with top-right dominance, the level of chiral suppression in the couplings of $X_\mu$ is controlled by $\mathcal{Z} \simeq \mu_{\Psi_3} (Y_\Psi^R)^{-1}$. The left plot of \cref{fig:tripmasslim} demonstrates that the variation in the limits on $m_X$ vary in the exact same way as \cref{fig:bidoubXmasses} and requiring complete helicity suppression in the limits on $m_X$ requires $\Vert \mathcal{Z} \Vert_F \lesssim 10^{-3}$ as before. 

Whereas in the bi-doublet scenario the smallness of $\mu_{\Psi_{22}}$ could be directly connected to the light neutrino masses, here we find additional physics is required for a phenomenologically-viable neutrino mass spectrum. Therefore the relationship between the bare mass term $\mu_{\Psi_3}$ and the light neutrino masses cannot be established without properly considering the possible hierarchies of parameters in the full neutrino mass matrix, which is itself model dependent. While we do not thoroughly analyse the neutrino mass matrix of a more complete model, we note that~\cite{Balaji:2019kwe} found that if an $SU(2)_R$ triplet, $\Psi_3$, was extended with additional fermionic singlets (such that the full Yukawa Lagrangian was given by a combination of \cref{eqn:PSYuk,eqn:YUK3}) a viable neutrino mass spectrum was recovered for $\Vert \mu_{\Psi_3} \Vert \lesssim 1$ GeV which would suggest that $\mathcal{Z} \ll 10^{-3}$ and therefore the limits on $m_X$ will be helicity suppressed to their lowest values. However we note that the limit $\mu_{\Psi_3} \rightarrow 0$ does not recover a global symmetry of the Lagrangian as it does in the bi-doublet case and therefore it is unlikely that the smallness of $\mu_{\Psi_3}$ can be related to the smallness of neutrino mass. It may be possible that there are multiple regions in which viably light neutrino masses are recovered, some of which require the entries of $\mu_{\Psi_3}$ to be large, which would not lead to a reduction in the limits on $m_X$. While the smallness of $\mu_{\Psi_3}$ cannot be guaranteed by the smallness of the active neutrino masses, as in the bi-doublet scenario, the work in~\cite{Balaji:2019kwe} at least indicates that there is a region of parameter space which recovers all SM fermion masses with $\mu_{\Psi_3}$ small enough to reduce fully the limits on $m_X$.

Therefore exotic PS fermion multiplets which introduce additional states which mix with the SM charged and neutral leptons are an attractive method for reducing the limits on PS breaking. As both multiplets considered contain both charged and neutral states, the mixings within both sectors are now coupled. Requiring a valid neutrino-mass spectrum leads to a chiral suppression in the limits on $m_X$. In the case of additional $SU(2)_{L/R}$ bi-doublets, a chiral suppression can be motivated through technical naturalness arguments. In the case of additional $SU(2)_{L/R}$ triplets, the usual PS scalar content is ruled out phenomenologically. However, if the scalar and fermion sectors are modified appropriately the desired masses and chiral suppression can be achieved. Although not linked to any technical naturalness arguments, previous studies with the triplets and more exotic scalar content suggest that a viable neutrino mass spectrum will still lead to a complete chiral suppression in the limits on $m_X$. Due to the fact that the mass hierarchy between the down quarks and charged leptons differs between the different generations, the correct masses are most naturally generated for the two fermion types by having their masses arise from different Yukawa couplings and vevs. In the two scenarios considered above, viable charged-lepton and down-quark masses implied that the masses of all quarks arise from couplings of $f_{L/R}$ to $\phi$ with vevs $v_1$ and $v_2$, whereas the lepton masses are related to couplings of the relevant fermions to $\rchi_L$ with vev $v_L$, suggesting the hierarchy $v_{1,2} \gg v_L \gtrsim 1$ GeV.

\subsection{$d-D$ mixing: Sextet fermions}
\label{subsubsec:fermionspec-dD}

Mixing between the down quarks and heavy exotic partners is similar to the scenarios above involving lepton mixing. A number of possible fermion extensions lead to $d-D$ mixing. Here we will focus on the feasibility of $d-D$ mixing alone in generating both a viable SM mass spectrum and a suppression of the mass limits on $m_X$ which occurs through the addition of the $SU(4)$ sextet fermions, but may also occur with higher dimensional PS multiplets. The results for down-quark mixing will follow very similarly to \cref{subsubsec:modtripcase}. As can be seen in \cref{eqn:sexmassmix}, $y_{\Psi_6}^R v_R$ appears in the bottom-left entry of $M_{dD}$.

Writing the down-quark mass mixing matrix similarly to the case of the charged leptons
\begin{align}
\label{eqn:geneDmix}
\mathcal{L}_{dD} = \begin{pmatrix}
\overline{d_L} & \overline{D_L}
\end{pmatrix} &\underbrace{\begin{pmatrix}
m_{dd} & m_{dD}\\
m_{Dd} & m_{DD}
\end{pmatrix}}\begin{pmatrix}
d_R\\
D_R
\end{pmatrix} + \text{H.c}\nonumber\\
& \qquad\, M_{dD}
\end{align}
and noting that the entries of \cref{eqn:sexmassmix} appear similarly to the case of $M_{eE}$ above allows us to draw the same conclusions which we will summarise. Limits on exotic quark states exceed $1$ TeV~\cite{Aad:2015voa} and therefore for phenomenological reasons a seesaw in $M_{dD}$ is required. As before the only two entries of $M_{dD}$ not tied to the electroweak scale are $y_{\Psi_6} v_R$ and $\mu_{\Psi_6}$ and, as discussed, a significant chiral suppression in limits on the PS breaking scale will only occur for the scenario where $\Vert y_{\Psi_6}^R v_R \Vert > \Vert m_d, y_{\Psi_6}^L v_L, \mu_{\Psi_6} \Vert$.

This hierarchy implies the down-quark masses are given upon diagonalisation by
\begin{equation}
m_\mathscr{d} \simeq y_{\Psi_6}^L v_L^* - m_e \,(y_{\Psi_6}^R v_R)^{-1} \mu_{\Psi_6}
\end{equation}
where the PS symmetry enforces that the singular values of $m_e$ give the correct charged-lepton masses at the appropriate scale in the case where there are no additional multiplets that induce $e-E$ mixing. For the same reasons as with the $SU(2)_{L/R}$ triplets, the correct down-quark masses cannot be recovered in the limit $y_{\Psi_6}^L v_L \rightarrow 0$ as this would imply
\begin{equation}
m_\mathscr{d} \simeq - m_e \,(y_{\Psi_6}^R v_R)^{-1} \mu_{\Psi_6}
\end{equation}
suggesting all three generations of down-quarks are lighter than the corresponding generation of charged-lepton, as discussed in \cref{subsec:gapprop}. In \cref{sec:sexPS} the constraint from baryon number violation suggested that imposing baryon number conservation required suppressing the Yukawa interactions of $\rchi_L$ to the fermions (for example by removing the scalar). However as phenomenologically the mass term $y_{\Psi_6}^L v_L$ is required in order to generate a viable charged-lepton and down-quark mass spectrum, this suggests that models involving just $d-D$ mixing through the introduction of $SU(4)$ sextets are likely ruled out by proton lifetime measurements. An analysis of all possible proton and neutrino decay diagrams in the region where the correct down-quark and charged-lepton masses arise is beyond the scope of this work. If the dominant decay channels are two-body decays which proceed via $d=6$ effective four-fermi interactions similar to \cref{figure:sectetPdecay}, estimates from~\cite{Helo:2019yqp} suggest that for TeV scale new physics the product of Yukawa couplings within the relevant loop diagram must satisfy $Y \lesssim 10^{-6}$. As $y_{\Psi_6}^L v_L$ is related to the down-quark masses in this case and for low scales of $v_R$, $y_{\Psi_6}^R$ is required to be close to unity such that the heavy $D$ states are sufficiently heavy na\"ively suggests difficulty in suppressing the decays. 

If a viable model involving the $SU(4)$ sextets exists which can suppress the dangerous proton decay diagrams whilst generating the appropriate charged-lepton and down-quark masses in the regime where the mass term $\vert y_{\Psi_6}^R \vert$ is dominant, the requirements for a chiral suppression in the limits on $m_X$ would follow almost identically to the $e-E$ mixing case above. For small values of $\mu_{\Psi_6}$ the couplings of $X_\mu$ to $d'_R$ and $e'_R$ will be suppressed and decreasing the experimental limits to their lowest value would roughly require 
\begin{equation}
\Vert \mathcal{Z} \Vert_F \simeq (Y_{\Psi_6}^R)^{-1} \mu_{\Psi{6}} \lesssim 10^{-3}.
\end{equation}
The only difference between $d-D$ and $e-E$ mixing will be in the deviation of unitarity, where $e-E$ mixing is constrained by deviation of unitarity in the PMNS whereas $d-D$ mixing is constrained by deviation within the measured CKM matrix,
\begin{equation}
V_\text{CKM} \simeq (U_L^u)^\dagger \left(\id -\frac{1}{2}\mathcal{X} \mathcal{X}^\dagger\right) O_L^d
\end{equation}
where
\begin{align}
\label{eqn:botleftintmassmixD}
d'_L &\simeq (O_L^d)^\dagger \left(\id -\frac{1}{2}\mathcal{X} \mathcal{X}^\dagger\right) d_L - (O_L^d)^\dagger\mathcal{X}\,D_L\nonumber\\
d'_R &\simeq -(O_R^d)^\dagger\mathcal{Z}^\dagger\,d_R + (O_R^d)^\dagger\left(\id -\frac{1}{2}\mathcal{Z}^\dagger \mathcal{Z}\right) D_R
\end{align}
and
\begin{align}
\mathcal{X} &\simeq \frac{1}{v_R} m_X \left(Y_\Psi^R\right)^{-1} + \frac{v_L}{v_R^2} Y_\Psi^L \mu_\Psi^\dagger [(Y_\Psi^R)^\dagger]^{-1}(Y_\Psi^R)^{-1}\nonumber\\
\mathcal{Z} &\simeq \frac{1}{v_R}(Y_\Psi^R)^{-1} \mu_\Psi + \frac{v_L}{v_R^2}(Y_\Psi^R)^{-1}[(Y_\Psi^R)^\dagger]^{-1}m_d^\dagger Y_\Psi^L.
\end{align}
In \cref{subsubsec:fermionspec-eE-dD} which involves coupled $e-E$ and $d-D$ mixing we find that limits from CKM unitarity deviation lead to very similar constraints on the level of mixing within the theory and therefore we find that $d-D$ mixing would be equally viable to the $e-E$ mixing above in the absence of proton decay issues.

\subsection{Coupled $e-E$ and $d-D$ mixing: Decuplet}
\label{subsubsec:fermionspec-eE-dD}

Coupled mass mixing, which we define as scenarios which involve the introduction of exotic PS multiplets which contain both $D$ and $E$ states within the same multiplet, occurs with the addition of a pair of opposite-chirality $SU(4)$ decuplets or with pairs of $SU(4)\otimes SU(2)_{L/R}$ bi-fundamentals with appropriate chiral structure. There is an additional subtlety for such coupled scenarios compared to the previous analyses. Due to the coupled nature of the exotic states $D$ and $E$ there exist additional relevant gauge interactions with the leptoquark $X_\mu$.

Consider the full fermion kinetic Lagrangian with the addition of $SU(4)$ decuplet fermions $\Psi^{10}_{L/R}$:
\begin{equation}
\mathcal{L}_{\textsc{kin}}^{\textsc{f}} = i \overline{f_{L/R}} \slashed{D} f_{L/R} + i\, \overline{\Psi^{10}_{L/R}} \,\slashed{D} \Psi^{10}_{L/R}
\end{equation}
with covariant derivative given by
\begin{equation}
\label{eq:decupletcovariantderiv}
\slashed{D} \Psi^{10}_{L/R} = \partial_\mu \Psi^{10}_{L/R} + i g_4 \hat{G}_\mu \Psi^{10}_{L/R} + i g_4\Psi^{10}_{L/R} (\hat{G}_\mu)^T
\end{equation}
for the decuplets, which transform as $\Psi^{10}_{L/R} \rightarrow U_4^{\vphantom{T}} \Psi^{10}_{L/R} U_4^T$. Expanding out \cref{eq:decupletcovariantderiv} and leaving only the interactions of interest gives
\begin{equation}
\mathcal{L}_{\textsc{kin}}^{\textsc{f}} \supset \frac{g_4}{\sqrt{2}} \left( \overline{d} \slashed{X} P_{L/R} e + \sqrt{2}\, \overline{D} \slashed{X} P_{L/R} E\right) + \text{H.c}
\end{equation}
and therefore the exotic states $D$ and $E$ within $\Psi^{10}_{L/R}$ have similar gauge interactions to $X_\mu$ as the states $d$ and $e$ within $f_{L/R}$. Due to the mass mixing present in $M_{dD}$ and $M_{eE}$ in \cref{eqn:decupmassmixE,eqn:decupmassmixD} between the states $f_{L/R}$ and $\Psi^{10}_{L/R}$, the physical mass states $e',\,E',\,d'$ and $D'$ are admixtures as before. Writing
\begin{equation}
\label{ch4:eq:decupmixingmatricesgeneralED}
\begin{pmatrix}
e\\
E
\end{pmatrix}_{L/R} = \begin{pmatrix}
V^e_{L/R} e_{L/R}' + W^e_{L/R} E_{L/R}'\\
X^e_{L/R} e_{L/R}' + Y^e_{L/R} E_{L/R}'
\end{pmatrix}\,\,\text{and}\,\,\begin{pmatrix}
d\\
D
\end{pmatrix}_{L/R} = \begin{pmatrix}
V^e_{L/R} d_{L/R}' + W^e_{L/R} D_{L/R}'\\
X^e_{L/R} d_{L/R}' + Y^e_{L/R} D_{L/R}'
\end{pmatrix}
\end{equation}
for the unitary diagonalisation matrices as usual leads to
\begin{equation}
\frac{g_4}{\sqrt{2}}\left(\overline{d_L'} \left((V_L^d)^\dagger V_L^e + \sqrt{2}(X_L^d)^\dagger X_L^e\right)\slashed{X} e_L'+ (L \leftrightarrow R )\right)
\end{equation}
for the gauge interactions between the physical SM-like states $d'$ and $e'$. We have once again neglected to write the similar gauge interactions which contain the heavy states $D'$ or $E'$ as by assumption the relevant mesons are kinematically forbidden from decaying into them.

Therefore the relevant mixing matrices are now given by
\begin{equation}
\label{eq:decupKLRde}
K_{L/R}^{de} = (V_{L/R}^{d})^\dagger V_{L/R}^e + \sqrt{2}\, (X_{L/R}^{d})^\dagger X_{L/R}^e.
\end{equation}
Due to the second term which now appears in \cref{eq:decupKLRde} (which arises from the components of $d'$ and $e'$ contained within the states $D$ and $E$), the resultant mixing matrix can only be made helicity-suppressed if and only if a hierarchy in both $V$ and $X$ occurs, e.g. $\Vert V_R^d \Vert \ll \Vert V_R^e \Vert$ \textit{and} $\Vert X_R^d \Vert \ll \Vert X_R^e \Vert$. This is unlike the previous scenarios where the additional multiplets included did not induce any new interactions between $X_\mu$ and exotic states with down-quark and charged-lepton quantum numbers. For scenarios where the $D$ and $E$ states are introduced through two different PS multiplets, no such gauge interactions will occur and the requirements for a helicity suppression in the relevant meson decays will be given by a combination of the results in \cref{subsubsec:fermionspec-eE,subsubsec:fermionspec-dD}. Therefore what follows will be specific to `coupled' scenarios where the new states $D$ and $E$ transform within the same multiplet.

Consider the mass mixing matrices between the down-quark and charged-lepton states, in the three-generational scenario, which arises from the inclusion of a pair of $SU(4)$ decuplets:
\begin{equation}
\label{ch4:eq:massmatricesrewritteneEdD}
M_{eE} = \begin{pmatrix}
m_F & \sqrt{2} \, \mathlarger{Y}_{\Psi_{10}}^L v_L^* \\[2pt]
\sqrt{2} \, \mathlarger{Y}_{\Psi_{10}}^R v_R &   \,\,\,\,\mathlarger{\mu}_{\Psi_{10}} - \sqrt{\frac{3}{2}} \,\mathlarger{Y}_{\Phi} v_{\Phi} 
\end{pmatrix}
\,\,\,\,\text{and}\,\,\,\,
M_{dD} =  \begin{pmatrix}
m_F & \mathlarger{Y}_{\Psi_{10}}^L v_L^* \\[2pt]
\mathlarger{Y}_{\Psi_{10}}^R v_R &  \,\,\,\,\mathlarger{\mu}_{\Psi_{10}} - \sqrt{\frac{1}{6}} \,\mathlarger{Y}_{\Phi} v_{\Phi}
\end{pmatrix}.
\end{equation}
Previously it was found that a hierarchy where $\Vert \mathlarger{Y}_{\Psi_{10}}^R v_R \Vert$ is larger than all other entries of $M_{dD}$ or $M_{eE}$ is required for a helicity suppression to occur, and therefore for a reduction in the experimental mass limits of $X_\mu$. If such a hierarchy is assumed for the entries of $M_{dD}$ and $M_{eE}$ above, the unitary block diagonalisation matrices are given by
\begin{equation}
\label{ch4:eq:decupletULRforDE}
U_L^{e,d} \simeq \begin{pmatrix}
\id_{3\times 3} - \frac{1}{2} \mathcal{X} \mathcal{X}^\dagger & \mathcal{X}\\
-\mathcal{X}^\dagger & \id_{3 \times 3} - \frac{1}{2} \mathcal{X}^\dagger \mathcal{X}
\end{pmatrix}\,\,\,\,\text{and}\,\,\,\,
U_R^{e,d} \simeq \begin{pmatrix}
-\mathcal{Z} & \id_{3\times 3} - \frac{1}{2} \mathcal{Z} \mathcal{Z}^\dagger\\
\id_{3\times 3} - \frac{1}{2} \mathcal{Z}^\dagger \mathcal{Z} & \mathcal{Z}^\dagger
\end{pmatrix}
\end{equation}
where the explicit forms for $\mathcal{X}$ and $\mathcal{Z}$ in the case of $U_{L/R}^e$ and $U_{L/R}^d$ can be easily derived from \cref{subsec:multiDsingval}. As always, due to the assumption that there is a seesaw where $\Vert \mathlarger{Y}_{\Psi_{10}}^R v_R \Vert$ is dominant in both $M_{dD}$ and $M_{eE}$, the parameters $\mathcal{X}$ and $\mathcal{Z}$ are small: $\Vert \mathcal{X},\,\mathcal{Z}\Vert \ll 1$. Comparing \cref{ch4:eq:decupletULRforDE} with \cref{ch4:eq:decupmixingmatricesgeneralED} implies
\begin{equation}
\Vert V^e_L \Vert\simeq \Vert V^d_L \Vert \simeq 1,\,\Vert V^e_R \Vert \simeq \Vert V_R^d \Vert \simeq 0
\end{equation}
and
\begin{equation}
\Vert X^e_L \Vert \simeq \Vert X^d_L \Vert \simeq 0,\,\Vert X^e_R \Vert \simeq X^d_R \simeq 1.
\end{equation}
Therefore under the assumption that the mass term proportional to the $SU(2)_R$ breaking is dominant for both $M_{dD}$ and $M_{eE}$ (which previously led to a helicity-suppression with $\Vert K_R^{de} \Vert \ll 1$ for cases of uncoupled $e-E$ or $d-D$ mixing), we find no suppression in the relevant mixing matrices:
\begin{equation}
\Vert K_L^{de} \Vert \simeq \Vert K_R^{de} \Vert \simeq 1
\end{equation}
which is due to the second contribution to $K_{L/R}^{de}$ appearing in \cref{eq:decupKLRde}. Physically it is very simple to see why this occurs. Assuming that the bottom-left entries of $M_{dD}$ and $M_{eE}$ are both dominant in a seesaw implies that the light mass states $d_L',\,d_R',\,e_L'$ and $e_R'$ are predominately composed of $d_L,\,D_R,\,e_L$ and $E_R$ respectively. However $d_L$ and $e_L$ are contained in the same PS multiplet \textit{and} $D_R$ and $E_R$ are also contained within a PS multiplet together. As such, the gauge leptoquark $X_\mu$ couples to both chiralities of fermion mass states with little suppression. If instead the bottom-right entries of $M_{dD}$ and $M_{eE}$ are assumed dominant, for the same reasons as presented in \cref{subsubsec:fermionspec-eE}, the desired helicity suppression in the relevant meson decays will not occur.

From the above arguments it is clear that, for scenarios with coupled mixing in the down-quark and charged-lepton sectors, a chiral-suppression in the relevant meson decays is not possible if both mixing matrices $M_{dD}$ and $M_{eE}$ in \cref{ch4:eq:massmatricesrewritteneEdD} are assumed to have the same seesaw structure. However, note that the entries of the two mixing matrices differ by combinations of group theoretic factors from the Yukawa couplings of the adjoint scalar $\Phi$. Therefore it is in principle possible for there to be sufficient tuning of the mass parameters, particularly the parameters in the second row of $M_{eE}$ and $M_{dD}$, such that the seesaw structure in the two mass matrices differs.

To demonstrate this for for one-generation of fermions, assume that the hierarchy of parameters in $\mu_{\Psi_{10}},\,y_{\Psi_{10}}^R v_R$ and $y_\Phi v_\Phi$ is such that
\begin{equation}
 \mu_{\Psi_{10}} - \sqrt{\frac{1}{6}} y_\Phi v_\Phi   \geq 10\,  y_{\Psi_{10}}^R v_R \geq 100\left( \sqrt{\frac{1}{2}} \mu_{\Psi_{10}} - \sqrt{\frac{3}{4}} y_\Phi v_\Phi\right)
\end{equation}
is satisfied (taking each combination to be real and positive). This corresponds to the hierarchy $ m_{DD}  >  m_{Dd}, m_{Ee}  >  m_{EE} $. Under this assumption, the \textit{bottom-left} entry of $M_{eE}$ is at least an order of magnitude larger than the other entries of the charged-lepton mass mixing matrix. In addition, the \textit{bottom-right} entry of $M_{dD}$ is dominant in the down-quark mass mixing matrix. This certainly requires a degree of tuning between the mass parameters $\mu_{\Psi_{10}}$ and $y_\Phi\,v_\Phi$ and also implies that the adjoint scalar $\Phi$ is required in such a theory in order to introduce the required Georgi-Jarlskog-like factors. Now the unitary block diagonalisation matrices differ from \cref{ch4:eq:decupletULRforDE}:
\begin{align}
\label{ch4:eq:decupletULRforDETUNEd}
&U_L^{e} \simeq \begin{pmatrix}
\id_{3\times 3} - \frac{1}{2} \mathcal{X} \mathcal{X}^\dagger & \mathcal{X}\\
-\mathcal{X}^\dagger & \id_{3 \times 3} - \frac{1}{2} \mathcal{X}^\dagger \mathcal{X}
\end{pmatrix},\,
U_R^{e} \simeq \begin{pmatrix}
-\mathcal{Z} & \id_{3\times 3} - \frac{1}{2} \mathcal{Z} \mathcal{Z}^\dagger\\
\id_{3\times 3} - \frac{1}{2} \mathcal{Z}^\dagger \mathcal{Z} & \mathcal{Z}^\dagger\end{pmatrix},\nonumber\\
&U_L^{d} \simeq \begin{pmatrix}
\id_{3\times 3} - \frac{1}{2} \mathcal{X} \mathcal{X}^\dagger & \mathcal{X}\\
-\mathcal{X}^\dagger & \id_{3 \times 3} - \frac{1}{2} \mathcal{X}^\dagger \mathcal{X}
\end{pmatrix},\,
U_R^{d} \simeq \begin{pmatrix}
\id_{3\times 3} - \frac{1}{2} \mathcal{Z}^\dagger \mathcal{Z} & \mathcal{Z}^\dagger\\
-\mathcal{Z} & \id_{3\times 3} - \frac{1}{2} \mathcal{Z} \mathcal{Z}^\dagger
\end{pmatrix}.
\end{align}
Consequently
\begin{equation}
\Vert V^e_L \Vert\simeq \Vert V^d_L \Vert \simeq 1,\,\Vert V^e_R \Vert \simeq 0,\, \Vert V_R^d \Vert \simeq 1
\end{equation}
and
\begin{equation}
\Vert X^e_L \Vert \simeq \Vert X^d_L \Vert \simeq 0,\,\Vert X^e_R \Vert \simeq 1,\, \Vert X_R^d \Vert \simeq 0
\end{equation}
and therefore $\Vert K_L^{de} \Vert \simeq 1$ and $\Vert K_R^{de} \Vert \simeq 0$ as required. Again this is simple to see given that in this case the light mass states are predominately composed of $e_L,\,E_R,\,d_L$ and $d_R$, due to the seesaw assumption. Therefore the leptoquark $X_\mu$ will strongly couple to left-handed states\footnote{The light left-handed mass states will therefore also behave correctly in weak interactions.} (as $e_L$ and $d_L$ are in the same PS multiplet) whereas there will be a suppression in its right-handed couplings (as $E_R$ and $d_R$ are not within the same PS multiplet). Therefore, scenarios with coupled mass mixing matrices between the down-quark and charged-lepton sectors are able to generate the desired chiral couplings of $X_{\mu}$ for a tuned selection of certain mass parameters within the theory. To reiterate, this kind of tuning is required only when the extra states, $D$ and $E$, reside in the same PS multiplet. A similar tuning was required in~\cite{Calibbi:2017qbu,Iguro:2021kdw}, such that the mass mixing matrices in the down-quark and charged-lepton sectors had different seesaw structures, in order to achieve a chiral suppression the limits on $m_X$ for the case where multiple pairs of bi-fundamental PS fermions were introduced. We therefore find this to be a general requirement for models with coupled $e-E$ and $d-D$ mixing, when a chiral suppression in the couplings of $X_\mu$ is desired.

Importantly, the addition of $\Psi^{10}_{L/R}$ implies the existence of a Dirac colour-sextet fermion with electric charge $1/3$. Its mass is given by 
\begin{equation}
m_\psi = \mu_{\Psi_{10}} + \sqrt{\frac{1}{6}} y_\Phi v_\Phi
\end{equation}
and therefore requiring that the sextet is sufficiently heavy leads to constraints on the singular values of $\mu_{\Psi_{10}}$. The mass limits for a colour sextet fermion vary from roughly $100$ GeV if stable on collider length scales~\cite{PhysRevD.98.030001} up to a TeV~\cite{Han:2010rf} for large couplings and possibly in the multi-TeV range~\cite{Sirunyan:2019vgj}. However if a hierarchy similar to
\begin{equation}
\mu_{\Psi_{10}} - \sqrt{\frac{1}{6}} y_\Phi v_\Phi \gg y_{\Psi_{10}}^R v_R \gg \sqrt{\frac{1}{2}} \mu_{\Psi_{10}} - \sqrt{\frac{3}{4}} y_\Phi v_\Phi
\end{equation}  
is required, $m_\psi$ is already required to be at least at the multi-TeV scale considering the limits on the $SU(2)_R$ breaking scale. Therefore regions of parameter space which lead to the desired chiral suppression of $X_\mu$ will naturally require the colour-sextet masses to be sufficiently large.

Let us estimate whether this hierarchy of mass parameters, though somewhat tuned, is able to successfully reproduce the correct SM fermion mass hierarchies as well as generate a chiral suppression in the couplings of $X_\mu$ in the multi-generational scenario. For simplicity, put the bottom-right entry of $M_{eE}$ to be exactly zero (the maximally tuned case), that is:
\begin{equation}
\mu_{\Psi_{10}} - \sqrt{\frac{3}{2}} Y_\Phi v_\Phi = 0_{3\times 3}.
\end{equation}
This implies that the bottom-right entry of $M_{dD}$ is given by
\begin{equation}
\mu_{\Psi_{10}} - \sqrt{\frac{1}{6}} Y_\Phi v_\Phi = \frac{2}{3}\mu_{\Psi_{10}}
\end{equation}
and consequently
\begin{equation}
\label{ch4:eq:sextetmassinextremetuning}
m_\psi = \frac{4}{3}\mu_{\Psi_{10}}.
\end{equation}
The  two mass mixing matrices, $M_{eE}$ and $M_{dD}$, now simplify to
\begin{equation}
\label{ch4:eq:massmatricesrewritteneEdDfullytuned}
M_{eE} = \begin{pmatrix}
m_F & \sqrt{2} \, \mathlarger{Y}_{\Psi_{10}}^L v_L^* \\[2pt]
\sqrt{2} \, \mathlarger{Y}_{\Psi_{10}}^R v_R &   0_{3\times 3}
\end{pmatrix}
\,\,\,\,\text{and}\,\,\,\,
M_{dD} =  \begin{pmatrix}
m_F & \mathlarger{Y}_{\Psi_{10}}^L v_L^* \\[2pt]
\mathlarger{Y}_{\Psi_{10}}^R v_R &  \frac{2}{3}\mathlarger{\mu}_{\Psi_{10}}
\end{pmatrix}.
\end{equation}
We require that $\Vert \sqrt{\frac{1}{2}}\,m_F,\, Y_{\Psi_{10}}^L v_L \Vert < \Vert Y_{\Psi_{10}}^R v_R \Vert < \frac{2}{3}\Vert \mu_{\Psi_{10}} \Vert$ in order for the dominant entries of each mass matrix to be different. This hierarchy leads to
\begin{equation}
\label{ch4:eq:decuplightlepmassformula}
m_\ell  \simeq \sqrt{2} \,Y_{\Psi_{10}}^L v_L
\end{equation}
for the charged-lepton mass matrix and
\begin{equation}
\label{ch4:eq:decuplightdownmassformula}
m_\mathscr{d} \simeq m_F - \frac{3}{2} v_L v_R Y_{\Psi_{10}}^L \left(\mu_{\Psi_{10}}\right)^{-1} Y_{\Psi_{10}}^R
\end{equation}
for the down-quark mass mixing matrix. 

As the dominant block is in the bottom-left entry, the relationship between the interaction and mass eigenstates for the SM-like charged leptons is given identically to \cref{subsubsec:modtripcase}:
\begin{align}
\label{eqn:botleftintmassmixEdecuplet}
e'_L &\simeq (O_L^e)^\dagger \left(\id -\frac{1}{2}\mathcal{X}^e (\mathcal{X}^e)^\dagger\right) e_L - (O_L^e)^\dagger\mathcal{X}^e\,E_L\nonumber\\
e'_R &\simeq -(O_R^e)^\dagger(\mathcal{Z}^e)^\dagger\,e_R + (O_R^e)^\dagger\left(\id -\frac{1}{2}(\mathcal{Z}^e)^\dagger \mathcal{Z}^e\right) E_R
\end{align}
with $\mathcal{X}^e \simeq m_F (Y_{\Psi_{10}} v_R)^{-1}$ and $\mathcal{Z}^e \simeq \frac{v_L}{v_R^2}(Y_{\Psi_{10}}^R)^{-1} ((Y_{\Psi_{10}}^R)^\dagger)^{-1} m_F Y_{\Psi_{10}}^L$ at leading order. The light down-quark mass eigenstates are instead given by
\begin{align}
\label{eqn:botrightintmassmixDdecuplet}
d'_L &\simeq (O_L^d)^\dagger \left(\id -\frac{1}{2}\mathcal{X}^d (\mathcal{X}^d)^\dagger\right) d_L - (O_L^d)^\dagger\mathcal{X}^d\,D_L\nonumber\\
d'_R &\simeq (O_R^d)^\dagger\left(\id -\frac{1}{2} (\mathcal{Z}^d)^\dagger \mathcal{Z}^d\right) d_R+ (O_R^d)^\dagger(\mathcal{Z}^d)^\dagger  D_R
\end{align}
with $\mathcal{X}^d \simeq \frac{3}{2} v_L Y_{\Psi_{10}}^L (\mu_{\Psi_{10}})^{-1}$ and $\mathcal{Z}^d \simeq \frac{3}{2} v_R (\mu_{\Psi_{10}})^{-1} Y_{\Psi_{10}}^R$. The physical mixing matrices are therefore given by
\begin{align}
\label{eqn:mixingmatricesdecupcase}
K_L^{de} &\simeq (O_L^d)^\dagger \left[\left(\id -\frac{1}{2}\mathcal{X}^d (\mathcal{X}^d)^\dagger\right)\left(\id -\frac{1}{2}\mathcal{X}^e (\mathcal{X}^e)^\dagger\right)+ \mathcal{X}^d(\mathcal{X}^e)^\dagger\right]O_L^e,\nonumber\\
&\simeq  (O_L^d)^\dagger  O_L^e\nonumber\\
K_R^{de} &\simeq (O_R^d)^\dagger \left[(\mathcal{Z}^d)^\dagger\left(\id - \frac{1}{2}(\mathcal{Z}^e)^\dagger \mathcal{Z}^e\right)- \left(\id - \frac{1}{2} (\mathcal{Z}^d)^\dagger \mathcal{Z}^d\right)\mathcal{Z}^e \right]O_R^e\nonumber\\
&\simeq (O_R^d)^\dagger \left( (\mathcal{Z}^d)^\dagger - \mathcal{Z}^e  \right) O_R^e\nonumber\\
V_{\text{CKM}} &\simeq (U_L^u)^\dagger \left(  \id -\frac{1}{2}\mathcal{X}^d (\mathcal{X}^d)^\dagger  \right)O_L^d \text{ and }U_\text{PMNS} \simeq N_\nu^\dagger \left(\id -\frac{1}{2}\mathcal{X}^e (\mathcal{X}^e)^\dagger\right) O_L^e
\end{align}
where the two terms appearing in $K_{L/R}^{de}$ are generated from the gauge interactions $\overline{D} \slashed{X} E$ and $\overline{d}\slashed{X} e$ as explained above. This features the $X_\mu$ coupling being predominately to $d_L'$ and $e_L'$ as expected, with a strong suppression of its coupling to $d_R'$ and $e_R'$. 

Additionally, the unitarity deviation in the PMNS mixing matrix is given almost identically to \cref{subsubsec:modtripcase}, where it was quantified by $\mathcal{X} \simeq \frac{1}{v_R} m_d (Y_{\Psi_3}^R)^{-1}$, compared to the current scenario where $\mathcal{X}^e \simeq \frac{1}{v_R} m_F (Y_{\Psi_{10}}^R)^{-1}$. Note also that combining \cref{ch4:eq:decuplightlepmassformula,ch4:eq:decuplightdownmassformula} leads to
\begin{equation}
m_{\mathscr{d}} \simeq m_F - \frac{3}{2\sqrt{2}}v_R \,m_\ell \left(\mu_{\Psi_{10}}\right)^{-1} Y_{\Psi_{10}}^R ,
\end{equation}
which, combined with $\Vert Y_{\Psi_{10}}^R v_R \Vert < \Vert \mu_{\Psi_{10}} \Vert$, implies that $m_F \simeq m_d$, since the singular values of $m_\ell$ are of similar order to $m_d$. Therefore the constraints on the $SU(2)_R$ breaking scale, $v_R$, from the deviation of the PMNS matrix follows essentially identically to \cref{subsubsec:modtripcase} and therefore \cref{fig:tripdeviation} is also valid for this scenario. Therefore for all values of $v_R$ which are at the TeV scale or higher, there are no constraints arising from the deviation of unitarity from the PMNS matrix. 

\begin{figure}[t]
\centering
{
  \includegraphics[width=0.58\linewidth]{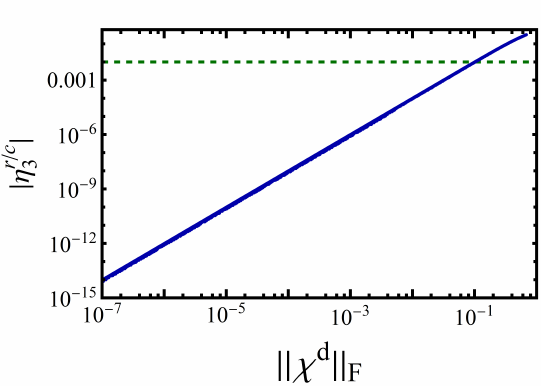}
}
\caption[Plot of the deviation of unitarity in the decuplet scenario.]{Plot of the deviation of unitarity in the CKM matrix as a function of $\Vert \mathcal{X}^d \Vert_F$ where the dashed green line roughly indicates the current experimental limit. The CKM matrix deviation is measured as the difference in the sum of the squares of each row and column. Even though it is currently the least constrained, the sum of the squares of the third row and column provide the strongest bounds on the deviation due to the large masses of the third generation of quarks compared to the first and second generation. This plot indicates that $\Vert \mathcal{X}^d \Vert_F \lesssim 10^{-1}$ is required. The limits for CKM matrix elements are taken from~\cite{Zyla:2020zbs}.}
\label{fig:decupdeviationunit}
\end{figure}

Due to the additional down-quark mixing, there are also constraints from the deviation from unitarity of the CKM matrix. In this case, the deviation is determined by the matrix $\mathcal{X}^d \propto v_L Y_{\Psi_{10}}^L (\mu_{\Psi_{10}})^{-1}$ where again $v_L Y_{\Psi_{10}}^L$ is proportional to the charged-lepton mass matrix, $m_\ell$. Now, $\mu_{\Psi_{10}}$ is related to the colour-sextet fermion mass scale from \cref{ch4:eq:sextetmassinextremetuning}. In \cref{fig:decupdeviationunit} we measure the level of deviation through the parameters
\begin{equation}
\eta^r_j = 1 - \sum_{i} |(V_{\text{CKM}})_{ji}|^2 \quad\text{ and }\quad \eta^c_k = 1 - \sum_{i} |(V_{\text{CKM}})_{ik}|^2.
\end{equation}
The strongest experimental limits on $\eta^{r/c}$ come from the sum of the squares of the first row or column of the CKM matrix. However, due to the much larger masses of the third generation of quarks, we find that deviations in the sum of the third row and column of the CKM matrix lead to stronger constraints on $\mathcal{X}^d$. We find that $\eta^r_3$ and $\eta^c_3$ place almost identical limits on the magnitude of $\Vert \mathcal{X}^d \Vert_F$ and therefore only plot one for brevity. Using~\cite{Zyla:2020zbs} places the limit 
\begin{equation}
-0.01 \lesssim \eta^{r/c}_3 \lesssim 0.09
\end{equation}
and therefore we conservatively set the limit $\vert \eta^{r/c}_3 \vert < 10^{-2}$ at the PS breaking scale. \Cref{fig:decupdeviationunit} therefore sets a rough limit of $\Vert \mathcal{X}^d \Vert_F \lesssim10^{-1}$ in order to prevent significant deviation occurring for the CKM matrix. This can be compared to the limits from PMNS deviation which currently lead to no derived limits on $v_R$.

The implications for the experimental limits on $m_X$ can also be very easily determined from the results derived in \cref{subsubsec:fermionspec-eE} and the explicit derivations of $K_{L/R}^{de}$ derived in \cref{eqn:mixingmatricesdecupcase}. From the results in \cref{fig:tripmasslim} it is clear that in order for a substantial chiral suppression to occur in the limits on $m_X$, where $K_R^{de} \simeq (O_R^d)^\dagger \mathcal{Z} O_R^e$, we roughly require $\Vert \mathcal{Z} \Vert \lesssim 10^{-2}$. The expression for $K_R^{de}$ given in \cref{eqn:mixingmatricesdecupcase} in this case is of a identical form with
\begin{equation}
K_R^{de} \simeq (O_R^d)^\dagger \left( (\mathcal{Z}^d)^\dagger - \mathcal{Z}^e \right) O_R^e \equiv (O_R^d)^\dagger \mathcal{Z}_{10} O_R^e.
\end{equation}
Therefore we clearly require that $\Vert \mathcal{Z}_{10} \Vert \lesssim 10^{-2}$ in order for the desired chiral suppression to occur. This will clearly be satisfied if both $\mathcal{Z}^d$ and $\mathcal{Z}^e$ individually satisfy this requirement, and we do not consider a scenario with a tuned cancellation between the two matrices. By inspection, from the definitions of $\mathcal{Z}^d$ and $\mathcal{Z}^e$ in \cref{eqn:botleftintmassmixEdecuplet,eqn:botrightintmassmixDdecuplet} this will be trivially satisfied by $\mathcal{Z}^e$, as it contains terms which are at most at the GeV scaled which are suppressed by terms which are at a minimum at the TeV scale, but in the case of $\mathcal{Z}^d$ roughly implies the hierarchy
\begin{equation}
\frac{\Vert v_R Y_{\Psi_{10}}^R \Vert}{\Vert \mu_{\Psi_{10}} \Vert} \lesssim 10^{-2}
\end{equation} 
and therefore we require the colour-sextet masses to be at least two order of magnitude larger than the mass term proportional to the $SU(2)_R$ breaking term, if a full suppression in the experimental limits on $m_X$ is desired. 

However note that here we are assuming that $\mu_{\Psi_{10}} = \sqrt{\frac{3}{2}} Y_\Phi v_\Phi$ where $v_\Phi$ is a vev which breaks $SU(4)$ and therefore contributes to the mass of $X_\mu$. Therefore the gauge leptoquark $m_X$ will roughly develop a mass at the same order as the colour-sextet fermion $m_\psi$, which will typically be a stronger limit on $m_X$ than the experimental constraints, for many choices of $K_R^{de}$. For example, if $v_R \simeq 5$ TeV, a large chiral suppression requires $\mu_{\Psi_{10}}$ to be roughly two order of magnitude larger than this, implying that $m_{\psi} \simeq m_X \gtrsim 500$ TeV. Therefore a mild suppression in the limits on $m_X$ is expected due to the required hierarchy between $Y_{\Psi_{10}} v_R$ and $\mu_{\Psi_{10}}$ (and therefore $Y_\Phi v_\Phi$). Only for situations where the mixing angles in $K_R^{de}$ are such that the experimental constraint on $m_X$ are initially close to their largest value, e.g. $K_R^{de} = \id_{3\times 3}$, will this lead to a suppression in the limits on $m_X$. We therefore find that it is possible for the leptoquark $X_\mu$ to develop a chiral coupling to the SM-like fermions and requires a tuning between the mass matrices $\mu_{\Psi_{10}}$ and $Y_{\Phi} v_\Phi$. However, while this may lead to a suppression in the decay rates of the relevant mesons, the large hierarchy required between $v_\Phi$ and $v_R$ requires $m_X \gtrsim 100$ TeV at a minimum.



\end{sloppypar}
\subsection{Comparison to the literature}
\label{subsubsec:compare-lit}

Attempts to lower the scale of Pati-Salam breaking have had a recent resurgence within the literature. Our  work focuses primarily on addressing two questions related to this. 

Firstly, we re-evaluated  the current experimental constraints on the PS breaking scale  in \cref{sec:mesonPS} in order to find the minimum allowed scale of PS breaking. Similar calculations have already appeared within the literature, most recently in~\cite{Smirnov:2007hv,Smirnov:2018ske} for the case where $X_\mu$ couples to both chiralities of SM fermions. Our analysis extends these results and calculates similar limits in the case of chiral ($V-A$ or $V+A$) couplings of $X_\mu$, establishing up-to-date lower bounds  on $m_X$, as presented in \cref{subsec:fermionmassdeg,table13,table14}. Secondly, we have studied  the interpolation between the two cases where $X_\mu$ couples either vectorially or chirally to $d$ and $e$, in order to establish how the experimental limits on $m_X$ decrease as the couplings of $X_\mu$ become more chiral-like, as presented in \cref{fig:bidoubXmasses,fig:tripmasslim}. This has implications for the phenomenological analyses within \cref{sec:fermionspec} and for  model building in similar scenarios.


The results in \cref{sec:exoticPS,sec:fermionspec} complement and extend past analyses on the same idea, as we now outline. To the best of our knowledge, the idea to break the down-isospin mass degeneracy with additional fermionic degrees of freedom was first introduced in~\cite{Foot:1997pb,Foot:1999wv} where additional $SU(2)_{L/R}$ bi-doublet fermions were introduced, similar to \cref{sec:biPS,subsubsec:bidoubcase,subsec:bidoubneutmass}. We build on this work through a detailed exploration of the charged-lepton and neutrino mass mixing matrices that arise. We prove that there is  only one hierarchy of parameters which prevents the active neutrinos from developing large masses as well as generating the correct down-quark and charged-lepton masses. We show that this hierarchy of parameters also  implies a full suppression in the experimental limits on $m_X$, therefore identifying an interesting connection between the helicity-suppressed PS limits and the smallness of neutrino mass. $SU(2)_{L/R}$ triplet fermions similar to \cref{sec:tripPS,subsubsec:modtripcase,subsec:tripneutmass} were first introduced in~\cite{Balaji:2018zna} for the left-right asymmetric version of $G_{\text{PS}}$, $SU(4)_c \otimes SU(2)_L \otimes U(1)_X$, in the context of the flavour  anomalies in $b\rightarrow s \mu\mu$. We find when extending this model to the left-right symmetric version of $G_{\text{PS}}$ that, without extra degrees of freedom, there is no parameter space which generates the correct neutrino, charged-lepton and down-quark masses simultaneously. This  complements the follow-up analysis of~\cite{Balaji:2019kwe}, where additional fermion singlets were introduced, such that the fermion content contained sub-GeV sterile neutrinos, allowing for $b \rightarrow c \tau \nu$ to be mediated by scalar leptoquarks within the theory. Our results indicate that this appears to be the simplest model with $SU(2)_{L/R}$ triplet fermions which is phenomenologically viable, suggesting that modifications to ${b \rightarrow c \tau \nu}$ naturally occur in this model. The analysis for the case of $SU(4)$ sextet and decuplet fermions in \cref{sec:sexPS,sec:decPS,subsubsec:fermionspec-dD,subsubsec:fermionspec-eE-dD} in the context of breaking the down-isospin mass degeneracy as well as lowering the PS breaking limits has not appeared in the literature, to the best of our knowledge. Finally, in \cref{sec:vectorPS} we find that two extra $SU(4)_c \otimes SU(2)_{L/R}$ bi-fundamental fermion multiplets are not phenomenologically viable in agreement with the results of~\cite{Calibbi:2017qbu,Iguro:2021kdw}. From the above results we conclude that models with PS breaking close to the TeV scale favour fermion multiplets which induce charged-lepton mixing. For cases where $D$ and $E$ states reside within the same PS multiplet, we show that lowering the PS scale occurs through a tuning of mass parameters requiring additional scalar fields.

A detailed examination of the link between the experimental limits on $m_X$ and the breaking of the down-isospin mass degeneracy is interesting, as Pati-Salam is one of the simplest models involving quark-lepton unification not constrained to have GUT scale breaking. For now we simply comment on the potential viability of $X_\mu$ as a mediator\footnote{As explained in~\cite{Hati:2019ufv}, combined explanations of $b \rightarrow s \mu \mu$ and $b \rightarrow c \tau \nu$ solely from the interactions of $X_\mu$ are highly disfavoured and prefer $m_X \lesssim 6$ TeV, which is in tension with the mass limits on $m_X$ derived in \cref{sec:mesonPS}.} of the $b \rightarrow s \mu\mu$ anomalies. Firstly, we find that the $SU(4)$ sextet appears to be not be viable phenomenologically if a helicity suppression in the relevent meson decays is desired, and therefore certainly cannot explain the anomalies in its minimal form. The $SU(4)$ decuplet can lead to a chiral suppression in the experimental limits of $X_\mu$. However, as we find that $m_X$ must exceed $100$ TeV at a minimum, in this scenario the gauge leptoquark is simply too heavy to provide a viable explanation. As mentioned, the triplet scenario has already been studied in this context~\cite{Balaji:2018zna,Balaji:2019kwe} and it was found that leptoquark masses within $12 \lesssim m_X \lesssim 31$ TeV are viable if $K_L^{de}$ has a structure similar to \cref{eqn:struct2}. Our evaluation of the experimental limits on $m_X$, for the matrix texture $K_L^{de}$ considered by~\cite{Balaji:2018zna,Balaji:2019kwe}, indicates that $m_X$ can be as low as $5$ TeV and therefore this scenario is viable. When introducing bi-doublet fermions, the gauge leptoquark $X_\mu$ now couples predominantly to right-handed leptons. However, the dominant effect appears to stem from couplings to left-handed leptons~\cite{DAmico:2017mtc} and, therefore, although we find that this scenario is the most attractive in lowering the PS limits, it appears unlikely to be a viable scenario if one wishes for (a light) $X_\mu$ to play a role in the anomalies. We therefore summarise that PS extended by $SU(2)_{L/R}$ triplets (and additional fermion singlets) appears to be the simplest UV model capable of explaining the B-meson anomalies if the gauge leptoquark predicted plays a dominant role in the $b \rightarrow s \ell \ell$ anomalies. 

While a gauge mediated solution is attractive, the scalar content of PS is less constrained and necessarily includes a number of scalar leptoquarks. These scalar leptoquarks alone are able to explain the anomalies, see e.g.~\cite{Heeck:2018ntp}, and therefore scenarios, such as with additional fermion bi-doublets, may still remain viable if the anomalies  arise from interactions of the scalar leptoquarks. A decrease in the experimental limit on $m_X$ is not necessarily required in this case, as scalar masses are not constrained by the PS breaking scale. However, we note that a helicity suppression in the PS limit is still favourable in these cases as it may help prevent the generation of a gauge hierarchy problem between a large PS breaking scale and the small scalar leptoquark masses assumed. We leave a full analysis of the ability of each model to explain the flavour anomalies, when the full particle spectrum is considered, to  future work.

\subsection{Connection to $SO(10)$}
\label{subsubsec:GUT connection}

While we are motivated by the experimentally attractive prospect of lowering the scale of PS breaking to be as low as possible, here we briefly consider how the exotic fermionic multiplets proposed could fit into an $SO(10)$ GUT and mention any obvious problems. We note the smallest dimensional multiplets of $SO(10)$ which contain each exotic multiplet as well as additional states this would predict. It is well known that two standard PS fermion multiplets fit into the spinorial
\begin{equation}
    \textbf{16}\rightarrow (\textbf{4},\textbf{1},\textbf{2}) \oplus (\textbf{4},\textbf{2},\textbf{1}) = f_L \oplus f_R
\end{equation}
of $SO(10)$. The fundamental representation of $SO(10)$,
\begin{equation}
    \textbf{10} \rightarrow (\textbf{1},\textbf{2},\textbf{2}) \oplus (\textbf{6},\textbf{1},\textbf{1}) = \Psi_{22} \oplus \Psi_{6},
\end{equation}
contains both the bi-doublet and sextet fermion states. While we find that the bi-doublet fermion is the most attractive candidate in lowering the PS breaking scale, the sextet fermion leads to undesirable proton decay diagrams. Some mechanism would therefore be desirable in order to significantly increase the masses of the $D$ states from the sextet, and not the $E$ and $N$ states from the bi-doublet, in order to prevent large proton decay widths. This cannot be through different bare mass terms, e.g. $\mu_{\Psi_{22}} \ll \mu_{\Psi_6} \simeq 10^{16}$ GeV, as the $SO(10)$ symmetry now enforces the two terms to be equal. 

The adjoint representation
\begin{equation}
    \textbf{45} \rightarrow (\textbf{1},\textbf{3},\textbf{1}) \oplus (\textbf{1},\textbf{1},\textbf{3}) \oplus (\textbf{15},\textbf{1},\textbf{1}) \oplus (\textbf{6},\textbf{2},\textbf{2}) \supset \Psi_{3_L} \oplus \Psi_{3_R}
\end{equation}
contains both the $SU(2)_L$ and $SU(2)_R$ triplet fermions. Interestingly, the additional $(\textbf{15},\textbf{1},\textbf{1})$ fermion contains a neutral state and therefore will contribute to the full neutrino mass matrix. As we found that the triplets alone are unable to generate a viable neutrino mass spectrum, due to the various mass equalities predicted, it is convenient that the minimal $SO(10)$ multiplet which contains these states also contains additional neutrino states which could possibly generate a viable spectrum of masses. The $(\textbf{6},\textbf{2},\textbf{2})$ fermion will likely require a similar mechanism to the $(\textbf{6},\textbf{1},\textbf{1})$ fermion in order to significantly increase its mass compared to the triplets within the theory. Finally the $SU(4)$ decuplets first appear in the $\textbf{120}$ representation of $SO(10)$. This representation also contains the bi-doublet fermion which will itself allow for the generation of chiral-like couplings of $X_\mu$ as well as generate neutrino masses, which the decuplet fermions do not. An interesting possible direction of future work is to analyse the feasibility of embedding such low-scale variants of PS into an $SO(10)$ GUT in such a way that the required fermion states do not develop GUT scale masses whereas any fermion in a given embedding which may mediate proton decay, or similar undesirable processes, have masses developed at the $SO(10)$ breaking scale.

\section{Conclusion}

We have studied the implications of introducing additional fermionic states to the usual Pati-Salam fermion multiplets. In particular we have focused on multiplets of relatively low dimensionality which contain partner states to the charged-leptons and/or down-quarks. The implications of mixing effects induced by these states were extensively studied, particularly with a focus on the feasibility of lowering the scale of PS breaking. We identified four multiplets in principle that allow for the possibility of a suppression in the PS limits where at most two extra fermion multiplets per generation are added. Of the four, the inclusion of an $SU(2)_{L/R}$ bi-doublet alone can lead to a valid mass spectrum for all SM particles and a significant reduction in the PS limits. The decuplet, requires the addition of extra singlet states (at a minimum) for neutrino mass, the inclusion of a scalar that induces a Georgi-Jarlskog like texture (such as $\Phi$) and is only able to reduce the limits on $m_X$ for some choices of $K_{L/R}^{de}$ and for a tuning between mass parameters. The remaining two multiplets, the $SU(2)_L$ and $SU(2)_R$ triplets, require both a modification of the scalar sector for a viable charged-lepton mass spectrum as well as additional fermionic states for a viable neutrino mass spectrum. A common feature of all scenarios was the existence of scalars $\rchi_L$ and $\rchi_R$ in order to generate a viable down-quark and charged-lepton mass spectrum.
Therefore three $SU(2)_L$ Higgs doublets are predicted.

The most attractive models, the $SU(2)_{L/R}$ bidoublet or triplets, contain only additional leptonic states, inducing both $e-E$ mixing as well as mixing within the neutrino sector. While both these models have already appeared in the literature, we have extended these analyses by rigorously studying the requirements for viable mass spectra of the down-quarks, charged-leptons and neutrinos as well as the implications for the PS breaking scale limits. We identify an attractive connection between the smallness of neutrino mass and the helicity suppression of the limits on $m_X$ which, particularly in the case of the bi-doublet, necessarily has a chiral-like coupling of $X_\mu$ to the SM fermions.

The sextet option leads to mixing in the down-quark sector alone but we find that this likely leads to large proton decay widths in the only region which can generate a viable mass spectrum. Finally the scenario with fermion decouplets, which leads to both $d-D$ and $e-E$ mixing, was considered. It also includes additional exotic states potentially discoverable at current and future colliders. The regime of interest, where the limits on PS breaking can be lowered, is very sensitive to future experimental limits on $SU(2)_R$ breaking scale $v_R$ and relies on a tuning of mass parameters.

We find that a chiral suppression in the PS breaking limits can reduce the usual limits of $81- 2467$ TeV, depending on the structure of $K_{L/R}^{de},$ down to as low as  $5.6 - 194$ TeV. An attractive property of models which lead to a chiral suppression is that only one of $K_{L/R}^{de}$ is required to have a specific structure. Without a chiral suppression, the lowest PS limits obtained required both of $K_{L/R}^{de}$ to have a specific structure which might suggest a parity symmetry in conflict with the assumed low-scale setup. 

A more in-depth examination of baryon number violation implications in the case with $SU(4)$ sextets is desired before it can be ruled out as a low-scale candidate. Additionally, due to the mass mixing induced, collider constraints for other particles predicted (such as the $W'$ and $Z'$) could lead to significantly smaller mass limits than what is usually expected. Exploring the reach of the LHC and future colliders for these extended PS scenarios with chiral-like gauge couplings may be required.

\acknowledgments
This work was supported in part by the Australian Research Council. TPD thanks John Gargalionis for helpful discussion and advice during the early parts of this work and Graham White for email correspondence.

\appendix
\section{Baryon number violation}
\label{sec:Appendix Z}

In order to assess the implications for baryon number violation arising from the unification of quarks and leptons within PS we consider the Yukawa and kinetic portion of the Lagrangian, and  the scalar potential separately. Consider first the electroweak sector of the Yukawa Lagrangian:
\begin{equation}
\mathcal{L}_{\textsc{yuk,ew}^{\vphantom{T}}} = \text{Tr} \left[ y_1^{\vphantom{T}} \overline{f_L} \phi \,(\mphant{f_R}{T})^T + y_2^{\vphantom{T}} \overline{f_L} \,\phi^c (\mphant{f_R}{T})^T \right] + \text{H.c.}
\end{equation}
 This Lagrangian is invariant under a single global $U(1)_J^{\vphantom{\dagger}}$ transformation 
\begin{equation}
f_L \rightarrow e^{i \theta J} f_L,\quad f_R \rightarrow e^{i \theta J} f_R \quad \text{and}\quad	 \phi \rightarrow \phi	\end{equation} 
and therefore the J charge of each field can be chosen such that
\begin{equation}
J(f_L) = J(f_R) = 1 \,\,\text{and}\,\,J(\phi) = 0.
\end{equation}
As $\phi$ is uncharged under the $SU(4)$ of Pati-Salam and the global symmetry $J$, the vev $\mvev{\phi}$ also does not break either symmetry e.g. $J(\mvev{\phi}) = T(\mvev{\phi}) = 0$, where T corresponds to the fifteenth generator of $SU(4)$ identified with $B-L$.
Baryon and lepton number can be identified as different linear combinations of $J$ and $T$
\begin{equation}
B = \frac{1}{4}(J + T) \,\,\,\text{and}\,\,\,L=\frac{1}{4}(J - 3 T)
\end{equation}
such that
\begin{equation}
B(f_L) = B(f_R) = \begin{pmatrix}
1/3 & 1/3\\
1/3 & 1/3\\
1/3 & 1/3\\
0 & 0\\
\end{pmatrix} \,\,\text{and}\,\,L(f_L) = L(f_R) = \begin{pmatrix}
0 & 0\\
0 & 0\\
0 & 0\\
1 & 1
\end{pmatrix}
\end{equation}
as required for the embeddings of SM fermion in $f_L$ and $f_R$ defined in \cref{eqn:fermionPSvanillaemb}. As the electroweak Yukawa Lagrangian is invariant under both $J$ and $T$ independently in both the broken and unbroken phase, it is clearly also invariant under a linear combination of the two and therefore all interactions conserve both $B$ and $L$.

Turning to the remaining terms in the Yukawa Lagrangian\footnote{Adopting the scalar and fermion particle content detailed in \cref{subsec:basicPS}.},
\begin{equation}
\mathcal{L}_{\textsc{yuk,ps}^{\vphantom{T}}} = \text{Tr} \left[ y_R^{\vphantom{T}} \overline{S_L} \rchi_R^{\dagger} f_R^{\vphantom{T}} +y_L^{\vphantom{T}} \overline{f_L} \rchi_L^{\vphantom{T}} (\mphant{S_L}{\dagger})^c\right] + \frac{1}{2}\mu_S^{\vphantom{T}} \overline{S_L^{\vphantom{c}}} S_L^c + \text{H.c}
\end{equation}
the global symmetry $U(1)_J^{\vphantom{\dagger}}$ is unbroken with the additional charge assignments
\begin{equation}
J(\rchi_L) = J(\rchi_R) = 1 \,\,\text{and}\,\,J(S_L) = 0
\end{equation}
where the $B$ and $L$ numbers of each component of $\rchi_{L/R}^{\vphantom{\dagger}}$ are identical to those of $f_{L/R}$. The scalars $\rchi_L$ and $\rchi_R$ are charged under both $J$ and $T$, such that $\frac{1}{4}(J+T)(\langle \rchi_{L/R} \rangle ) = B(\langle \rchi_{L/R} \rangle ) = 0$, whereas $\frac{1}{4}(J-3T)(\langle \rchi_{L/R} \rangle ) = L(\langle \rchi_{L/R} \rangle ) \neq 0$. Therefore $B$ is conserved by the Yukawa Lagrangian at all scales whereas lepton number is spontaneously broken at the scale of $SU(2)_\text{R}$ breaking. Baryon number being an accidental symmetry of the PS Yukawa Lagrangian is a feature which appears to be insensitive to the choice of scalars used to break the PS symmetry. If the scalars $\Delta_{L/R}^{\alpha}$ defined in \cref{eqn:PSscalar1} were present instead of $\rchi_{L/R}$ (often considered in high-scale PS models) identical conclusions are reached with baryon number remaining an accidental symmetry of the Yukawa sector while lepton number is spontaneously broken.

The kinetic portion of the Lagrangian
\begin{equation}
\mathcal{L}_{\textsc{kin}} = i \sum_{F}\overline{F} \slashed{D} F + \sum_{S} (D_\mu S)^\dagger (D^\mu S),
\end{equation}
where $F = (f_L,\,f_R,\,S_L)$ and $S = (\rchi_L,\,\rchi_R,\,\phi,\,\Phi)$, does not violate $B - L = T$ as it is a gauge symmetry. However it is easy to show that it also conserves a global $B+L = \frac{1}{2}(J - T)$ symmetry where 
\begin{equation}
(B+L)(\hat{G}_\mu) = \begin{pmatrix}
0 & 0 & 0 & -2/3\\
0 & 0 & 0 & -2/3\\
0 & 0 & 0 & -2/3\\
2/3 & 2/3 & 2/3 & 0 \, ,
\end{pmatrix}
\end{equation}
as every gauge field is uncharged under $U(1)_J$ and $\hat{G}_\mu$ corresponds to the $SU(4)_\text{c}$ gauge fields. The gauge fields of $SU(2)_\text{L}$ and $SU(2)_\text{R}$ are uncharged under $J$ and $T$ and therefore uncharged under $B$ and $L$. As the gauge interactions conserve both $B- L$ and $B+L $ simultaneously they necessarily conserve $B$ and $L$ separately and therefore there is no gauge-mediated baryon- or lepton-number violation predicted by PS, though these may appear if PS is embedded into some GUT at a higher scale, with the effects suppressed by the relevant unification scale (see e.g.~\cite{Croon:2018kqn}).

A comprehensive analysis of the scalar potential including minimisation of the potential is beyond the scope of this work. Of all the possible gauge invariant terms within the scalar potential, we find only one term\footnote{Additionally there are two extra possible terms ${\widetilde{\lambda}_{L}(\rchi_L)^4 = \widetilde{\lambda}_{L}\,(\rchi_L)^{Aa}(\rchi_L)^{Bb} (\rchi_L)^{C c} (\rchi_L)^{D d} \epsilon_{ABCD}\epsilon_{ab}\epsilon_{c d}}$ and ${\widetilde{\lambda}_{R}(\rchi_R)^4 = \widetilde{\lambda}_{R}\,(\rchi_R)^{A\alpha}(\rchi_R)^{B\beta} (\rchi_R)^{C \gamma} (\rchi_R)^{D \delta} \epsilon_{ABCD}\epsilon_{\alpha\beta}\epsilon_{\gamma \delta}}$ which would also break $U(1)_J$ however we find them to be identically zero once contracted.} which will violate $U(1)_J$ for the charge assignments imposed by the Yukawa Lagrangian:
\begin{equation}
\label{eqn:scalarpotbaryonV}
V(\phi,\Phi,\rchi_L,\rchi_R)\supset \widetilde{\lambda}_{L R}\,(\rchi_L)^{Aa}(\rchi_L)^{Bb} (\rchi_R)^{C \alpha} (\rchi_R)^{D \beta} \epsilon_{ABCD}\epsilon_{ab}\epsilon_{\alpha \beta},
\end{equation}
where $(A,B,\dots)/(a,b,\dots)/(\alpha,\beta,\dots)$ correspond to $SU(4)/SU(2)_L/SU(2)_R$ indices respectively. As $J(\rchi_{L/R}) =1$ is required by the Yukawa sector, the existence of this term in the scalar potential violates $J$ by four units and therefore violates $B$ by one unit.

Therefore proton and neutron decay diagrams can exist within low-scale PS and will involve a combination of the couplings $y_R$, $y_L$ and $\widetilde{\lambda}_{L R}$. This requires the existence of both $\rchi_L$ and $\rchi_R$. Models with only $\rchi_R$ included (required for PS breaking) have exact proton stability assuming no additional particle content. Additionally, baryon number can be easily imposed when both scalars are present by setting $\widetilde{\lambda}_{L R} \rightarrow 0$ which is not constrained by any other phenomenology and is technically natural and therefore insensitive to quantum corrections. Constraining the allowed size of $\widetilde{\lambda}_{L R}$ from the current experimental constraints on rare proton and neutron decays is beyond the scope of this work. However, this was briefly looked at in~\cite{Foot:1999wv} for a similar model where they found $\widetilde{\lambda}_{L R} \leq 10^{-5}$ as a constraint arising from $N \rightarrow ee\nu$. PS models therefore are relatively unconstrained by baryon number violating decays compared to GUT scenarios and other experimental measurements (or lack thereof) are required in order to constrain the scale of PS breaking.

We note that the scalar potential must satisfy additional phenomenological bounds beyond the suppression of baryon number violation. In particular, due to the multiple $SU(2)_L$ doublets which appear in the theory, the masses of the physical scalars which arise after minimisation of the scalar potential must be sufficiently heavy as they mediate various flavour-changing processes at tree-level, for example neutral-meson mixing. These constraints are identical for \textit{all} left-right symmetric extensions of the SM which contain, at a minimum, a scalar bi-doublet and have been extensively studied~\cite{Gilman:1983bh,Mohapatra:1983ae,Gilman:1983ce,Barenboim:2001vu,Pospelov:1996fq}. Recent studies on the masses of the additional physical scalar fields~\cite{Bertolini:2014sua} with flavour changing couplings puts a lower bound on its mass scale of $\mathcal{O}(20)$ TeV. Although this limit requires masses to be somewhat larger than the electroweak scale, it has been pointed out~\cite{Pospelov:1996fq} that quartic couplings in the scalar potential can generate mass contributions proportional to the $SU(2)_R$ breaking scale for these scalars and therefore large masses can naturally be generated. Regardless, while such constraints are important for low-scale left-right symmetric models, of which PS is an example, they have no impact on the experimental limits on the scale of PS breaking and therefore we do not consider them further.

\section{Pseudoscalar meson decay calculations}
\label{sec:Appendix A}

The leptoquark $X_\mu$ couples to the fermions embedded in $f_{L/R}$ through their kinetic terms in the Lagrangian
\begin{equation}
\label{eqn:kinfermPS}
\mathcal{L}_{\textsc{kin}}^{\textsc{f}} = i \overline{f_L} \slashed{D} f_L + i \overline{f_R} \slashed{D} f_R
\end{equation}
where the covariant derivatives for $f_{L/R}$ are defined similarly to $\rchi_{L/R}$ given in \cref{eqn:covderivPS}. Expanding out \cref{eqn:kinfermPS} explicitly with the fermion multiplets given in~\cref{eqn:fermionPSvanillaemb} and leaving only the interactions of interest gives  
\begin{equation}
\label{eqn:elldXgauge}
\mathcal{L}_{\textsc{kin}}^{\textsc{f}}  \supset \frac{g_4}{\sqrt{2}} \left( \overline{d}  \slashed{X} P_L e + \overline{d} \slashed{X} P_R e \right) + \text{ H.c},
\end{equation}
where colour and generational indices have been suppressed. Here the fields $d$ and $e$ represent the gauge eigenstates. Rotating to the mass eigenstates  leads to
\begin{equation}
\label{eqn:elldXmass}
\mathcal{L}_{Xd e} = \frac{g_4}{\sqrt{2}}\left( \overline{d'_i}\, (K^{d e}_L)_{ij}\, \slashed{X} P_L e'_j + \overline{d'_i} \,(K^{d e}_R)_{ij} \,\slashed{X}P_R e'_j \right) + \text{ H.c.}
\end{equation}
where now the generational indices are explicitly shown and primed fields represent mass eigenstates\footnote{Where there is no chance of confusion between gauge and mass eigenstates, fields will remain unprimed for both basis.}. The matrices $K_{L/R}^{d e}$ are CKM-like mixing matrices between the down-type quarks and charged leptons. The existence of both left- and right-handed couplings to $X_\mu$ leads to two different mixing matrices which, without a parity symmetry, are not necessarily equal. The matrices can be expressed in terms of the unitary matrices used to diagonalise the mass matrices e.g. $(U^i_L)^\dagger M_i^{\vphantom{d}} U^i_R = M_i^d = \text{diag}(\dots)$. For the case of $G_{\text{PS}}$ in total there are eight different physical mixing matrices (compared to the two of the SM) given by
\begin{alignat}{2}
\label{eqn:physicalmixingmatrices}
V^{\textsc{ckm}}_L &= (U^u_L)^\dagger \, U^d_L, \,\,\,\,\,\,V^{\textsc{ckm}}_R = (U^u_R)^\dagger \, U^d_R,\qquad K^{d e}_L &&= (U^d_L)^\dagger \,U^e_L, \,\,\,\,\, K^{d e}_R = (U^d_R)^\dagger \,U^e_R\nonumber\\[3pt]
U^{\textsc{lept}}_L &= (U^\nu_L)^\dagger \,U^e_L, \,\,\,\,\, U^{\textsc{lept}}_R = (U^\nu_R)^\dagger \,U^e_R,\quad\,\,\,\,\, K^{u \nu}_L &&= (U^u_L)^\dagger  U^\nu_L, \,\,\,\,\, K^{u \nu}_R = (U^u_R)^\dagger\,  U^e_R \nonumber\\
\end{alignat}
where $U^\nu_{L/R}$ is a $3 \times (3+n)$ matrix due to the possible seesaw nature of the neutrino sector and the upper-left $3 \times 3$ block of $U^{\textsc{lept}}_L$ is given by the PMNS matrix.

For a pseudo-scalar meson $M_{pq}$, where $q_p\text{ and }q_q$ correspond to the valence quarks of the meson e.g. $B_d^0 = \mphant{M_{bd}}{T}$, the partial width for the two-body decay $M_{pq} \rightarrow \ell_i^+ \ell_j^-$ is given by~\cite{Goodsell:2017pdq,Smirnov:2008zzb}
\begin{equation}
\label{eqn:twobodydecaywidth}
\Gamma_{M_{pq} \rightarrow \ell_i^+ \ell_j^-} = \frac{m_{M_{pq}}}{16 \pi} \,\lambda (m_{M_{pq}},m_{\ell_i},m_{\ell_j})\sum_{h} \vert \mathcal{M}_{pq,ij} \vert^2
\end{equation}
where 
\begin{equation}
\label{eqn:kallenfn}
\lambda (m_{M_{pq}},m_{\ell_i},m_{\ell_j}) = \sqrt{\left[1-\left(\mu_{\ell_i}+\mu_{\ell_j}\right)^2\right] \left[1-\left(\mu_{\ell_i} - \mu_{\ell_j}\right)^2\right]}
\end{equation}
and $\mu_X^{\vphantom{T}} = m_{X}/m_{M_{pq}}$. For the decay of a scalar to two fermions the sum over helicity states is given by
\begin{align}
\label{eqn:twobodyhelsum}
\sum_{h} \vert \mathcal{M}_{pq,ij} \vert^2 = \left(1-\mu_{\ell_i}^2-\mu_{\ell_j}^2\right)&\Big[\,\vert \mathcal{M}_{pq,ij}^L \vert^2 \, +\, \vert \mathcal{M}_{pq,ij}^R \vert^2\Big]\nonumber\\
 &- 2\, \mu_{\ell_i} \mu_{\ell_j} \Big[\mathcal{M}_{pq,ij}^L \left(\mathcal{M}_{pq,ij}^R\right)^* + \mathcal{M}_{pq,ij}^R \left(\mathcal{M}_{pq,ij}^L\right)^* \Big]
\end{align}
where
\begin{equation}
\label{eqn:twobodyamp}
\mathcal{M}_{pq,ij} \equiv \mathcal{M}_{pq,ij}^L\, \overline{u}(p_{\ell_j}) P_L\, v(p_{\ell_i}) + \mathcal{M}_{pq,ij}^R\, \overline{u}(p_{\ell_j}) P_R\, v(p_{\ell_i}).
\end{equation}
The matrix elements of the axial and pseudoscalar currents for the relevant mesons are given by~\cite{Valencia:1994cj,Smirnov:2008zzb}
\begin{align}
\label{eqn:mesonmatrixel}
\langle 0 \vert \overline{d_p} \gamma^\mu \gamma^5 d_q \vert M_{pq} \rangle &= i f_{M_{pq}} (p_{\ell_j}^\mu + p_{\ell_i}^\mu)\nonumber\\
\langle 0 \vert \overline{d_p} \gamma^5 d_q \vert M_{pq} \rangle\,\,\, &= - i f_{M_{pq}}  \overline{m}_{pq}
\end{align}
where $f_{M_{pq}}$ is the meson's decay constant and $\overline{m}_{pq} = m^2_{M_{pq}}/(m_{q_p}+m_{q_q})$. Combining these matrix elements with \cref{figure:pseudoscalarmesondecay} and \cref{eqn:elldXmass} leads to
\begin{equation}
\label{eqn:MLMR}
\mathcal{M}^{L/R}_{pq,ij} = f_{M_{pq}} \left[\,R_{pq}\,\overline{m}_{pq} \big(M^{L/R}_P\big)_{pq,ij} - \left\{m_{\ell_j} \big(M^{L/R}_A\big)_{pq,ij} - m_{\ell_i} \big(M^{R/L}_A\big)_{pq,ij}\right\} \right] 
\end{equation}
where
\begin{align}
\label{eqn:MAMP}
&\left( M^{L/R}_A\right)_{pq,ij} = \mp \frac{g_4^2}{4 m_V^2} \left(K^{d e}_{L/R} \right)_{pi} \left(K^{d e}_{L/R} \right)_{qj}^*\nonumber\\
&\left( M^{L/R}_P\right)_{pq,ij} = \mp \frac{g_4^2}{2 m_V^2} \left(K^{d e}_{L/R} \right)_{pi} \left(K^{d e}_{R/L} \right)_{qj}^*
\end{align}
and $R_{pq}$ is a factor introduced to account for the running of the strong coupling from the high scale, $\mu\sim m_X$, down to the relevant hadron mass scale, $\mu\sim m_{M_{pq}}$. This leads to an enhancement in the pseudoscalar-current matrix element in \cref{eqn:mesonmatrixel}~\cite{Kuznetsov:1994tt,Smirnov:2008zzb} but no enhancement for the axial current which does not run due to the Ward identity of QCD~\cite{Cai:2017wry}.

To demonstrate, for the leptonic decays of $K_L^0$ the correction factor is given by
\begin{equation}
\label{eqn:RK} 
R_{K_L^0}(m_{K_L^0},\,m_X) = R(m_{K_L^0},\,m_c;\,3) R(m_c,\,m_b;\,4) R(m_b,\,m_t;\,5) R(m_t,\,m_X;\,6)
\end{equation}
where
\begin{equation}
R(\mu_1,\,\mu_2;\,n_f) = [g_c(\mu_1)/g_c(\mu_2)]^{8/b(n_f)},
\end{equation} 
$b(n_f) = 11 - (2/3)n_f$ and $n_f$ corresponding to the number of active quark flavours in each energy regime.

As $X_\mu$ couples to both left- and right-handed quarks and leptons, two different types of contributions can be seen in \cref{eqn:MLMR}. The first term does not depend on the masses of the final state leptons and corresponds to the helicity-unsuppressed contribution and only exists if both $K_{L/R}^{de}$ exist, which is only possible if $X_\mu$ couples to both fermion chiralities. The last two terms are proportional to the final state charged lepton masses. This corresponds to the helicity-suppressed contribution which arises from a mass insertion and only requires one of $K_{L/R}^{de}$ to exist. It is forbidden in the limit of massless final-state leptons, in complete analogy to weak meson decays in the SM. If the final state particle masses are ignored, the decay width is given purely by the helicity-unsuppressed contribution, and can be simplified to~\cite{Smirnov:2007hv,Smirnov:2008zzb,Smirnov:2018ske}
\begin{equation}
\label{eqn:HUcont}
\Gamma^{\text{HU}}_{M_{pq} \rightarrow \ell_i^+ \ell_j^-} = \frac{\mphant{m_{M_{pq}}}{T} \left[g_4^{\vphantom{T}}(\mphant{m_X}{T})\right]^4 f_{M_{pq}}^2 \overline{m}_{pq}^2}{64 \,\pi\, m_X^4} R_{pq}^2 \left(\left|K^{d e}_L\right|^2_{pi} \left| K^{d e}_{R} \right|^2_{qj} + \left|K^{d e}_R \right|^2_{pi} \left| K^{d e}_L \right|^2_{qj} \right).
\end{equation}
In the limit where $X_\mu$ couples to only one chirality of fermions\footnote{This is a special case of the general case which occurs for example by setting ${K_{R}^{de} = 0_{3 \times 3}}$ but $K_L^{d e}$ remains unitary.} similar to the weak force, $M_P^{L/R} = 0$ but one of $M_A^{L/R}$ remains nonzero. This results in the total decay width depending only on the helicity-suppressed terms appearing in~\cref{eqn:MLMR}. The usual result for helicity-suppressed meson decays is recovered
\begin{align}
\Gamma^{\text{HS}}_{M_{pq} \rightarrow \ell_i^+ \ell_j^-}  = \frac{\mphant{m_{M_{pq}}}{T} \left[g_4^{\vphantom{T}}(\mphant{m_X}{T})\right]^4 f_{M_{pq}}^2}{256\, \pi\, m_X^4} \sqrt{\left[1-\left(\mu_{\ell_i}+\mu_{\ell_j}\right)^2\right] \left[1-\left(\mu_{\ell_i} - \mu_{\ell_j}\right)^2\right]}& \nonumber\\[5pt]
\times\left[ \left(1 - \mu_{\ell_i}^2 - \mu_{\ell_j}^2 \right)\left(m_{\ell_i}^2 + m_{\ell_j}^2 \right) - 4 \mu_{\ell_i} \mu_{\ell_j} m_{\ell_i} m_{\ell_j} \right]&
\end{align}
which, in the limit $m_{\ell_j} > m_{\ell_i}$, is well approximated by
\begin{align}
\label{eqn:HScont}
\Gamma^{\text{HS}}_{M_{pq} \rightarrow \ell_i^+ \ell_j^-}  \simeq \frac{\mphant{m_{M_{pq}}}{T} \left[g_4^{\vphantom{T}}(\mphant{m_X}{T})\right]^4 f_{M_{pq}}^2 }{256\, \pi\, m_X^4}\, m_{\ell_j}^2 \left(1 - \frac{m_{\ell_j}^2}{m_{M_{pq}}^2}\right)^2\Biggl(\left|K^{d e}\right|^2_{pi} \left| K^{d e} \right|^2_{qj}\Biggr),
\end{align}
where $K^{d e}$ corresponds to either $K_L^{de}$ or $K_R^{de}$ depending on which chirality of fermions $X_\mu$ couples to. Comparing \cref{eqn:HUcont,eqn:HScont} unsuprisingly suggests that generically the helicity-unsuppressed contribution is expected to dominate:
\begin{align}
\label{eqn:ratioHUHS}
R_{\text{HU/HS}} = \frac{\Gamma^{\text{HU}}}{\Gamma^{\text{HS}}} \simeq \frac{4\, \overline{m}_{pq}^2 R_{pq}^2}{m_{\ell_j}^2} &\underbrace{\left(\frac{\left|K^{d e}_L\right|^2_{pi} \left| K^{d e}_{R} \right|^2_{qj} + \left|K^{d e}_R \right|^2_{pi} \left| K^{d e}_L \right|^2_{qj}}{\left|K^{d e}_L\right|^2_{pi} \left| K^{d e}_{L} \right|^2_{qj} + \left|K^{d e}_R \right|^2_{pi} \left| K^{d e}_R \right|^2_{qj}}\right)}.\nonumber\\
&\qquad\qquad\qquad\quad\,\,\,\,\,\kappa
\end{align}
For example, for $K_L^0 \rightarrow \mu e$ we roughly find $R_{\text{HU/HS}} \simeq 10^4 \,\kappa$ where $\kappa$ corresponds to the combination of mixing matrices involving $K^{d e}_{L/R}$ above. However the helicity unsuppressed contribution can be sub-dominant if the couplings are strongly suppressed for one chirality over the other e.g. $|K^{d e}_L|_{pi/qj} \neq 0$ and $|K^{d e}_R|_{pi/qj}\simeq 0$. In such scenarios the limits for the vector leptoquark mass (and the therefore the PS breaking scale) will be significantly decreased due to a reduction in the decay rate with $R_{\text{HU/HS}} < 1$. However this would require the hierarchy
\begin{equation}
\frac{4\, \overline{m}_{pq}^2 R_{pq}^2}{m_{\ell_j}^2} < \frac{1}{\kappa}
\end{equation}
implying a difference in the couplings of $X_\mu$ to $f_L$ and $f_R$ of several order of magnitude at least.

In order to numerically find the lower-bound mass range for $X_\mu$ we fix the values of the total decay width of each relevant meson to the experimentally observed central values
\begin{equation}
\label{eqn:widths}
\Gamma^{\text{\tiny TOT}}_{K_L^0} = 1.29 \times 10^{-17}\text{ GeV},\,\,\,\, \Gamma^{\text{\tiny TOT}}_{B_d^0} = 4.33 \times 10^{-13}\text{ GeV}\,\,\mand \,\,\Gamma^{\text{\tiny TOT}}_{B_s^0} = 4.36 \times 10^{-13}\text{ GeV},
\end{equation}
which we take from the PDG~\cite{Zyla:2020zbs}. The parameter $R_{pq}$ which appears in the helicity-unsuppressed contribution to a given decay and is defined in \cref{eqn:RK} requires running from the scale $\mu = m_X$ down to $\mu = m_{M_{pq}}$ and the gauge coupling constant $g_4$ is related to the strong coupling constant at the scale of PS breaking:
\begin{align}
\label{eqn:runnsparam}
g_4(m_X) = g_c(m_X) &= 2\sqrt{\pi} \left(\frac{17}{2} + \frac{7}{2\pi}\log \left[\frac{m_X}{90}\right]\right)^{-1/2}\nonumber\\
R_{K_L^0} &\simeq 0.51 \left(\frac{17}{2} + \frac{7}{2\pi}\log \left[ \frac{m_X}{90}\right] \right)^{4/7}\nonumber\\
R_{B_s^0} \simeq R_{B_d^0} & \simeq 0.37 \left(\frac{17}{2} + \frac{7}{2\pi}\log \left[ \frac{m_X}{90}\right] \right)^{4/7}
\end{align}
where $m_X$ is in units of GeV and for simplicity we assume the one-loop SM running of the gauge coupling constant $g_c$ which can be found in \cref{sec:Appendix C}. We calculate lower bound limits on the leptoquark mass by numerically solving \cref{eqn:twobodydecaywidth,eqn:runnsparam} as a function of the leptoquark mass and comparing to the experimental limits listed in \cref{table1}.

\section{Various Seesaw Properties}
\label{sec:Appendix B}

\subsection{Singular Values in One Generation}
\label{subsec:oneDsingval}
Consider the matrix $M \in M_{2}(\mathbb{C})$ given by
\begin{equation}
\label{eqn:2x2mat}
M = \begin{pmatrix}
a & b\\
c & d\\
\end{pmatrix}.
\end{equation}
This matrix can be diagonalised via its singular value decomposition: 
\begin{equation}
\label{eqn:2x2svd}
U_L^\dagger\, M \,\mphant{U_R}{\dagger} = M_{\mathscr{d}} =\begin{pmatrix}
m_{\mathscr{l}} & 0\\
0 & m_{\mathscr{h}}
\end{pmatrix} .
\end{equation}
There exist exact analytic expressions for $U_L,\,U_R,\,m_\mathscr{l}$ and $m_\mathscr{h}$ for this $2 \times 2$ matrix, however in the seesaw limit where the absolute value of one entry of \cref{eqn:2x2mat} is significantly larger than the others, these expressions can be significantly simplified by writing them as perturbation series. For example, in the limit where $\vert d\vert \gg \vert a\vert,\vert b\vert,\vert c\vert$ we find for the unitary-diagonalisation matrices  
\begin{align}
\label{eqn:2x2unitdiag}
U_L &= \underbrace{\begin{pmatrix}
1-\frac{1}{2}\left|\frac{ac^*+bd^*}{|d|^2}\right|^2 &\frac{ac^*+bd^*}{|d|^2}\\
-\left(\frac{ac^*+bd^*}{|d|^2}\right)^* & 1-\frac{1}{2}\left|\frac{ac^*+bd^*}{|d|^2}\right|^2
\end{pmatrix}}_{Q_L}\underbrace{\begin{pmatrix}
\sqrt{\frac{m_{\mathscr{l}}}{\vert m_{\mathscr{l}}\vert}} & 0\\
0 & \sqrt{\frac{m_{\mathscr{h}}}{\vert m_{\mathscr{h}} \vert}}
\end{pmatrix}}_{K_L} \,\,+\,\, \mathcal{O}(\epsilon^{3})\nonumber\\
U_R &= \underbrace{\begin{pmatrix}
1-\frac{1}{2}\left|\frac{ab^*+cd^*}{|d|^2}\right|^2 & \left(\frac{ab^*+cd^*}{|d|^2}\right)^*\\
-\frac{ab^*+cd^*}{|d|^2} & 1-\frac{1}{2} \left|\frac{ab^*+cd^*}{|d|^2}\right|^2
\end{pmatrix}}_{Q_R}\underbrace{\begin{pmatrix}
\sqrt{\frac{m^*_{\mathscr{l}}}{\vert m_{\mathscr{l}}\vert}} & 0\\
0 & \sqrt{\frac{m^*_{\mathscr{h}}}{\vert m_{\mathscr{h}} \vert}}
\end{pmatrix}}_{K_R} \,\,+\,\, \mathcal{O}(\epsilon^{3})
\end{align}
where $Q_{L/R}$ removes the off-diagonal entries of $M$, $K_{L/R}$ is required to ensure the remaining diagonal entries are real and positive and $\epsilon \sim \frac{1}{\vert d\vert}$. The singular values of $M$ written up to sub-subleading order are
\begin{align}
\label{eqn:2x2singval}
m_{\mathscr{l}} &= \left| a - \frac{bc}{d} - \frac{a}{2}\left(\frac{|b|^2+|c|^2}{|d|^2}\right) \right| \,\,+ \mathcal{O}(\epsilon^{3})\nonumber\\
m_{\mathscr{h}} &= \left| d + \frac{1}{2}\left(\frac{|b|^2+|c|^2}{d^*}\right) + \frac{a^* bc}{|d|^2} \right|\,\, + \mathcal{O}(\epsilon^3) .
\end{align}
Note that in the type-I seesaw scenario ($a=0,\,b=c=m_D,\,d=m_R$), \cref{eqn:2x2unitdiag} implies $U_L^\dagger = U_R^T$ and therefore the (now symmetric) matrix M can be diagonalised either by $U_R^T M U_R^{\vphantom{T}}$ or $U_L^\dagger M U_L^{*\vphantom{\dagger}}$ with the singular values of \cref{eqn:2x2singval} simplifying to 
\begin{align}
m_{\mathscr{l}} &= \left| \frac{m_D^2}{m_R} \right| + \mathcal{O}(\epsilon^2)\nonumber\\
m_{\mathscr{h}} &= \left| m_R + \frac{\,\,\vert m_D\vert^2}{m_R}\right| + \mathcal{O}(\epsilon^2)
\end{align}
as expected.

The results above can be applied to find similar expressions for $U_L,\,U_R,\,m_\mathscr{l}$ and $m_\mathscr{h}$ in scenarios where different entries of $M$ are the largest in the seesaw through appropriate permutations. For example consider when $\vert c\vert \gg \vert a\vert,\vert b\vert,\vert d\vert$, as before there exist $U_L$ and $U_R$ which diagonalise $M$, however we note that \cref{eqn:2x2svd} can be rewritten,
\begin{equation}
\label{eqn:2x2perm}
U_L^\dagger \begin{pmatrix}
a & b\\
c & d
\end{pmatrix}\mphant{U_R}{\dagger} = U_L^\dagger \begin{pmatrix}
b & a\\
d & c
\end{pmatrix}\tau_1\,\mphant{U_R}{\dagger} = (U'_L)^\dagger \begin{pmatrix}
b & a\\
d & c
\end{pmatrix} U'_R = \text{diag}(m_{\mathscr{l}},\,m_{\mathscr{h}})
\end{equation}
where $\tau_1$ corresponds to the first Pauli matrix.

The results in \cref{eqn:2x2unitdiag,eqn:2x2singval} can now be applied to \cref{eqn:2x2perm} as $c$ is assumed to be the largest entry of $M$. Therefore $m_\mathscr{l}$ and $m_\mathscr{h}$ are given by the expressions in \cref{eqn:2x2singval} with the permutations $a \leftrightarrow b$ and $c \leftrightarrow d$, and the unitary diagonalisation matrices are given by $U'_L=U_L^{\vphantom{T}}$ and $U'_R = \tau_1\, U_R^{\vphantom{T}}$ taken from \cref{eqn:2x2unitdiag} with similar permutations. Applying a similar procedure for $\vert b\vert \gg \vert a\vert,\vert c\vert,\vert d\vert$ we find that $U'_L = \tau_1\, U_L^{\vphantom{T}}$ and $U'_R = U_R^{\vphantom{T}}$ and the results from \cref{eqn:2x2unitdiag,eqn:2x2singval} are valid with permutations $a \leftrightarrow c$ and $b \leftrightarrow d$. Finally when $\vert a\vert \gg \vert b\vert,\vert c\vert,\vert d\vert$ we find $U'_L = \tau_1\, U_L^{\vphantom{T}}$ and $U'_R = \tau_1\, U_R^{\vphantom{T}}$ with the permutations $a \leftrightarrow d$ and $b \leftrightarrow c$ applied to \cref{eqn:2x2unitdiag,eqn:2x2singval}.

\subsection{Singular Values for Multiple Generations}
\label{subsec:multiDsingval}

Consider now a general matrix $M_\textsc{B} \in \mathbb{C}^{m\times n}$ which can be partitioned into a $2 \times 2$ block matrix of the form
\begin{equation}
\label{eqn:blockmatgen}
M_\textsc{B} = \begin{pmatrix}
A_{m_1 \times n_1} & B_{m_1 \times l_1}\\
C_{l_1 \times n_1} & D_{l_1 \times l_1}
\end{pmatrix}
\end{equation}
where the subscripts on the matrix blocks correspond to their dimensions, $m_1 + l_1 = m$, $n_1 + l_1 = n$ and $D$ is restricted to be nonsingular. This matrix can be \textit{block} diagonalised by  
\begin{equation}
\label{eqn:blocksvd}
Q_L^\dagger M_\textsc{B} \,\mphant{Q_R}{\dagger} = M_\mathscr{D} = \text{diag}(m_L,\,m_H) = \begin{pmatrix}
m_L & 0_{m_1\times l_1}\\
0_{l_1 \times n_1} & m_H
\end{pmatrix}
\end{equation}
where $m_L$ and $m_H$ now correspond to $m_1 \times n_1$ and $l_1 \times l_1$ block matrices respectively and $0_{n_1\times l_1}$ corresponds to a matrix of zeroes with dimension $n_1 \times l_1$. Unlike in the $2 \times 2$ case, exact analytical expressions for the block matrices $m_L$ and $m_H$ or the diagonalisation matrices $Q_L$ and $Q_R$ do not exist. However in the seesaw limit, where the lightest singular value of the square block $D$ is assumed to be significantly larger than the largest singular value of all other blocks, a similar perturbative expansion can be derived. This is done by assuming, as an ansatz, that the block matrices $U_L$ and $U_R$ have a similar form to the matrices appearing in \cref{eqn:2x2unitdiag}\footnote{In other words that $U_L$ and $U_R$ can be written perturbatively in terms of an expansion parameter $\epsilon$ related to the seesaw in $M$.} in combination with the conditions $U_L^\dagger (M_\textsc{B}^{\vphantom{\dagger}} M_\textsc{B}^\dagger)\, U_L^{\vphantom{\dagger}}=\text{diag}(m_L^{\vphantom{\dagger}}m_L^\dagger,\,m_H^{\vphantom{\dagger}}m_H^\dagger)$ and $U_R^\dagger (M_\textsc{B}^\dagger M_\textsc{B}^{\vphantom{\dagger}})\, U_R^{\vphantom{\dagger}}=\text{diag}(m_L^\dagger m_L^{\vphantom{\dagger}},\,m_H^\dagger m_H^{\vphantom{\dagger}})$ which allows $U_{L/R}$ to be solved separately from each other.

\begin{sloppypar}
In this limit, $||D|| \gg ||A,B,C||$, we find $M_\textsc{B}$ can be \textit{fully} diagonalised at sub-subleading order by
\begin{align}
\label{eqn:blocksvddiag}
U_L &\simeq 
\underbrace{\begin{pmatrix}
\id_{n\times n} - \frac{1}{2} X X^\dagger & X\\
-X^\dagger & \,\id_{n\times n}- \frac{1}{2} X^\dagger X
\end{pmatrix}}_{Q_L}\underbrace{\begin{pmatrix}
V_L & 0\\
0 & W_L
\end{pmatrix}}_{K_L}  \nonumber\\
U_R &\simeq 
\underbrace{\begin{pmatrix}
\id_{n\times n}\, - \,\frac{1}{2} Z^\dagger Z & Z^\dagger\\
-Z & \id_{n\times n}\,- \,\frac{1}{2} Z Z^\dagger
\end{pmatrix}}_{Q_R}\underbrace{\begin{pmatrix}
V_R & 0\\
0 & W_R
\end{pmatrix}}_{K_R} 
\end{align}
where 
\begin{align}
\label{eqn:blocksvddiag2}
X &= (A C^\dagger + B D^\dagger)(D D^\dagger)^{-1}\simeq BD^{-1}\nonumber\\
Z &= (D^\dagger D)^{-1}(B^\dagger A + D^\dagger C)\simeq D^{-1}C.
\end{align}
Similar to the $2 \times 2$ case, $Q_{L/R}$ remove the off-diagonal entries of $M_\textsc{B}$, $K_{L/R}$ diagonalise the blocks $m_{L/H}$ via $(V/W)_L^\dagger\,m_{L/H}\,(V/W)_R^{\vphantom{\dagger}} = m_{L/H}^d = \text{diag}(\dots)$ and the expression for $V_{L/R}$ and $W_{L/R}$ depends on the structure of the submatrices $m_L$ and $m_H$.
\end{sloppypar}
Applying \cref{eqn:blocksvddiag,eqn:blocksvddiag2} to \cref{eqn:blocksvd} gives the light and heavy submatrices at sub-subleading order as per
\begin{alignat}{2}
\label{eqn:blocksvdsingblocks}
m_{L} &\simeq \left( A - B D^{-1} C - \frac{1}{2}\left(  B D^{-1} (D^\dagger)^{-1} B^\dagger A + A\, C^\dagger (D^\dagger)^{-1} D^{-1} C\right)\right) \nonumber\\
m_{H} &\simeq D + \frac{1}{2}\left( C C^\dagger (D^\dagger)^{-1} + (D^\dagger)^{-1}B^\dagger B\right) \nonumber\\
&\qquad\,\,+ \frac{1}{2}\left((D^\dagger)^{-1}D^{-1}C A^\dagger B + C A^\dagger B D^{-1} (D^\dagger)^{-1}  \right).
\end{alignat}
Note that \cref{eqn:blocksvdsingblocks} agrees with \cref{eqn:2x2singval} in the special case where $M_\textsc{B} \in \mathbb{C}^{2 \times 2}$, i.e. all the blocks of $M_\textsc{B}$ are just complex numbers, $A\rightarrow a \dots D \rightarrow d$. 

In the special case of \cref{eqn:blockmatgen} where $B = C^T = m_D$ (implying $n_1 = m_1$), $A = \emptyset_{n_1\times n_1}$ and $D = M_R$ is a symmetric matrix, which occurs in the n-dimensional type-I seesaw model of neutrino mass, $U_L^\dagger = U_R^T$ and $m_L$ and $m_H$ simplify to at sub-subleading order
\begin{alignat}{2}
m_L  &\simeq - \,&& m_D^{\vphantom{T}} M_R^{-1} m_D^{T} \nonumber\\
m_H  &\simeq && M_R + \frac{1}{2}\left( m_D^T m_D^* (M_R^*)^{-1} + (M_R^*)^{-1} m_D^{\dagger} m_D^{\vphantom{\dagger}}\right)
\end{alignat}
where $M_R^{\vphantom{T}} = M_R^T$ has been used. Therefore the usual results are recovered in this case and we find them to be in full agreement\footnote{We also find agreement in the case where $A = A^T \neq 0_{n \times n}$ which for example is relevant in hybrid type-I+II models.} with~\cite{Grimus:2000vj} which derived an algorithm to find the light and heavy mass matrices in neutrino seesaw models at arbitrary order.

\begin{sloppypar}

\begin{table}[t]
\begin{center}
{\renewcommand{\arraystretch}{1.15}
\scalebox{1.0}{
\begin{tabular}{ccc}
\toprule
Dominant Block & \,Permutations\, & $(U'_L,\,U'_R)$	 \\ 
\midrule 
\midrule
$A$ & $A\leftrightarrow D\,\,\&\,\,B\leftrightarrow C$  & $(\slashed{\id}(l_1,m_1)\,U_L,\,\slashed{\id}(n_1,l_1)\,U_R)$ \\
$B$ & $A \leftrightarrow C\,\,\&\,\,B\leftrightarrow D$  & $(\slashed{\id}(l_1,m_1)\,U_L,\,U_R)$ \\
$C$ & $A \leftrightarrow B\,\,\&\,\,C\leftrightarrow D$ & $(U_L,\,\slashed{\id}(n_1,l_1)\, U_R)$\\
$D$ &    $-$    & $(U_L,\,U_R)$ \\
\bottomrule
\end{tabular}
}
}
\end{center}
\caption[Permutations and rotations required for different block entires dominant.]{Required rotations on $U_L$ and $U_R$ and permutations of block elements for the formulas in \cref{eqn:blocksvddiag,eqn:blocksvddiag2,eqn:blocksvdsingblocks} to be valid for different block elements assumed to be the nonsingular-square dominant block of $M_\textsc{B}$ in seesaw scenarios.  $\slashed{\id}(l_1,m_1)$ is defined in \cref{eqn:blockmatrot} and here $l_1$ is taken to be the dimension of the dominant square matrix block and $(m_1,n_1)$ corresond to the dimensions of the matrix block diagonally opposite.} 
\label{table6}
\end{table}

In a similar way to the case with $M \in \mathbb{C}^{2 \times 2}$, the results above can be extended to find $m_L$, $m_H$, $U_L$ and $U_R$ in situations where the norm of different blocks in \cref{eqn:blockmatgen} are dominant\footnote{This is with the caveat that $M_\textsc{B}$ can be partitioned such that the dominant block has a well defined inverse.}. Once again consider a scenario where $||C|| \gg ||A,B,D||$ (and C is square) and note that 
\begin{equation}
\label{eqn:blockmatrot}
M_\textsc{B} = \begin{pmatrix}
A_{m_1 \times l_1} & B_{m_1 \times n_1}\\
C_{l_1 \times l_1} & D_{l_1 \times n_1}
\end{pmatrix} = \begin{pmatrix}
B_{m_1 \times n_1} & A_{m_1 \times l_1}\\
D_{l_1 \times n_1} & C_{l_1 \times l_1}
\end{pmatrix}\underbrace{\begin{pmatrix}
0_{n_1 \times l_1} & \id_{n_1 \times n_1}\\
\id_{l_1 \times l_1} & 0_{l_1 \times n_1}
\end{pmatrix}}_{\slashed{\id}(n_1,l_1)}.
\end{equation}
Therefore \cref{eqn:blocksvddiag,eqn:blocksvddiag2,eqn:blocksvdsingblocks} are valid if ${U'_L=U_L}$, ${U'_R=\slashed{\id}(n_1,l_1) \,U_R}$ and the permutations $A \leftrightarrow B$ and $C \leftrightarrow D$ are performed, where ${(U'_L)^\dagger\,M_\textsc{B}\,U'_R=\text{diag}(m_L,\,m_H)}$. This can be performed for any of the four blocks of $M_\textsc{B}$ being dominant and the results for this are summarised in \cref{table6}.
\end{sloppypar}

\subsection{Gap properties between the charged leptons and down quarks}
\label{subsec:gapprop}

Below we briefly prove the claim that, in the triplet scenario, the charged-lepton mass matrices defined in \cref{eqn:tripchargelepmixR} imply that the SM-like charged lepton masses must be lighter than the down-quark masses for all generations. This arises due to the zero block appearing in their mass matrices diagonally opposite the dominant block in the seesaw. Unlike the previous section which relied on an approximate block diagonalisation technique with neglected higher-order terms, the results below are exact statements that do not rely on any perturbative arguments. We closely follow the work presented in~\cite{Besnard:2016tcs} which proved a similar gap property specifically for the type-I seesaw symmetric mass matrix, which we will generalise to an arbitrary complex matrix relevant to our charged-lepton seesaw. 

We state without proof three matrix properties related to the Courant-Fischer-Weyl min-max theorem:
\begin{itemize}
\item For two hermitian $n \times n$ matrices $A$ and $B$, if $C = A+B$ then 
\begin{equation}
a_k + b_1 \leq c_k \leq a_k + b_n
\end{equation}
where $x_i$ corresponds to the i-th largest eigenvalue of the matrix X and $k \leq n$. For $k=1$ this reduces to the special case
\begin{equation}
\label{eqn:prop1}
\text{min}(A) + \text{min}(B) \leq \text{min}(A+B)
\end{equation}
where $\text{min}(X) = x_1$.

\item For all $A \geq 0$ and $B \in M_n(\mathbb{C})$, if $C = B^{\dagger} A B$ then
\begin{equation}
\label{eqn:prop2}
a_k \,\text{min}(B^\dagger B) \leq c_k.
\end{equation}

\item If $Q$ is an $n\times n$ submatrix of the $N \times N$ matrix $M$ then
\begin{equation}
\label{eqn:prop3}
m_k \leq q_k \leq m_{N-n+k}
\end{equation}
for every $k \leq n$, this is known as the Cauchy interlacing theorem.
\end{itemize}
Additional details including proofs for some of these properties can be found in~\cite{Besnard:2016tcs}.

We take the mass mixing matrix $M_{eE}$ to have the same structure as the case of an $SU(2)_R$ triplet defined in \cref{eqn:tripchargelepmixR}:
\begin{equation}
\label{eqn:gengapmat}
M_{eE} = \begin{pmatrix}
m_{ee} & 0_{3\times 3}\\
m_{Ee} & m_{EE}
\end{pmatrix},
\end{equation}
where we are assuming three generations of the exotic triplet fermion and therefore each block is $3 \times 3$ and we further assume $m_{Ee}$ is nonsingular. As discussed in \cref{subsubsec:fermionspec-eE}, in order to achieve the desired helicity-suppression in the PS limits we require the hierarchy $\Vert m_{Ee} \Vert > \Vert m_{ee},\,m_{EE}\Vert$ to be satisfied which implies
\begin{equation}
\sigma_n(m_{ee}),\,\sigma_n(m_{EE}) < \sigma_1(m_{Ee})
\end{equation}
where $\sigma_i(X)$ corresponds to the i-th largest singular value of the matrix $X$ and $n=3$. This defines a seesaw in $M_{eE}$ with $m_{Ee}$ as the dominant block.

The bottom-right sub-matrix of
\begin{equation}
\left( M_{eE} M_{eE}^\dagger \right)^{-1} = \begin{pmatrix}
\cdot & \cdot \\
\cdot & m_{EE}^{-1} m_{Ee} m_{ee}^{-1} (m_{ee}^\dagger)^{-1} m_{Ee}^\dagger (m_{EE}^\dagger)^{-1} + m_{EE}^{-1}(m_{EE}^\dagger)^{-1}
\end{pmatrix}
\end{equation}
which we will label as $X$ satisfies the property defined in \cref{eqn:prop3}, where the dots correspond to sub-matrices which are not relevant. As the sub-matrix X has dimensions $n \times n$ and the full matrix has dimensions $2n \times 2n$ the interlacing theorem implies
\begin{equation}
\label{eqn:eigenvalueequality}
x_{n-k} \leq m_{2n -k }
\end{equation}
where we have chosen $k = n-k$ in \cref{eqn:prop3}. As $M_{eE} (M_{eE}^\dagger)^{-1}$ is Hermitian, its eigenvalues correspond to the squared singular values of $M_{eE}$. Furthermore, from simple properties of eigenvalues, the i-th \textit{largest} eigenvalue of an $n \times n$ matrix $M^{-1}$ corresponds to the (n-i)-th \textit{smallest} eigenvalue of $M$. Therefore \cref{eqn:eigenvalueequality} implies
\begin{equation}
\label{eqn:eigensingequal}
x_{n-k} \leq \frac{1}{\sigma_k(M_{eE})^2}.
\end{equation}
Applying \cref{eqn:prop1} to $x_{n-k}$ leads to the inequality
\begin{equation}
\label{eqn:iterrimproof}
\left[m_{EE}^{-1} m_{Ee} m_{ee}^{-1} (m_{ee}^\dagger)^{-1} m_{Ee}^\dagger (m_{EE}^\dagger)^{-1}\right]_{n-k} + \text{min}(m_{EE}^{-1}(m_{EE}^\dagger)^{-1}) \leq x_{n-k}
\end{equation}
and now with repeated applications of \cref{eqn:prop2} to the first term of \cref{eqn:iterrimproof} leads to
\begin{align}
\left(\left[ m_{ee}^{-1} (m_{ee}^\dagger)^{-1}\right]_{n-k} \,\text{min}\left(m_{Ee} m_{Ee}^\dagger\right) + \id\right)\text{min}(m_{EE}^{-1}(m_{EE}^\dagger)^{-1}) \leq x_{n-k}
\end{align}
which implies
\begin{equation}
\frac{1}{\sigma_n(m_{EE})^2}\left(\frac{\sigma_1(m_{Ee})^2}{\sigma_k(m_{ee})^2}\right) \leq \frac{1}{(\sigma_k(M_{eE}))^2}
\end{equation}\
and therefore
\begin{equation}
\label{eqn:oneineq}
\sigma_k(M_{eE}) \leq \sigma_k(m_{ee})\frac{\sigma_n(m_{EE})}{\sqrt{\sigma_1(m_{Ee})^2 + \sigma_k(m_{ee})^2}} < \sigma_k(m_{ee})\frac{\sigma_n(m_{EE})}{\sigma_1(m_{Ee})}
\end{equation}
where we have used the seesaw assumption that $\sigma_n(m_{ee},\,m_{EE}) < \sigma_1(m_{Ee})$.

The exact same procedure can be applied to the matrix $(M_{eE}^\dagger)^{-1} M_{eE} $ which instead leads to the inequality
\begin{equation}
\label{eqn:twoineq}
\sigma_k(M_{eE}) \leq \sigma_k(m_{EE})\frac{\sigma_n(m_{ee})}{\sqrt{\sigma_1(m_{Ee})^2 + \sigma_k(m_{EE})^2}} < \sigma_k(m_{EE})\frac{\sigma_n(m_{ee})}{\sigma_1(m_{Ee})}.
\end{equation}
Combining \cref{eqn:oneineq,eqn:twoineq} leads to an upper-bound on the k-th largest \textit{light} singular value of $M_{eE}$
\begin{equation}
\label{eqn:lightmassineq}
\sigma_k(M_{eE}) < \frac{\text{min}\left[\sigma_k(m_{ee})\sigma_n(m_{EE}),\sigma_k(m_{EE})\sigma_n(m_{ee})\right]}{\sigma_1(m_{Ee})}\leq \frac{\sigma_n(m_{ee})\sigma_n(m_{EE})}{\sigma_1(m_{Ee})}.
\end{equation}
Comparing \cref{eqn:lightmassineq} to the charged-lepton mass matrix in the case of the $SU(2)_R$ triplet in \cref{eqn:tripchargelepmixR} leads to the strict inequality
\begin{equation}
\sigma_k(m_\mathscr{l}) < \sigma_k(m_d) \left(\frac{\sigma_3(\mu_\Psi)}{v_R\,\sigma_1(Y_\Psi )}\right)
\end{equation}
and as the singular values of each matrix correspond to the physical masses, each light lepton mass of a given generation must be strictly lighter than the corresponding down-quark mass of the same generation. This is due to the seesaw assumption which enforces the fraction in brackets to be strictly smaller than one. As shown in \cref{table3} the charged-lepton and down-quark mass hierarchies change between generations at low scales, therefore due the PS symmetry, a matrix of the form in \cref{eqn:gengapmat} is phenomenologically ruled out.

Although not relevant in our analysis, a similar procedure can be applied to the Hermitian matrices $(M_{eE}^\dagger) M_{eE} $ and $M_{eE} (M_{eE}^\dagger)  $ to derive inequalities on the masses of the $n$ heavy states of $M_{eE}$ and for example leads to
\begin{equation}
\sigma_{n+1}(M_{eE}) \geq \text{max}\left(\sqrt{\sigma_1(m_{Ee})^2 + \sigma_1(m_{ee})^2},\sqrt{\sigma_1(m_{Ee})^2 + \sigma_1(m_{EE})^2}\right) > \sigma_1(m_{Ee})
\end{equation}

Therefore the strict hierarchy implied by these inequalities leads to the `gap property'
\begin{equation}
\sigma_1(M_{eE}) \leq \mydots \leq \sigma_n(M_{eE}) \leq \frac{\sigma_n(m_{ee})\sigma_n(m_{EE})}{\sigma_1(m_{Ee})} \ll  \sigma_{1}(m_{Ee}) < \sigma_{n+1}(M_{eE}) \leq \mydots \leq \sigma_{2n}(M_{eE})
\end{equation}
which does not rely on any perturbative arguments and we emphasise that these results are only valid for `type-I like' scenarios where the entry diagonally opposite the dominant block is zero. Our results agree with the gap properties derived in~\cite{Besnard:2016tcs} in the type-I limit which assumes $m_{ee} = (m_{EE})^T = m_D$ and $m_{Ee} = m_R$ and recovers the usual type-I hierarchy.

\subsection{Lepton masses with additional bi-doublets}
\label{subsec:bidoubneutmass}

The neutrino mass matrix which arises with the addition of an $SU(2)_{L/R}$ bi-doublet to the usual PS fermions $f_{L/R}$ is given by
\begin{equation}
\label{eqn:appbineutmass}
\mathcal{L}_{\nu N}=\frac{1}{2}\begin{pmatrix}
\overline{\nu_L} & \overline{\nu_R^c} & \overline{N_L} & \overline{N_R^c} 
\end{pmatrix} \begin{pmatrix}
0 & m_u & 0 & \mathlarger{y}_{\Psi_{22}}^R v_R\\
m_u & 0 & 0 & \mathlarger{y}_{\Psi_{22}}^L v_L^*\\
0 & 0 & 0 & \mu_{\Psi_{22}}  \\
\mathlarger{y}_{\Psi_{22}}^R v_R & \mathlarger{y}_{\Psi_{22}}^L v_L^* & \mu_{\Psi_{22}}  & 0
\end{pmatrix}
\begin{pmatrix} 
\nu_L^c\\
\nu_R\\
N_L^c\\
N_R
\end{pmatrix}
\end{equation}
where we have labelled the fields as in \cref{eqn:biemb}. A phenomenologically-viable mass spectrum for the heavy charged lepton states requires either $y_\Psi^R v_R$ or $\mu_\Psi$ to have masses at a TeV or above as discussed in \cref{subsubsec:fermionspec-eE} for the charged-lepton mass mixing matrix
\begin{equation}
\label{eqn:appbidoubchargelepmix}
\mathcal{L}_{eE} = \begin{pmatrix}
\overline{e_L} & \overline{E_L}
\end{pmatrix} \begin{pmatrix}
m_d & \mathlarger{y}_{\Psi_{22}}^R v_R\\
\mathlarger{y}_{\Psi_{22}}^L v_L^*  & \mu_{\Psi_{22}} 
\end{pmatrix}
\begin{pmatrix}
e_R\\
E_R
\end{pmatrix}.
\end{equation}
The parameter $m_u$ is fixed by the up-quark masses due to the PS symmetry, $m_d$ is fixed by the down-quark masses, $y_{\Psi_{22}}^L v_L$ is tied to the electroweak scale, $y_{\Psi_{22}}^R v_R$ can at most be the size of $SU(2)_R$/PS breaking and $\mu_{\Psi_{22}}$ is unconstrained. 

Different hierarchies between the parameters appearing in \cref{eqn:appbineutmass} lead to different neutrino mass phenomenology and here we show that only one possibility allows for a viable mass spectrum for the charged leptons as well as sufficiently light neutrino masses assuming a low scale PS breaking. We will use a one-generational example for illustrative purposes as the same statements hold true for the more complicated multi-generational scenario. As discussed in \cref{subsubsec:fermionspec-eE} at least one of $y_{\Psi_{22}}^R v_R$ or $\mu_{\Psi_{22}}$ must have masses at least an order of magnitude larger than the electroweak scale such that the charged lepton partners are sufficiently heavy phenomenologically.

Firstly in the scenario where $\vert \mu_\Psi \vert \gg \vert y_{\Psi_{22}}^R v_R,\,y_{\Psi_{22}}^L v_L,\,m_u \vert$ the light, charged-lepton masses are given by
\begin{equation}
m_\ell \simeq m_d - v_L v_R \,y_{\Psi_{22}}^L (\mu_{\Psi_{22}})^{-1} y_{\Psi_{22}}^R
\end{equation} 
and diagonalising \cref{eqn:appbineutmass} leads to two different pairs of pseudo-Dirac neutrinos with masses given at first order by
\begin{align}
m_{1,2} &\simeq \mu_{\Psi_{22}} &&\nu_{1,2} \simeq \frac{1}{\sqrt{2}} N_L \pm \frac{1}{\sqrt{2}} N_R^c \nonumber\\
m_{3,4} &\simeq m_u &&\nu_{3,4} \simeq \frac{1}{\sqrt{2}} \nu_L \pm \frac{1}{\sqrt{2}} \nu_R^c.
\end{align}
Therefore the lightest neutrino masses would be predicted to have masses and splittings similar to the up-quark sector. The above equation is true irrespective of the hierarchy between the non-dominant parameters $y_{\Psi_{22}}^R v_R,\,y_{\Psi_{22}}^L v_L$ and $m_u$. Therefore we find that this hierarchy of couplings does not lead to the desired seesaw required to explain the lightness of the active neutrino masses for any scale of PS breaking.

Turning to the alternate possibility where $\vert y_{\Psi_{22}}^R v_R \vert \gg \vert \mu_{\Psi_{22}},\,y_{\Psi_{22}}^L v_L,\,m_u \vert$ the charged lepton masses are now given by
\begin{equation}
\label{eqn:appbidoubchargelepmass}
m_\ell \simeq  y_{\Psi_{22}}^L v_L - \frac{m_d \mu_{\Psi_{22}}}{y_{\Psi_{22}}^R v_R} .
\end{equation} 
Here we find two different neutrino mass regimes depending on the hierarchy between the non-dominant parameters $\vert \mu_{\Psi_{22}} \vert $ and $\vert y_{\Psi_{22}}^L v_L\vert$.

For the hierarchy $\vert y_{\Psi_{22}}^R v_R \vert \gg \vert \mu_{\Psi_{22}} \vert > 2 \vert y_{\Psi_{22}}^L v_L \vert$ the physical neutrino states have masses given at first order by
\begin{align}
m_{1,2} &\simeq y_{\Psi_{22}}^R v_R &&\nu_{1,2} \simeq \frac{1}{\sqrt{2}} \nu_L \pm \frac{1}{\sqrt{2}} N_R^c \nonumber\\
m_{3,4} &\simeq \frac{m_u}{y_{\Psi_{22}}^R v_R}\left(\mu_{\Psi_{22}} \pm y_{\Psi_{22}}^L v_L\right) &&\nu_{3,4} \simeq \frac{1}{\sqrt{2}} N_L \pm \frac{1}{\sqrt{2}} \nu_R^c
\end{align}
where the above results are insensitive to the relative size of $\vert m_u\vert$ to $\vert y_{\Psi_{22}}^L v_L \vert$ and $\vert\mu_{\Psi_{22}}\vert$. Sufficiently light neutrinos now implies the ratio
\begin{equation}
\label{eqn:bidoubratiocase}
\frac{m_{\nu}}{m_u} \simeq \frac{\mu_{\Psi_{22}}}{y_{\Psi_{22}}^R v_R} \simeq 10^{-11}
\end{equation}
where we have estimated $m_\nu \lesssim 10^{-9}$ GeV from known neutrino upper mass limits and $m_u \simeq 170$ GeV in the case of the top quark. This consequently implies that 
\begin{equation}
m_\ell \simeq y_{\Psi_{22}}^L v_L
\end{equation} 
for the charged-lepton masses as the second term in \cref{eqn:appbidoubchargelepmass} will be highly suppressed. Therefore in order to recover the correct charged-lepton masses we require $y_{\Psi_{22}}^L v_L \simeq 1$ GeV for the tau lepton and as we have assumed $\vert \mu_{\Psi_{22}} \vert > \vert y_{\Psi_{22}}^L v_L \vert$ this necessarily implies 
\begin{equation}
\vert \mu_{\Psi_{22}} \vert> 1 \text{ GeV}
\end{equation}
and therefore 
\begin{equation}
\vert y_{\Psi_{22}}^R v_R\vert > 10^{11} \text{ GeV}
\end{equation}
using \cref{eqn:bidoubratiocase}. Therefore for $\vert \mu_{\Psi_{22}} \vert > \vert y_{\Psi_{22}}^L v_L \vert$, sufficiently light neutrino masses are only possible for very large $SU(2)_R$/PS breaking scales.

Alternatively the hierarchy $\vert y_{\Psi_{22}}^R v_R \vert \gg \vert y_{\Psi_{22}}^L v_L \vert > \frac{1}{2}\vert \mu_{\Psi_{22}} \vert$ leads to 
\begin{alignat}{2}
m_{1,2} &\simeq v_R \, y_{\Psi_{22}}^R  &&\nu_{1,2} \simeq \frac{1}{\sqrt{2}} \nu_L \pm \frac{1}{\sqrt{2}} N_R^c \nonumber\\
m_3 &\simeq \frac{v_L}{v_R} \frac{2 y_{\Psi_{22}}^L m_u}{y_{\Psi_{22}}^R} &&\nu_3 \simeq \nu_R^c \nonumber\\
m_4 &\simeq \frac{1}{v_L v_R} \frac{\mu_{\Psi_{22}}^2 m_u}{2\, y_{\Psi_{22}}^L y_{\Psi_{22}}^R}\qquad\qquad\qquad\qquad\qquad &&\nu_4 \simeq N_L
\end{alignat}
where again the results are insensitive to the overall size of $m_u$ as long as it is smaller than $y_{\Psi_{22}}^R v_R$. This scenario therefore leads to the neutrino mass hierarchy
\begin{equation}
m_{\nu_4} \ll m_{\nu_3} \ll m_{\nu_{1,2}}
\end{equation}
Sufficiently light neutrinos implies
\begin{equation}
\frac{m_\nu}{m_u} y_{\Psi_{22}}^L v_L \simeq 10^{-11}\text{ GeV}\simeq \frac{1}{2}\frac{\mu_{\Psi_{22}}^2}{y_{\Psi_{22}}^R v_R}
\end{equation}
for the same estimates. Once again this implies that the second term in \cref{eqn:appbidoubchargelepmass} will be subdominant and therefore the charged lepton masses are given by $m_\ell \simeq y_{\Psi_{22}}^L v_L$. Now light neutrinos are possible for low PS breaking scales for similar reasons to the usual inverse seesaw mechanism, for example for $y_{\Psi_{22}}^R v_R \simeq 10$ TeV we can estimate
\begin{equation}
\mu_{\Psi_{22}} \simeq 4.4 \times 10^{-3}\text{ GeV}.
\end{equation}
Therefore we find for a mass matrix of the form in \cref{eqn:appbineutmass}, only one hierarchy of parameters leads to a viable neutrino mass spectrum for low scales of PS breaking. As this requires the term $\mu_{\Psi_{22}}$ to be very small, this also conveniently leads to a helicity-suppression in the PS mediated rare meson decays, and therefore allows the limits for PS breaking to be significantly lowered as discussed in \cref{subsubsec:fermionspec-eE}.

In the multi-generational scenario we find the block diagonalisation of \cref{eqn:appbineutmass}, assuming the hierarchy $\Vert Y_{\Psi_{22}}^R v_R \Vert \gg \Vert Y_{\Psi_{22}}^L v_L \Vert > \Vert \mu_{\Psi_{22}} \Vert$\footnote{More accurately, the hierarchy between each matrix required for a low scale spectrum is given by $\Vert \mu_{\Psi_{22}} (y_{\Psi_{22}}^R v_R)^{-1} m_u\Vert < \Vert  v_L v_R \left(Y_{\Psi_{22}}^L (y_{\Psi_{22}}^R v_R)^{-1} m_u + [Y_{\Psi_{22}}^L (y_{\Psi_{22}}^R v_R)^{-1} m_u]^T\right) \Vert$ for non-commuting matrices. This should in general imply the hierarchy $\Vert \mu_{\Psi_{22}} \Vert < \Vert Y_{\Psi_{22}}^L v_L \Vert$ however. }, is given at first order by
\begin{align}
m_{1,2} &\simeq v_R \, Y_{\Psi_{22}}^R  \nonumber\\
m_3 &\simeq \frac{v_L}{v_R} \left( Y_{\Psi_{22}}^L (Y_{\Psi_{22}}^R)^{-1} m_u + m_u^T \left[Y_{\Psi_{22}}^L (Y_{\Psi_{22}}^R)^{-1}\right]^T\right) \nonumber\\
m_4 &\simeq \frac{1}{v_L v_R} \mu_{\Psi_{22}} \left[(Y_{\Psi_{22}}^R)^T (m_u^T)^{-1} Y_{\Psi}^L + (Y_{\Psi_{22}}^L)^T m_u^{-1} Y_{\Psi_{22}}^R \right]^{-1} \mu_{\Psi_{22}}^T
\end{align}
for each mass block where $Y_{\Psi_{22}}^{L/R},\,m_u$ and $\mu_{\Psi_{22}}$ are now matrices of appropriate dimension.

We estimate the relationship between the mass and interaction eigenstates for the relevant light neutrinos states to be
\begin{align}
\nu_4 \simeq &\,\,\,U_N^\dagger\left[\mu_{\Psi_{22}} (y_{\Psi_{22}}^R)^{-1}\right]^\dagger \nu_L^c -\frac{1}{2}U_N^\dagger\left[Y_{\Psi_{22}}^L(y_{\Psi_{22}}^R)^{-1}([y_{\Psi_{22}}^R]^\dagger)^{-1}\mu_{\Psi_{22}}^\dagger\right]^\dagger\nu_R\nonumber\\
&\qquad\qquad\qquad\qquad\qquad\qquad+U_N^\dagger\left[ \id -\frac{1}{2}\mu_{\Psi_{22}} (Y_{\Psi_{22}}^R)^{-1}([Y_{\Psi_{22}}^R]^\dagger)^{-1}\mu_{\Psi_{22}}^\dagger\right]^\dagger N_L^c\nonumber\\
\nu_3 \simeq & \,\,\,O_3^\dagger\left[Y_{\Psi_{22}}^L (Y_{\Psi_{22}}^R)^{-1}\right]^\dagger \nu_L^c -\frac{1}{2}O_3^\dagger\left[Y_{\Psi_{22}}^L(Y_{\Psi_{22}}^R)^{-1}([Y_{\Psi_{22}}^R]^\dagger)^{-1}\mu_{\Psi_{22}}^\dagger\right]^\dagger N_L^c + O_3^\dagger\left[m_u^T (Y_{\Psi_{22}}^R)^{T-1}\right]^\dagger N_R\nonumber\\
&\quad+ O_3^\dagger\left[ \id -\frac{1}{2}\left(Y_{\Psi_{22}}^L (Y_{\Psi_{22}}^R)^{-1}([Y_{\Psi_{22}}^R]^\dagger)^{-1}(Y_{\Psi_{22}}^L)^\dagger + m_u^T (Y_{\Psi_{22}}^R)^{T-1}([Y_{\Psi_{22}}^R]^*)^{-1}m_u^*\right)\right]^\dagger\nu_R
\end{align}
using a perturbative seesaw expansion where $U_N^\dagger$ and $O_3$ diagonalise the lightest and next-to-lightest mass blocks of the neutrino mass matrix after block diagonalisation. Therefore for the lightest neutrino states, $\nu_4$, the deviation from unitarity between its couplings to the SM charged leptons, which are predominantly made up of $E_L$ which weakly couples to $N_L$, is given at lowest order by the matrix
\begin{equation}
\eta \simeq \frac{1}{2} \mu_{\Psi_{22}}(Y_{\Psi_{22}}^R)^{-1} \left(\mu_{\Psi_{22}}(Y_{\Psi_{22}}^R)^{-1}\right)^\dagger.
\end{equation}
This determines the degree of non-unitarity in the physical PMNS matrix which stems from mixing effects within the neutral sector. As the fermion bi-doublet also introduced additional exotic charged-lepton states, there will additionally be mixing effects within the charged sector which leads to additional non-unitarity effects for the PMNS. Interestingly, as shown in \cref{subsubsec:bidoubcase} we find the deviation of non-unitarity within the charged lepton sector to be the same as in the neutral sector above implying
\begin{equation}
U_{\text{PMNS}} \simeq \left(\id - \eta\right)U_N^\dagger \left(\id - \eta\right)O_L^e.
\end{equation}

\begin{figure}[t]
\centering
{
  \includegraphics[width=0.4\linewidth]{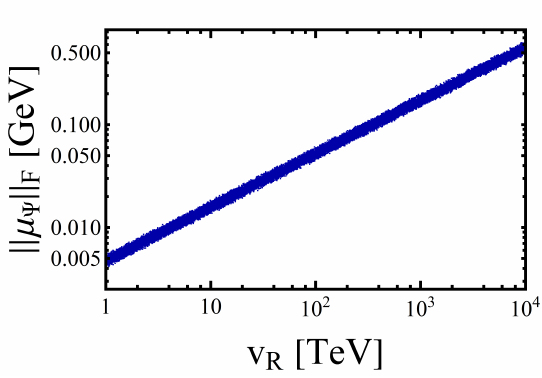}\hfill
}
{
  \includegraphics[width=0.48\linewidth]{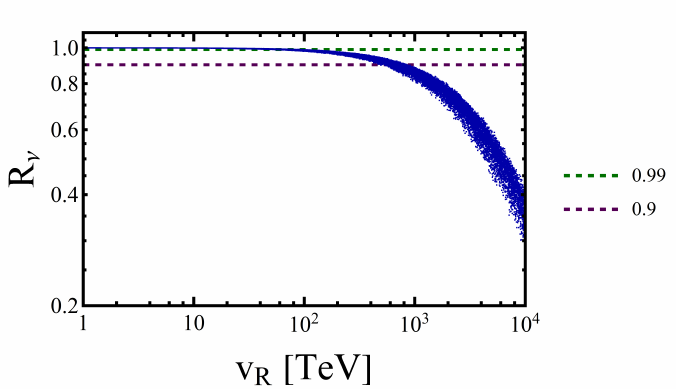}
}
\caption[Plot of generated $\mu_{\Psi_{22}}$ as a function of $v_R$ and its implications for the light neutrino masses.]{The Frobenius norm of the matrix $\mu_{\Psi_{22}}$ generated through a Casas-Ibarra parametrisation as a function of $v_R$ \textbf{(left)} and $R_\nu$ as a function of the same scale \textbf{(right)}, where $R_\nu$ is defined in \cref{eqn:bidoubrnudef}. Here we have randomly scanned over a degenerate spectrum for $Y_{\Psi_{22}}^R$ such that the heavy charged lepton partners have masses at around the scale $v_R$. For larger values of $v_R$ the generated $\mu_{\Psi_{22}}$ develops singular values larger than the required hierarchy $\Vert Y_{\Psi_{22}}^L \Vert > \Vert \mu_{\Psi_{22}} \Vert$ where $Y_{\Psi_{22}}^L v_L$ is fixed to give the correct SM charged-lepton masses and the resultant neutrino masses begin to disagree with experimental values.}
\label{fig:biddoubcasas}
\end{figure}

\Cref{fig:biddoubcasas} plots the norm of the required matrix $\mu_{\Psi_{22}}$ for a given choice of $v_R$ and $Y_{\Psi_{22}}^R$ which leads to a viable mass spectrum for the active neutrinos where we have fixed the lightest neutrino mass to 
\begin{equation}
m_{\nu_1} = 10^{-10} \text{ GeV}
\end{equation}
and assumed a normal ordering for the remaining masses. The matrix $Y_{\Psi_{22}}^R$ was randomly scanned over for a degenerate spectrum such that the singular values of the matrix were within an order of magnitude to each other. The $SU(2)_R$ breaking scale $v_R$ was scanned between $1-10^4$ TeV and $\mu_{\Psi_{22}}$ was fixed by
\begin{equation}
\mu_{\Psi_{22}} = U_\nu^* m_\nu^{1/2} A \left[Y_L m_u^{-1} Y_{\Psi_{22}}^R + (Y_{\Psi_{22}}^R)^T (m_u^T)^{-1} Y_L \right]^{1/2}
\end{equation}
through a Casas-Ibarra parametrisation~\cite{Casas:2001sr}. Here $U_\nu$ corresponds to the unitary matrix which diagonalises the lightest neutrino mass block, $m_\nu$ is a diagonal matrix composed of the assumed neutrino masses and $A$ is a random orthogonal matrix. The left plot of \cref{fig:biddoubcasas} shows that as $v_R$ increases, the singular values of $\mu_{\Psi_{22}}$ must increase for a fixed neutrino mass scale. As a result at around $10^3$ TeV, the singular values of $\mu_{\Psi_{22}}$ become larger than $Y_{\Psi_{22}}^L$ spoiling the required hierarchy of scales for light neutrinos at low scales. The right plot of \cref{fig:biddoubcasas} plots the same scale as a function of $R_\nu$ where
\begin{equation}
\label{eqn:bidoubrnudef}
R_\nu = \prod_{i} \frac{m_i}{m_i^{\text{input}}}
\end{equation}
where $m_i$ corresponds to the mass of the i-th generation of the three active neutrinos numerically calculated after diagonalising the mass matrix and $m_i^{\text{input}}$ to the masses inputed via the Casas-Ibarra parameterisation. Clearly larger scales for $v_R$ are unable to reproduce the correct neutrino masses. For all points considered the light states from the charged-lepton mass matrix reproduce the charged-lepton masses with $m_e \simeq Y_{\Psi_{22}}^L$ to a high degree of accuracy. Therefore the neutral- and charged-lepton mass matrices generated in the scenario with additional fermionic bi-doublets are viable candidates for a low-scale PS setup provided that 
\begin{equation}
y_{\Psi_{22}}^R v_R \lesssim 10^3 \text{ TeV.}
\end{equation}

\subsection{Lepton masses with additional triplets}
\label{subsec:tripneutmass}

The neutrino mass matrix which arises with the addition of $SU(2)_{L/R}$ triplets is given by the usual linear/inverse seesaw:
\begin{equation}
\label{eqn:apptripneutmass}
\mathcal{L}_{\nu N}=\frac{1}{2}\begin{pmatrix}
\overline{\nu_L} & \overline{\nu_R^c}  & \overline{N_R^c} 
\end{pmatrix} \begin{pmatrix}
0 & m_u  & \mathlarger{y}_{\Psi_3}^L v_L\\
m_u & 0  & \mathlarger{y}_{\Psi_3}^R v_R^*\\
\mathlarger{y}_{\Psi_3}^L v_L & \mathlarger{y}_{\Psi_3}^R v_R^* & \mu_{\Psi}  
\end{pmatrix}
\begin{pmatrix} 
\nu_L^c\\
\nu_R\\
N_R
\end{pmatrix}
\end{equation}
where we have labelled the fields as in \cref{eqn:tripemb} and the PS symmetry enforces that the singular values of $m_u$ are given by the up-quark masses. As discussed in \cref{sec:tripPS,subsubsec:fermionspec-eE}, both $\mathlarger{y}_{\Psi_3}^L v_L$ and $\mathlarger{y}_{\Psi_3}^L v_R$ are required in order to generate a phenomenologically valid charged lepton mass spectrum. Both terms will not be present if the usual scalars $\rchi_{L/R}$ are assumed and therefore require a more exotic choice of scalars. For example if $\rchi_L \sim (\textbf{4},\textbf{2},\textbf{1})$ is replaced with $\rchi' \sim (\textbf{4},\textbf{2},\textbf{3})$ in the case of the $SU(2)_R$ triplet as was done in~\cite{Balaji:2019kwe}, the missing mass term in the Yukawa Lagrangian will now be generated. We will therefore assume in what follows that the scalar spectrum is such that all the terms appearing in \cref{eqn:apptripneutmass} are present. Consequently the charged lepton mass matrix will be given by 
\begin{equation}
\label{eqn:apptripchargelepmixR}
\mathcal{L}_{eE} = \begin{pmatrix}
\overline{e_L} & \overline{E_L}
\end{pmatrix} \begin{pmatrix}
m_d & \sqrt{2}\, y_{\Psi_3}^L v_L \\
\sqrt{2}\, y_{\Psi_3}^R v_R^* & \,\,\,\mu_{\Psi_3}
\end{pmatrix}
\begin{pmatrix}
e_R\\
E_R
\end{pmatrix}
\end{equation}
where now the top-right entry is non-zero and therefore a valid spectrum of charged-lepton masses is possible, unlike before.

Unlike the usual low-scale PS setup, as described in \cref{subsec:YukPS}, the neutral lepton mass mixing matrix is related to a charged lepton mass mixing matrix and a viable neutrino mass spectrum may not be possible whilst simultaneously generating a viable charged-lepton mass spectrum. As in \cref{subsec:bidoubneutmass} we analyse all possible hierarchies of parameters in the two mass matrices in order to establish whether a viable charged-lepton and neutrino mass spectrum can be simultaneously generated for the SM-like states for low scales of PS breaking. 

In the scenario where $\mu_{\Psi_3}$ is the dominant block in each mass matrix, there are multiple different possible neutrino mass regimes depending on the hierarchy between the non-dominant parameters.

Firstly if $\vert \mu_{\Psi_3} \vert > \vert m_u \vert > \vert \frac{(y_{\Psi_3}^L v_L)^2}{\mu_{\Psi_3}}, \frac{(y_{\Psi_3}^R v_R)^2}{\mu_{\Psi_3}} \vert$ is satisfied the mass states are given at first order by 
\begin{align}
m_{1} &\simeq \mu_\Psi &&\nu_{1,2} \simeq  N_R^c \nonumber\\
m_{2,3} &\simeq m_u &&\nu_{3,4} \simeq \frac{1}{\sqrt{2}} \nu_L \pm \frac{1}{\sqrt{2}} \nu_R^c
\end{align}
and therefore this hierarchy is ruled out as the light neutrinos would be pseudo-Dirac with masses comparable to the up-quark sector.

If instead $\vert \mu_{\Psi_3} \vert > \vert \frac{(y_{\Psi_3}^R v_R)^2}{\mu_{\Psi_3}} \vert > \vert m_u \vert $ is satisfied the mass spectrum is now given by 
\begin{align}
\label{eqn:invmudom}
m_{1} &\simeq \mu_{\Psi_3} &&\nu_{1,2} \simeq  N_R^c \nonumber\\
m_{2} &\simeq  \frac{(y_{\Psi_3}^R)^2 v_R^2}{\mu_{\Psi_3}} &&\nu_{2} \simeq  \nu_R^c \nonumber\\
m_{3} &\simeq \left\vert \frac{1}{v_R^2}\left(\frac{m_u}{y_{\Psi_3}^R}\right)^2 \mu_{\Psi_3} - \frac{2v_L}{v_R}\frac{y_{\Psi_3}^L  m_u}{y_{\Psi_3}^R } \right\vert&&\nu_{3} \simeq \nu_L 
\end{align}
which is rather interesting as the lightest mass state contains a term linearly (rather than inversely) proportional to the dominant term in the seesaw but is still small due to the hierarchy in the non-dominant parameters. The SM charged lepton has a mass given by 
\begin{equation}
m_\ell \simeq m_d - v_L v_R\frac{y_{\Psi}^L y_{\Psi}^R}{\mu_\Psi}
\end{equation}
by diagonalising \cref{eqn:apptripchargelepmixR}. These equations can be solved for $y_{\Psi_3}^L v_L$ and $y_{\Psi_3}^R v_R$ as a function of $\mu_{\Psi_3}$ for the required lepton masses. In the case of the third generation where $m_d \simeq 4.2$ GeV, $m_u \simeq 173$ GeV, $m_\ell \simeq 1.7$ GeV and setting $m_\nu \simeq 10^{-10}$ GeV leads to 
\begin{equation}
\frac{(y_{\Psi_3}^L v_L)^2}{\mu_{\Psi_3}} \simeq 10^{-14} \text{ GeV} \qquad\text{and}\qquad\frac{(y_{\Psi_3}^R v_R)^2}{\mu_{\Psi_3}} \simeq 10^{14} \text{ GeV}
\end{equation}
which implies that $\mu_\Psi > 10^{14}$ GeV by the initial seesaw assumption, therefore requiring both a viable charged-lepton and neutrino mass spectrum with $\mu_\Psi$ dominant is only possible for high scale PS breaking scenarios. The final possible hierarchy where $\vert \mu_{\Psi_3} \vert > \vert \frac{(y_{\Psi_3}^L v_L)^2}{\mu_{\Psi_3}} \vert > \vert m_u \vert $ is also unviable for similar arguments.

The alternative scenario where $\vert y_{\Psi_3}^R v_R \vert$ is the dominant term in the mass matrix recovers the usual inverse/linear seesaw
\begin{align}
\label{eqn:invyrdom}
m_{1,2} &\simeq  y_{\Psi_3}^R v_R &&\nu_{1,2} \simeq  \frac{1}{\sqrt{2}} N_R^c \pm \frac{1}{\sqrt{2}} \nu_R^c \nonumber\\
m_{3} &\simeq \left\vert \frac{1}{v_R^2}\left(\frac{m_u}{y_{\Psi_3}^R}\right)^2 \mu_{\Psi_3} - \frac{2v_L}{v_R}\frac{y_{\Psi_3}^L m_u}{y_{\Psi_3}^R} \right\vert &&\nu_{3} \simeq \nu_L.
\end{align}
and the lightest charged-lepton mass is given by
\begin{equation}
m_\ell \simeq v_L y_{\Psi}^L - \frac{1}{v_R}\frac{m_d \mu_\Psi }{y_{\Psi}^R}.
\end{equation}
The lightest neutrino state is now suppressed by the scale $y_{\Psi_3}^R v_R$ as usual, it is quite interesting that the lightest neutrinos in \cref{eqn:invmudom,eqn:invyrdom} have identical masses for completely different hierarchies of parameters.

For three generations of each fermion multiplet the light mass blocks are given by
\begin{equation}
\label{eqn:apptripneutmassinv}
m_\nu \simeq \frac{1}{v_R^2} m_u [(Y_{\Psi_3}^R)^T]^{-1} \mu_{\Psi_3} (Y_{\Psi_3}^R)^{-1} m_u^T - \frac{v_L}{v_R} m_u [(Y_{\Psi_3}^R)^T]^{-1} (Y_{\Psi_3}^L)^T - \frac{v_L}{v_R}  Y_{\Psi_3}^L (Y_{\Psi_3}^R)^{-1} m_u^T
\end{equation}
for the neutrinos and
\begin{equation}
\label{eqn:apptripchargemassinv}
m_\ell \simeq Y_{\Psi_3}^L v_L - \frac{1}{v_R} m_d (Y_{\Psi_3}^R)^{-1} \mu_{\Psi_3}
\end{equation}
for the charged leptons.

\begin{figure}[t]
\centering
{
  \includegraphics[width=0.41\linewidth]{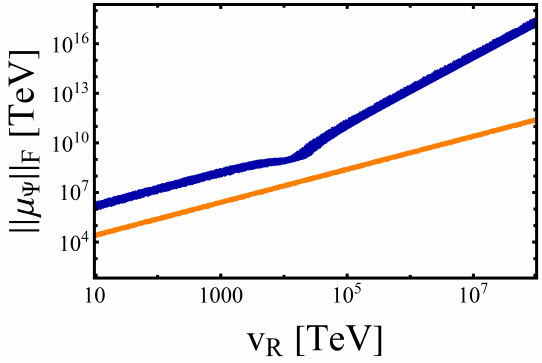}\hfill
}
{
  \includegraphics[width=0.48\linewidth]{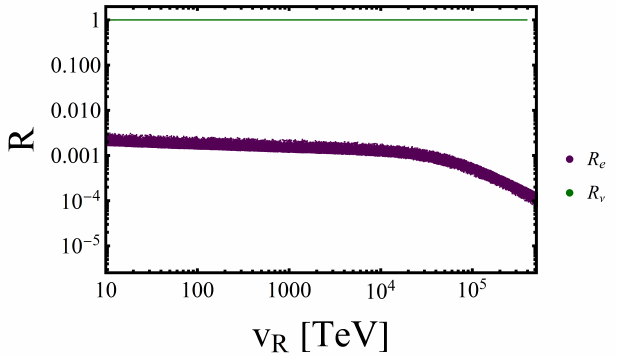}
}
\caption[Required $\mu_{\Psi_3}$ as a function of $v_R$ and the implications for the charged-lepton and light-neutrino masses.]{Plot of the Frobenius norm of $\mu_{\Psi_3}$ generated by solving \cref{eqn:apptripneutmass,eqn:apptripchargelepmixR} \textbf{(left)} as a function of the scale $v_R$ with a random scan over $Y_{\Psi_3}^R$ with degenerate singular values and a plot of $R_{\nu/e}$ \textbf{(right)}, defined in \cref{eqn:triprdef}, as a function of the same scale. The orange line in the left plot corresponds to $\Vert \mu_{\Psi_3} \Vert_F = \Vert Y_{\Psi_3}^R \Vert_F v_R$ and therefore at all scales considered the required size of $\mu_{\Psi_3}$ spoils the assumed hierarchy of parameters in order to achieve an inverse/linear seesaw. Though $R_\nu \simeq 1$, the resultant charged-lepton masses after diagonalisation are signficantly differnet to their input values and therefore we find that additional triplets to the usual PS fermions alone are not sufficient to lead to a viable mass spectrum for all SM fermions (due to the quark-lepton symmetry assumed). }
\label{fig:invlinplots}
\end{figure}

For a given choice of $v_R Y_{\Psi_3}^R$, the unknown matrices $v_L Y_\Psi^L$ and $\mu_\Psi$ are fixed by solving the above equations. Due to the complexity of solving the above matrix equations we solve \cref{eqn:apptripneutmassinv,eqn:apptripchargemassinv} numerically where we assume a normal ordering for the light neutrino masses and fix the lightest neutrino mass to
\begin{equation}
m_{\nu_1} = 10^{-10} \text{ GeV}.
\end{equation}
The left plot in \cref{fig:invlinplots} plots the Frobenius norm of the required matrix $\mu_{\Psi_3}$ (in blue) required in order to solve the two above equations for different choices of $v_R Y_{\Psi_3}^R$. The orange line corresponds to $\Vert \mu_{\Psi_3} \Vert_F = \Vert v_R Y_{\Psi_3}^R \Vert_F $ and for any choice of $v_R Y_{\Psi_3}^R$ the required size of $\mu_{\Psi_3}$ in order to solve the above equations is larger, destroying the hierarchy assumed in deriving \cref{eqn:apptripneutmassinv,eqn:apptripchargemassinv}. The right plot of \cref{fig:invlinplots} plots
\begin{equation}
\label{eqn:triprdef}
R_{\nu/e} = \prod_{i} \frac{m_i}{m_i^{\text{input}}} 
\end{equation} 
where $m_i$ corresponds to the numerically computed mass for the i-th generation of lepton and $ m_i^{\text{input}}$ corresponds to the experimental values for the masses inputted (run up to the appropriate scale). The charged lepton masses which result are significantly departed from their SM values whilst the neutrino values are well predicted, this is a coincidence due to the light neutrino mass block having the same formula regardless of the hierarchy\footnote{$\Vert Y_{\Psi_3}^R v_R \Vert > \Vert m_u \Vert$ is always satisfied in our scan.} between $\Vert \mu_{\Psi_3} \Vert$ and $\Vert Y_{\Psi_3}^R v_R \Vert$ as can be seen in \cref{eqn:invmudom,eqn:invyrdom}. 

The light, charged-lepton mass block however is different depending on the hierarchy and we therefore find that due to the SM parameters $m_u$ and $m_d$ entering \cref{eqn:apptripneutmass,eqn:apptripchargelepmixR} as well as requiring viable correct charged-lepton and neutrino masses upon diagonalisation is not possible for low seesaw scales (for the entries of $Y_{\Psi_3}^R v_R $ which we scanned over). Though unlikely, it may be possible that special textures within $Y_{\Psi_3}^R v_R $ could lead to valid charged-lepton and neutrino masses for the appropriate hierarchy of block matrices. Therefore the addition of triplets to the usual PS multiplets will in general require additional physics, in~\cite{Balaji:2019kwe} for example an additional fermionic singlet $S_L$ was introduced and a viable mass spectrum for the leptons was recovered. The addition of both an $SU(2)_L$ and $SU(2)_R$ triplet was found to lead to an unviable mass spectrum for the scalars $\rchi_L$ and $\rchi_R$, if more exotic scalars were introduced such that all mass terms in the Yukawa sector were generated, a viable mass spectrum would likely be generated for a specific hierarchy of parameters which may or may not lead to a chiral suppression of the $X_\mu$ mass limits as with the bi-doublet. In this case however the neutrino mass matrix would not be given by an inverse/linear seesaw and would be similar to \cref{eqn:tripneutlepmixL&R} with additional entries.

\section{Running of Gauge and Yukawa couplings}
\label{sec:Appendix C}

In order to estimate the running of the relevant SM parameters up to the potential scale of PS breaking we solve the simultaneous differential equation
\begin{equation}
\label{eqn:SMrunning}
\mu \frac{d x}{d \mu} = \frac{1}{(4\pi)^2}\beta_x^{(1)} + \dots
\end{equation}
where for simplicity we restrict ourselves to the one-loop renormalisation group equations. We make two reasonable assumptions on the running of each parameter: the exotic PS particle content does not significantly affect the running of the SM parameters below the scale of PS breaking we are interested in (due to their masses) and secondly as a simplifying assumption, we only include the contribution from the top-quark Yukawa coupling and gauge couplings in the evaluation of all relevant parameters due to its dominant role.

The beta functions $\beta_x^{(i)}$ have been extensively studied for the SM~\cite{Machacek:1983fi,Balzereit:1998id,Luo:2002ey,JuarezWysozka:2002kx}, at one-loop the beta functions which are non-zero when only the gauge and top-quark couplings are included\footnote{Additionally as in~\cite{Balzereit:1998id}, terms smaller than $\mathcal{O}(10^{-3})$ in $\beta_x/x$ were approximated to zero for simplicity.}   are
\begin{align}
\label{eqn:SMrunningparam}
\beta_{g_i}^{(1)} &= b_i \big[ g_i(\mu)\big]^3\nonumber\\
\beta_{y_{d,i}}^{(1)} &= \mphant{\mathlarger{y}_{d,i}}{\dagger}(\mu) \left( a_i \big[\mphant{\mathlarger{y_{u,3}}}{\dagger}(\mu)\big]^2 - \left\{\frac{1}{4}\big[g_1(\mu)\big]^2 + \frac{9}{4}\big[g_2(\mu)\big]^2 + 8\big[g_3(\mu)\big]^2 \right\}  \right)\nonumber\\
\beta_{y_{\ell,i}}^{(1)} &= \mphant{\mathlarger{y_{\ell,i}}}{\dagger}(\mu)\,\left( 3 \,\big[\mphant{\mathlarger{y_{u,3}}}{\dagger}(\mu)\big]^2  - \frac{9}{4}\left\{ \big[g_1(\mu)\big]^2 + \big[g_2(\mu)\big]^2\right\}\,\right)\nonumber\\
\beta_{y_{u,i}}^{(1)} &= \mphant{\mathlarger{y}_{u,i}}{\dagger}(\mu)\left((3a_i-2c_i)\big[\mphant{\mathlarger{y_{u,3}}}{\dagger}(\mu)\big]^2 - \left\{\frac{17}{20}\big[g_1(\mu)\big]^2 + \frac{9}{4}\big[g_2(\mu)\big]^2 + 8\big[g_3(\mu)\big]^2 \right\}\right)\nonumber\\
\beta_{\theta_i}^{(1)} &= \theta_i(\mu) \left( c_i \,\big[\mphant{\mathlarger{y_{u,3}}}{\dagger}(\mu)\big]^2 \right).
\end{align}
Here $y_{d,i}= (y_d,\,y_s ,\,y_b )$ and similarly for the charged-lepton and up-type Yukawa couplings, $\theta_i   = (\theta_{13} ,\,\theta_{23} ,\,\theta_{12})$ are the CKM mixing angles and
\begin{equation}
a_i = \begin{dcases}
		3 & i=1,2\\
		\frac{3}{2} & i=3\\
	 \end{dcases},\qquad b_i = \begin{dcases}
		\hphantom{-}\frac{41}{10} & i=1\\
		-\frac{19}{6} & i=2\\
		\,\,\,-7 & i=3
	 \end{dcases}\quad\text{and}\quad c_i = \begin{dcases}
		3 & i=1,2\\
		0 & i=3\\
	 \end{dcases}
\end{equation}
where the gauge coupling $g_1$ is given as it usually is in $SU(5)$ normalisation: $g_1^2 = 5/3 g_Y^2$. The beta functions (at one-loop) for the CP-violating CKM parameter $\delta_{CKM}$ and all the parameters related to the neutrino sector are zero for our stated assumptions and therefore will not run. The parameters in \cref{eqn:SMrunningparam} are run up from $\mu = m_Z$ with initial values taken from~\cite{Antusch:2013jca} which utilised the Mathematica package RunDec to evolve the QCD parameters up to $m_Z \simeq 91.19$ GeV in the $\overline{\text{MS}}$ scheme and are summarised in \cref{table7}.

\begin{figure}[t]
\centering
{
  \includegraphics[width=0.48\linewidth]{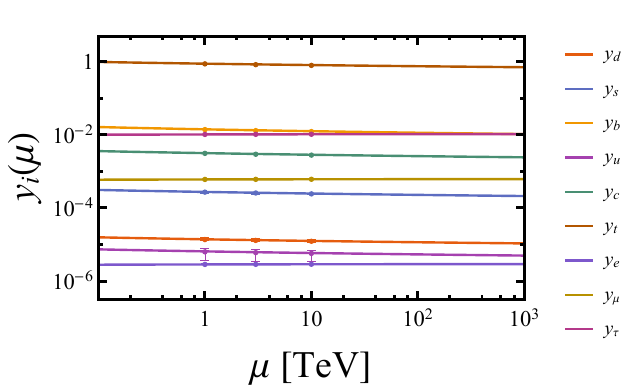}
}
{
  \includegraphics[width=0.48\linewidth]{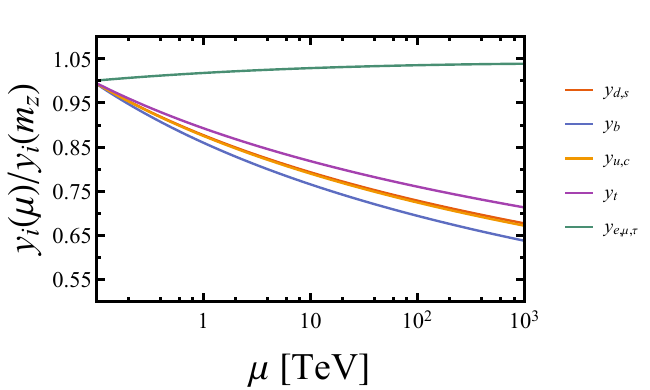}
}\\
{
  \includegraphics[width=0.48\linewidth]{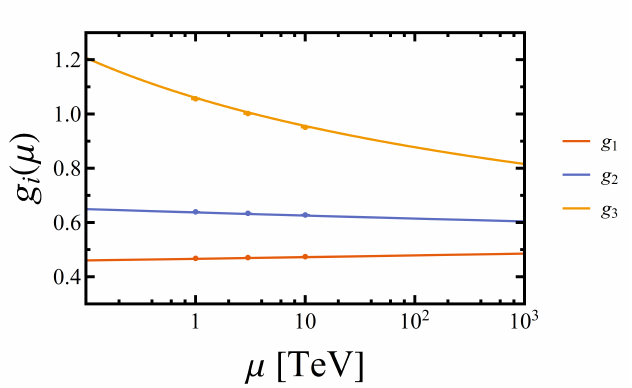}
}
{
  \includegraphics[width=0.48\linewidth]{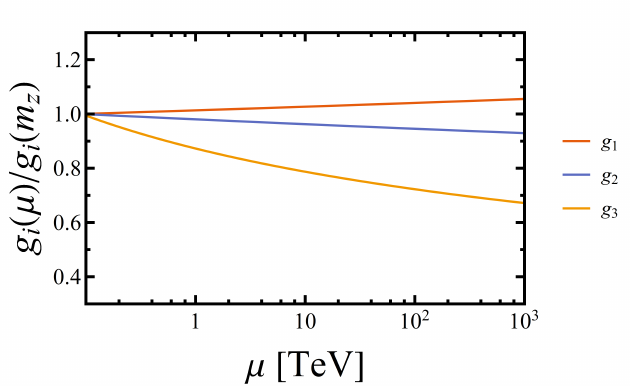}
}\\
{
  \includegraphics[width=0.48\linewidth]{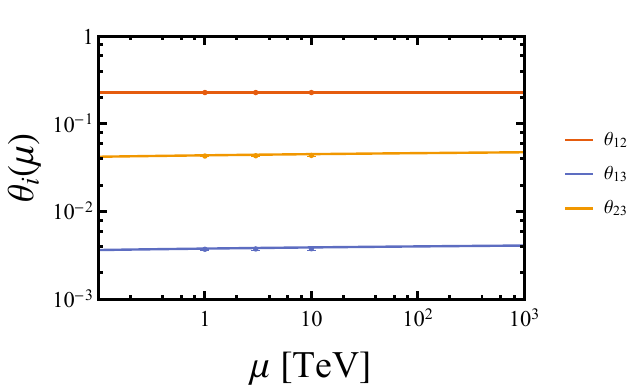}
}
{
  \includegraphics[width=0.48\linewidth]{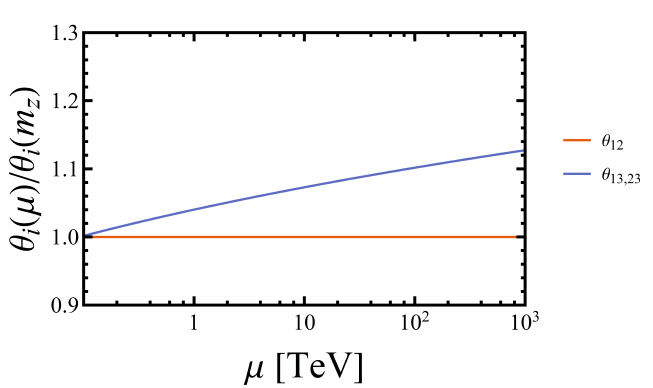}
}
\caption[Plot of the running of SM parameters as a function of $\mu$ using one-loop RGE equations.]{Plot of the numerical running of various SM parameters \textbf{(left)} (run up from $\mu = m_Z$) at one-loop as a function of the energy scale $\mu$. The points correspond to values taken from~\cite{Antusch:2013jca} at $1,\,3$ and $10$ TeV with their respective $1\sigma$ uncertainties, where the running was performed to two-loops to demonstrate the accuracy of our one-loop approximations. Also shown are the ratios \textbf{(right)} of the run parameter against its initial value at $\mu = m_Z$.}
\label{figure:runningSMparameters}
\end{figure}

The results of \cref{eqn:SMrunning,eqn:SMrunningparam} are shown in \cref{figure:runningSMparameters} for the Yukawa couplings, gauge couplings and relevant CKM parameters run from $\mu = m_Z$ up to $\mu = 1000$ TeV. The points on the left plots correspond to the results obtained using RunDec at two-loops~\cite{Antusch:2013jca} at $1,\,3$ and $10$ TeV in the SM. The plots on the right correspond to the ratio of the running parameter to its initial value at $\mu = m_Z$. The quark Yukawa couplings all significantly decrease as the energy scale increases whereas the charged-lepton Yukawa modestly increase. Due to the approximations we assume in \cref{eqn:SMrunningparam}, the ratios $\beta_{x}^{(1)}/x$ are equal for the first and second generations of quark Yukawas, all three generations of charged-lepton Yukawas as well as between $\theta_{13}$ and $\theta_{23}$.

As is often chosen, we work in a basis where the Yukawa coupling matrix for the up-type quarks and charged leptons is diagonal. Therefore the running of each relevant Yukawa coupling matrix, for the SM fermions, at a given energy scale is given by
\begin{align}
Y_u(\mu) &= \text{diag}\big[y_u(\mu),\,y_c(\mu),\,y_t(\mu)\big]\nonumber\\
Y_d(\mu) &= V_{\textsc{CKM}}^\dagger\big[\theta_{12}^q(\mu),\,\theta_{23}^q(\mu),\,\theta_{13}^q(\mu),\delta_{\textsc{CKM}}(\mu)\big]\,\text{diag}\big[y_d(\mu),\,y_s(\mu),\,y_b(\mu)\big]\nonumber\\
Y_\ell(\mu) &= \text{diag}\big[y_e(\mu),\,y_\mu(\mu),\,y_\tau(\mu)\big]
\end{align}
where we use the numerical results from solving \cref{eqn:SMrunning,eqn:SMrunningparam} using \cref{table7} as the initial values.

\begin{table}[t]
\begin{center}
{\renewcommand{\arraystretch}{1.15}
\scalebox{1.0}{
\begin{tabular}{cc}
\toprule
 & $\mu = m_Z$ 	 \\ 
\midrule 
\midrule
$g_1$ & $0.461425^{+0.000044}_{-0.000043}$ \\
$g_2$ & $0.65184^{+0.00018}_{-0.00017}$ \\
$g_3$ & $1.2143^{+0.000044}_{-0.000043}$ \\
\bottomrule
\end{tabular}\,\,\,
\begin{tabular}{cc}
\toprule
 & $\mu = m_Z$ 	 \\ 
\midrule 
\midrule
$y_u$ & $7.4^{+1.5}_{-3.0} \times 10^{-6}$\\
$y_c$ & $3.6^{+0.11}_{-0.11} \times 10^{-3}$\\
$y_t$ & $0.9861^{+0.0086}_{-0.0087} $\\
\midrule
$y_d$ & $1.58^{+0.23}_{-0.10} \times 10^{-5}$\\
$y_s$ & $3.12^{+0.17}_{-0.16} \times 10^{-4}$\\
$y_b$ & $1.639^{+0.015}_{-0.015} \times 10^{-2}$\\
\midrule
$\theta_{12}^q$ & $0.22735^{+0.00072}_{-0.00071}$\\
$\theta_{23}^q$ & $4.208^{+0.064}_{-0.064}\times 10^{-2}$\\
$\theta_{13}^q$ & $3.64^{+0.13}_{-0.13}\times 10^{-3}$\\
$\delta_{\textsc{CKM}}$ & $1.208^{+0.054}_{-0.054}$\\
\bottomrule
\end{tabular}
\,\,\,
\begin{tabular}{cc}
\toprule
 & $\mu = m_Z$ 	 \\ 
\midrule 
\midrule
$y_e$ & $2.79^{+0.000015}_{-0.000016} \times 10^{-6}$\\
$y_\mu$ & $5.90^{+0.000019}_{-0.000018} \times 10^{-4}$\\
$y_\tau$ & $1.00^{+0.00090}_{-0.00091} \times 10^{-2}$\\
\midrule
$\theta_{12}^\ell$ & $0.59^{+0.0136}_{-0.0133}$\\
$\theta_{23}^\ell$ & $0.84^{+0.0192}_{-0.0332}$\\
$\theta_{13}^\ell$ & $0.15^{+0.0023}_{-0.0023}$\\
$\delta_{\textsc{PMNS}}$ & $3.87^{+0.6632}_{-0.4887}$\\
\midrule
$\Delta m_{21}^2/\text{eV}^2$ & $7.39^{+0.21}_{-0.20} \times 10^{-5}$\\
$\Delta m_{31}^2/\text{eV}^2$ & $2.523^{+0.032}_{-0.030} \times 10^{-3}$\\
\bottomrule
\end{tabular}
}
}
\end{center}
\caption[Values of SM parameters used at initial boundary of $\mu=m_Z$ found using $\overline{\text{MS}}$ scheme.]{Values used for the SM parameters at $\mu = m_Z$ in the $\overline{\text{MS}}$ scheme taken from table 1 of~\cite{Antusch:2013jca} for the gauge couplings, quark and charged-lepton parameters with their respective $1\sigma$ uncertainty. The neutrino parameters with their $1\sigma$ uncertainty are taken from~\cite{Esteban:2018azc}, which are not measured at $\mu = m_Z$, however as the PMNS parameters do not significantly run at low energies~\cite{Antusch:2005gp} (in the SM) we take these values to be valid at this scale.}
\label{table7}
\end{table}

\FloatBarrier
\bibliographystyle{JHEP}
\bibliography{references}

\end{document}